\documentclass[12pt,cite,epsfig]{article}
\usepackage{times,epsfig}
\usepackage{caption}
 \usepackage{amsmath}	   % faire des espaces negatif \!,\!!!
\newcommand{\nc}{\newcommand}

\nc{\bfEPJC}{ }

\nc{\bb}{\bibitem}
\nc{\be}{\begin{equation}}
\nc{\ee}{\end{equation}}
\nc{\pa}{\partial}
\nc{\parsym} {\stackrel{\leftrightarrow}{\pa}}
\nc{\ra}{\rightarrow}
\nc{\la}{\leftarrow}
\nc{\etp}{{\eta^\prime}}
\nc{\epspr}{\epsilon^\prime}
\nc{\omg}{\omega}
\nc{\ggam}{\gamma \gamma}
\nc{\cgam}{\gamma \gamma \gamma}
\nc{\gam}{\gamma }
\nc{\hvar}{\epsilon}
\nc{\epsProd} {$\epsilon \epsilon^\prime$}
\nc{\epsProdRel} {$\epsilon \epsilon^\prime=-\epsilon_0^2 \sin{2 \theta_P}$}
\nc{\tlambda}{\widetilde{\lambda}}
\nc{\bea}{\begin{eqnarray}}
\nc{\eea}{\end{eqnarray}}
\nc{\beas}{\begin{eqnarray*}}
\nc{\eeas}{\end{eqnarray*}}
\nc{\non}{\nonumber}
\nc{\second}{{\prime\prime}}
\nc{\ep}{$e^+e^-\ra\pi^+\pi^-\;$}
\nc{\indentB}{\indent \indent}
\def\hhhd{\rule[-3.mm]{0.mm}{2.mm}}
\def\hhhe{\rule[-3.mm]{0.mm}{3.mm}}

\def\hhhv{\rule[-3.mm]{0.mm}{9.mm}}
\def\hhhw{\rule[-1.mm]{0.mm}{5.mm}}
\def\hhhq{\rule[-5.mm]{0.mm}{12.mm}}

\newcommand{\epo}{\;.}

\newcommand{\ba}{\begin{eqnarray}}
\newcommand{\ea}{\end{eqnarray}}

\nc{\E}{{\rm e} }
\nc{\cF}{{\cal F} }
\nc{\cM}{{\cal M} }
\newcommand{\semis}{\;; }
\newcommand{\crn}{\nonumber \\}

\begin{document}
\begin{titlepage}
\vbox{~~~ \\
%                                   \null \hfill LPNHE 2010--XX\\
                                   \null \hfill DESY 21--079\\
                                   \null \hfill HU-EP 21/13\\
				       
\title{BHLS$_2$ Upgrade~: $\tau$ spectra, muon HVP \\ and the [$\pi^0,~\eta,~\etp$] System 
%\\DRAFT 2.1 
}
 \author{
M.~Benayoun$^a$, L.~DelBuono$^a$, F.~Jegerlehner$^{b,c}$ \\
\small{$^a$ LPNHE des Universit\'es Paris VI et Paris VII, IN2P3/CNRS, F--75252 Paris, France }\\
\small{$^b$ Humboldt--Universit\"at zu Berlin, Institut f\"ur Physik, Newtonstrasse 15, D--12489 Berlin,
Germany }\\
\small{$^c$ Deutsches  Elektronen--Synchrotron (DESY), Platanenallee 6, D--15738 Zeuthen, Germany}
%%%% {\bf to be submitted to EPJ C.} \\
}
\date{\today}
\maketitle
\begin{abstract}

%Original tex file and Maple codes in /pbs/home/benayoun/MAP/$HLS2019$
%Ongoing manuscript in /pbs/home//benayoun/latex.files//COVI-19/final-Covi19.tex

The generic Hidden Local Symmetry (HLS) model has recently given rise to its
BHLS$_2$ variant, defined by introducing symmetry breaking  mostly 
in  the vector meson sector; the central mechanism is a modification of the 
covariant derivative at the root of the HLS approach. However,  the description
of the $\tau$ dipion spectra, especially the Belle one, is not fully satisfactory
whereas the simultaneous dealing with its annihilation sector 
($e^+ e^- \ra \pi^+ \pi^-/\pi^+ \pi^-\pi^0/ \pi^0 \gam/\eta \gam/K^+ K^-/K_L K_S$)
is optimum. We show that this issue is solved  by means of an additional breaking term 
which also allows to consistently include the mixing properties of the $[\pi^0,\eta,\etp]$
system within this extended BHLS$_2$ (EBHLS$_2$) scope. 
This mechanism, an extension of the  usual 't~Hooft determinant term, only
affects the  kinetic energy part of the BHLS$_2$ Lagrangian. 
One thus obtains a fair account for the $\tau$ dipion spectra which complements 
the fair account of the annihilation channels already reached. 
The Belle dipion spectrum is found
to provide evidence in favor of a violation of CVC in the $\tau$ lepton decay; this
evidence is enforced by imposing the conditions 
$<0|J_\mu^q |[q^\prime \overline{q^\prime}](p)>=ip_\mu f_q \delta_{q q^\prime },
 \{ [q \overline{q}], q=u,d,s\}$ on  EBHLS$_2$ axial current matrix elements.
EBHLS$_2$ is found to recover the usual (completed) 
formulae for the [$\pi^0,~\eta,~\etp$] mixing parameters  and the global
fits return mixing parameter values in agreement with expectations and 
better uncertainties. Updating the muon HVP,  one also 
argues that the strong tension 
between the  KLOE and BaBar pion form factors imposes to provide  two solutions~:
$a_\mu^{HVP-LO}({\rm KLOE})=687.48 \pm 2.93$
and $a_\mu^{HVP-LO}({\rm BaBar})=692.53 \pm 2.95$,
in units of $10^{-10}$, rather than some combination of these. Taking into account 
common systematics, their differences with the experimental BNL-FNAL average value
exhibit  significances  $> 5.4\sigma$ (KLOE) and  $> 4.1\sigma$ (BaBar) with fit probabilities
favoring the former.

\end{abstract}
}
\end{titlepage}
\tableofcontents
\newpage
\section{Introduction}
\label{Introduction}
\indentB 
The Standard Model   provides the accepted
framework which embodies the strong, electromagnetic and weak interactions;
it accounts accurately for the  observable values  reported from
  low energies up to the highest ones reached
at the LHC. A very few specific  measurements look, however,  
borderline enough to rise a hint of a physics beyond the  
Standard Model. Among these, the very precisely measured muon anomalous
magnetic moment $a_\mu$ plays a special role; it has generated 
-- and still generates -- an important experimental and theoretical
activity  related with its measurement by the E821 Experiment 
at BNL \cite{BNL}  $a_\mu ({\rm BNL})=[11659209.1 \pm 6.3]\times 10^{-10}$,
for its latest update; this value is at variance with expectations by 
$3.5 \sigma$ to  $ 4.5\sigma$, 
depending on the various predictions, essentially differing by their estimates of the
leading order hadronic vacuum polarisation (HVP-LO) as reported in \cite{WhitePaper_2020}
and displayed therein in Figure 44. 

Actually, the significance just quoted refers to using  
 HVP-LO  evaluations derived by means of various Dispersion Relation (DR) methods.
In contrast, the Lattice QCD (LQCD) Collaboration BMW \cite{BMW_amu_final}
recently published an estimate for the HVP-LO   \cite{BMW_amu_final}  claiming
a 0.8\% precision and very close
to what is needed to match  the BNL measurement; accordingly, the BMW HVP-LO is
 at  variance by more
than $2.6 \sigma$ with any of the reported DR evaluations of the HVP-LO.
 
The Muon $g-2$ Experiment running at FNAL has very recently published
its  first results \cite{FNAL:2021} and found them in excellent accord with the
previous  BNL measurement \cite{BNL}; the consistency of these two measurements 
allowing for their combination, the experimental reference value becomes~:
	$$a_\mu ({\rm Exp})=[11659206.1 \pm 4.1]\times 10^{-10}~~.$$
Compared to the BNL measurement, this weighted average	
provides a noticeably improved uncertainty  (30\% reduction)  and a  
downward shift by  $3 \times 10^{-10}$. This value still favors numerically
the BMW estimate \cite{BMW_amu_final} for the HVP-LO over any of the DR ones.

Hence, the puzzle which was "DR versus Data" may become
\footnote{One should  remind that $\tau$ based evaluations of the
muon HVP may not be in contradiction with the  BMW estimate as can be
seen in the recent \cite{Miranda:2020wdg} (see also \cite{WhitePaper_2020}
for previous evaluations).  This, however,
supposes that the significant Isospin breaking mechanisms involved in the
 pion form factor in the $\tau$ decay and in  $e^+e^-$ annihilations -- which can well
  differ  -- have all
been clearly identified.  This is one of the issues addressed in this article.
}   
"DR versus LQCD"
for which some different kind of physics beyond the Standard Model could have to
 be invoked. Indeed,
if the HVP-LO as derived by DR methods provides a  good global EW fit, the change
suggested by the BMW evaluation severely impacts the goodness of
the global EW fit \cite{Crivellin:2020}, 
except if the changes can be localized at low enough energy -- Reference \cite{Crivellin:2020}
quotes 1.94 GeV, assuming the change in the cross-sections to be a mere global rescaling.

As long as the missing piece may  spread out across the whole non-perturbative region of QCD which,
as shown by the KEDR data \cite{KEDR2016}, does not  extend much above
 $\simeq 2$ GeV, one has at hand a somewhat large lever arm. However, the recent KMPS 
 study
 \cite{Keshavarzi:2020} has shown that the missing contribution to the HVP-LO should
 come from the energy region  below  $\simeq 0.7$ GeV 
 to accomodate the global EW  fit. This locates the requested missing part of the
 annihilation cross-section  $\sigma (e^+e^- \ra {\rm Hadrons})$
 to a very limited energy region  widely explored by several 
 independent groups having collected  data using
  different detectors and colliders; this missing hadronic cross-section piece is expected
 to contribute an additional 
  $\delta a_\mu \simeq (15 \div 20) \times 10^{-10}$  to the muon
  HVP-LO, larger than the light-by-light (LbL) contribution to the HVP.

\vspace{0.5cm}
 
 Thus, the  energy region to scrutinize is  located well inside the realm of the Effective 
 Resonance Lagrangian 
 Aproaches (RLA) which have extended the scope of the Chiral Perturbation Theory (ChPT).
 The inclusion of  resonances has given rise to the Resonance
 Chiral Perturbation Theory (R$\chi$PT) formulation and to the Hidden Local 
 Symmetry (HLS) Model which have been proved equivalent \cite{Ecker1,Ecker2}.
 Beside other processes, the HLS Model \cite{HLSOrigin} encompasses the  non-anomalous 
 annihilation
 channels  $e^+e^- \ra \pi^+\pi^-/K^+K^-/K_LK_S$; it can be complemented by
 its  anomalous sector \cite{FKTUY} which allows to also  cover the 
 $e^+e^- \ra \pi^+\pi^-\pi^0/\pi^0 \gam/\eta \gam$ annihilation channels.
 Because it deals with only the lowest lying resonance nonet, the validity 
 range of the HLS  Lagrangian naturally extends up to the $\phi$ mass  region
 and is thus quite appropriate to explore the faulty energy
 region in accordance with QCD which is at the root of the different RLA.
 
 As the BMW evaluation of the muon HVP-LO questions the annihilation data 
 in the energy region up to the $\phi$ meson mass, it is worthwhile testing 
 our understanding of
 its physics by means of such RLA, in particular the HLS model, which allows
 to explore this region already well covered by a large number of  data samples  
 in all the significant channels and thus shrink the window of possibilities
 to find a non negligible missing $\delta a_\mu$.
\vspace{0.5cm}
 
 As the original HLS model\footnote{See \cite{HLSRef} for a comprehensive review
 covering the anomalous and non-anomalous sectors.}
 assumes U(3) symmetry in both the vector (V) and pseudoscalar (P or PS) sectors, 
 it should obviously be complemented by symmetry breaking inputs in order 
 to  account for the rich amount of data samples it is supposed to cover.
 A first release named BHLS \cite{ExtMod3,ExtMod4} essentially based on the BKY
 breaking mechanism \cite{BKY,Heath}, extended \cite{Hashimoto} to account
 for Isospin breaking effects, was proven to perform satisfactorily; 
 however, it exhibited some difficulty  to manage with the threshold and $\phi$ 
 regions for resp. the dipion and the 3 pion annihilation channels. In
 order to solve this  issue, the breaking procedure has been deeply
 revisited and gave rise to  BHLS$_2$ under two variants \cite{ExtMod7}
 named respectively the Basic Solution (BS) and the Reference Solution (RS).
 Both BHLS$_2$ variants are derived by complementing the BKY breaking mechanisms
 at work in the ${\cal L}_A$ and ${\cal L}_V$ sectors of the non-anomalous HLS Lagrangian
 by additional breaking schemes affecting solely the vector meson fields.
  
  Regarding the vector sector of the BHLS$_2$ BS variant, the new breaking input 
  -- named   Covariant Derivative (CD) Breaking -- turns out to perform the
  substitution\footnote{To avoid lengthy repetitions, we refer the reader 
  to the companion paper \cite{ExtMod7} for details. The  present study is  
  essentially an  extension of BHLS$_2$ which  endorses  its content,
 particularly  its vector sector.} 
  $  V \Longrightarrow V + \delta V$ in the covariant 
  derivative which is a fundamental ingredient of the HLS model; the aim of 
  $\delta V$  is to break the $U(3)_V$  symmetry for the components along the basis matrices
  $T_0=I/\sqrt{6}$, $T_3$ and $T_8$  of the canonical Gell-Mann $U(3)$ algebra.
  The $ V \Longrightarrow V + \delta V$ rule  naturally propagates  to
   the anomalous sector -- {\it i.e.} the $VVP$ and $VPPP$ Lagrangian 
   pieces\footnote{As one assumes the FKTUY \cite{FKTUY}  intrinsic parameters
   $c_3$ and $c_4$ to be equal, $VAP$ couplings ($A$ being the electromagnetic field)
   identically vanish.}.

  Regarding the PS sector of the BS variant of BHLS$_2$~: Besides the BKY breaking
  associated with the ${\cal L}_A$ part of the non--anomalous Lagrangian, the symmetry
  has been reduced by including the so--called 't~Hooft determinant term \cite{tHooft}; 
  for our purpose, this
  turns out to add  the singlet term $\lambda/2 \pa_\mu \eta_0 \pa^\mu \eta_0 $ to the kinetic energy
   of the PS fields.
   
   However, if this canceled out the difficulties met at the dipion threshold and in the $\phi$
   mass region, the account for the dipion spectra  collected in the $\tau$ decay
   was not fully satisfactory as can be seen in Table 3 of \cite{ExtMod7} (see also the
   discussion in Section 17 herein); a closer look indicated that it is the description
   of the high statistics Belle sample \cite{Belle} which is faulty. This issue
   was circumvented by introducing an additional breaking mechanism (the Primordial
   Mixing) which defines the RS Solution.
   \vspace{0.5cm}

   The original aim of the present study is to reexamine the issue actually raised 
   by the    $\tau$ dipion spectra and to figure out which kind of breaking 
   can solve the $\tau$ problem met by the BS variant of BHLS$_2$. 
   In order to motivate
   this new breaking scheme --  a generalization of the usual 't~Hooft determinant 
   term -- Section \ref{tau_spectra} proceeds to a thorough study of the various 
   dipion $\tau$ spectra and of their impact on the other channels involved
   in the BHLS$_2$ framework, especially on the pion form factor in the 
   timelike and spacelike regions.
   
    Once motivated, this  extended breaking scheme
   is precisely defined and analyzed in Sections \ref{thooft-kin} and \ref {PhysicalPSfields}. 
   The purpose of
   Section  \ref{tau_sector} is to  address the modifications generated in the
   non--anomalous
   BHLS$_2$ Lagrangian derived in \cite{ExtMod7} by this newly introduced kinetic breaking.
   Special  emphasis  is on the pion form factor involved
   in the decay of the $\tau$ lepton $F_\pi^\tau(s)$ compared to its partner in $e^+e^-$
   annihilations $F_\pi^e(s)$; it is shown that, while $F_\pi^e(0)=1$ is still 
   fulfilled\footnote{Under conditions on $\delta V$ discussed in Sections 4 and 
   7 of \cite{ExtMod7}.},
   the CVC assumption is violated in the $\tau$ sector as $F_\pi^\tau(0) \ne 1$. As 
   the ALEPH \cite{Aleph} and Cleo \cite{Cleo} spectra accomodate easily a modelling with
   either of the
   $F_\pi^\tau(0) \ne 1$ and $F_\pi^\tau(0) = 1$ constraints, this result emphasizes   
   the interest of having another $\tau$ dipion spectrum with a statististics
   comparable to those of Belle \cite{Belle} or larger.
  
   However, allowing for a violation of CVC  within BHLS$_2$ cannot be solely localized 
   in the $\tau$ sector of the BHLS$_2$ Lagrangian and it  propagates to the anomalous 
   Lagrangian pieces as noted
   in Section \ref{anomalous_pieces} and developped in the various Appendices.
   Hence, the description of processes as important as the 
   $e^+ e^- \ra \pi^+ \pi^- \pi^0/\pi^0 \gam/\eta \gam$ annihilations
   is deeply modified and so should be tested versus data, as well as the $P \ra \gam \gam$ 
    decay modes  and  those involving $\etp V \gam$ couplings.
\vspace{0.5cm}
   
   Prior to this exercise, our set of reference data samples has to be updated
   to account for new data samples \cite{BESIII_3pi,SND20}  or 
   updated ones \cite{BES-III,BESIII-cor}.  This is done in two steps. First, the
   purpose of Section \ref{BESIII} is to deal with the newly issued three pion data sample
   collected by the BESIII Collaboration \cite{BESIII_3pi}. It is shown that the BESIII 
   spectrum energies should be appropriately recalibrated to match the common energy scale
    of the   other data samples included in our reference data set, especially
    in the $\omg$  and  $\phi$ peak locations. 
    
   On the other hand, dealing with the dipion spectra is of course an important --
    and controversial --
   issue  because of the long standing discrepancy between some of the available
   high statistics data  samples. Therefore, we take profit of the newly published SND dipion
   spectrum \cite{SND20} to revisit in Section \ref{SND20txt} the consistency analysis
   of the different dipion samples to illustrate the full picture and motivate the way we
   deal with strong tensions when evaluating physics quantities of importance, especially
   the muon HVP-LO.
   
   Having updated our reference set of data samples, Section \ref{global-fits} reports
   on global fits performed under various conditions, updating the results derived with the
   BS and RS variants of the former version of BHLS$_2$ and those obtained using the
   extended formulation which
   is the subject of the present study;  this extension will be named EBHLS$_2$ for clarity.
   Section \ref{muon-hvp} addresses a key topic of the broken HLS model within 
    the  EBHLS$_2$ context.  Indeed, the question of supposedly uncontrolled uncertainties 
    associated with using fit results  based on an Effective Lagrangian may
    cast some shadow on this kind of methods. To definitely address this issue, the best
    is to quantify the effect by comparing the estimates for the muon HVP derived 
     from EBHLS$_2$ with  those derived using more traditional 
    methods {\it under similar conditions}.  Subsection \ref{amu_pipi} illustrates for 
    the dipion contribution 
   to the muon HVP that specific biases attributable to using EBHLS$_2$ are negligible compared
   to {\bf i/} the way the systematics, especially the normalization uncertainty of the various
   spectra,   are dealt
   with by the various authors, {\bf ii/} the sample content used to derive one's estimates. 
   These two major sources of uncertainty are, on the other hand, common to any of the
   reported evaluations.
      
   With these conclusions at hand, our evaluations of the muon HVP-LO are derived.
   Our favored
   result which excludes from the fit the dipion spectra from KLOE08  \cite{KLOE08}
   and BaBar \cite{BaBar,BaBar2} is examined in Subsection \ref{HVP-HLS};  an
   alternative solution where all KLOE data samples are discarded in favor of the BaBar 
   one is also presented in Subsection \ref{HVP-BaBar}. The full HVP-LO is constructed
   (Subsection \ref{HVP-total}) and compared with the other currently reported  evaluations
   in Subsection   \ref{hvp-compare}.
\vspace{0.5cm}

   Equiped with the kinetic breaking mechanism defined in Subsection \ref{XA-BRK},
   EBHLS$_2$  is well suited to address
   the mixing properties of the $[\pi^0,\eta,\etp]$ system more precisely than was 
   done with a similar -- but much less sophisticated -- modelling in \cite{WZWChPT}.
   The final aim is to rely on the results of the EBHLS$_2$ fit  over the largest set of  
   data samples ever used
   to derive the corresponding mixing parameter values with optimum accuracy.
   
   The derivation of the axial currents is the subject of Section \ref{Axial_currents}.
   Section \ref{octet-singlet} addresses the singlet-octet basis parametrization defined
   by Kaiser and Leutwyler \cite{Kaiser_2000, leutw,leutwb}; it is shown that 
   EBHLS$_2$ allows to recover the expected Extended ChPT relations. In Section
   \ref{quark-flavor}, a similar exercise is performed within the quark flavor basis
   developped by Feldmann, Kroll and Stech (FKS) in \cite{feldmann_1,feldmann_2,feldmann_3}
   and one also yields the expected results. This clearly represents a valuable piece of information
   about the dealing of EBHLS$_2$ in its PS sector. 
   
   The aim of Sections \ref{kroll_brk1}  and \ref{IB_brk} is to push  EBHLS$_2$ a step further~: 
   One focuses on how Isospin symmetry breaking shows up in  the axial currents $J_\mu^q$
    associated to light quark pairs $\{ [q \overline{q}], q=u,d,s\}$ when expressed in terms
    of PS bare fields -- a leading order approximation.     The Kroll Conditions
    \cite{Kroll:2005}~:
   $$<0|J_\mu^q |[q^\prime \overline{q\prime}](p)>=ip_\mu f_q \delta_{q q\prime }$$
      are then examined in detail and shown to exhibit -- at ${\cal O}(\delta)$ in breakings -- 
      unexpected constraints among the various components of the  kinetic breaking term. 
      In particular, satisfying the Kroll Conditions implies that a kinetic breaking
      with solely a $\pa_\mu\eta_0 \pa^\mu\eta_0$ term is not consistent and should be extended
      in order to involve $\pa \pi^0,\pa \eta_0$ and $\pa \eta_8$ quadratic contributions. Whether
      this  property  is inherent to only EBHLS$_2$ may look unlikely.
      
       In Section \ref{BHLS-mixing}, one reports on additional EBHLS$_2$ fits suggested
       by the Kroll Conditions and tabulates the fit parameter values. The short Section 
       \ref{side-results} reports on side consequences on some physics parameters, especially
         the muon HVP-LO.
        Sections \ref{Mixing-param} and \ref{IB-kroll} report on the numerical
      evaluation of the  $[\pi^0,\eta,\etp]$ mixing parameters and compare with 
       available results from other groups.
 
 	Finally, Section \ref{conclusions} collects the conclusions of this work, an almost 100\%
	COVID19 lockdown work.
   
%\newpage
\section{Preamble~: On the Free Parameters of the BHLS$_2$ Model}
\label{param_free} 
\indentB
Significant  (anti-)correlations between $\Sigma_V$, $z_V$ and the
specific HLS parameter $a$
have been reported in our study  \cite{ExtMod7}; this topic was the purpose 
 of its Subsection 20.1. As parameter correlations may easily be of pure numerical 
origin\footnote{ For instance, the coefficients of the subtraction polynomial of any given
 loop function, may undergo significant correlations in  minimization procedures;
 they are accounted for in the parameter error covariance matrix returned by the fits.},
 we did not go beyond analyzing the issue numerically but emphasized that the physics
 conclusions were safe, {\it i.e.} not shadowed by these correlations. 
 
 Actually, one can go a step further. Indeed, it can be remarked that the three parameters 
 $\Sigma_V$, $z_V$ and $a$ are involved only in the ${\cal L}_V$ piece of the  
 non--anomalous BHLS$_2$ Lagrangian
 ${\cal L}_{HLS}={\cal L}_A + a{\cal L}_V$ and do not occur in its 
 anomalous FKTUY pieces \cite{FKTUY,HLSRef}. 
  Let us consider  the  pieces inherited from $a{\cal L}_V $  named 
 here and in \cite{ExtMod7} ${\cal L}_{VMD}$ and ${\cal L}_\tau$ and perform therein 
the following parameter redefinition~:
\be
\begin{array}{lll} 
\displaystyle a \longrightarrow  a^\prime = a(1+ \Sigma_V ) ~~, &
\displaystyle z_V \longrightarrow  z_V^\prime = z_V/(1+ \Sigma_V ) \simeq  z_V (1- \Sigma_V ) 
\end{array}
\label{AA12}
\ee
where $\Sigma_V$ and $z_V$ are introduced by the $X_V$ breaking matrix affecting ${\cal L}_V $
which  actually writes
$X_V={\rm diag} (1+\Sigma_V/2,1-\Sigma_V/2,\sqrt{z_V})$ in the BHLS$_2$ 
framework\footnote{Note the missing square root symbol in the definition for $X_V$ given
in \cite{ExtMod7}.}.

One can then  check that the dependency upon $\Sigma_V$ drops out everywhere 
except in the $W^\pm$ mass term shown in the  ${\cal L}_\tau$ Lagrangian 
piece (see Equation (\ref{AA5}) below). 
Obviously, this mass term has no influence on the phenomenology 
we address and thus is discarded. 

 It follows herefrom that  the {\it actual} values for $a$,  $z_V$  and $\Sigma_V$
 are in fact out of reach and that the single quantities which can be accessed using the data are their
$a^\prime$,  $z_V^\prime$ combinations. Practically,  fitting within 
 the BHLS$_2$  framework --  having fixed $\Sigma_V=0$ --  reduces the parameter freedom 
 and the parameter
 correlations without any loss in the physics insight, being understood that
 the derived $a$ and $z_V$ are nothing but   $a^\prime$ and  $z_V^\prime$ 
  just defined.

\vspace{0.5cm}

On the other hand, specific parameters are involved in order to deal 
with the $[\pi^0,~\eta,~\etp]$ system. They come from
the transformation leading from the renormalized fields -- those which  diagonalize
the PS kinetic energy term -- to the physically observable $[\pi^0,~\eta,~\etp]$ states
and can be found\footnote{ See also Subsection 4.3 in \cite{ExtMod3}.}  in Section 
\ref{PhysicalPSfields}. These parameters  have been named $\theta_P$, $\epsilon$ and 
$\epsilon^\prime$ in accordance with the usual custom \cite{feldmann_1,Kroll:2005}. 

In our previous works on the HLS model, in particular  \cite{ExtMod3,ExtMod7}, one of 
the ($\eta,~\etp$) mixing angles \cite{Kaiser_2000,leutw} has been constrained
($\theta_0 \equiv 0$)   following a former study \cite{WZWChPT}.
This turns out to impose the mixing angle $\theta_P$  be algebraically related
to the BKY parameter $z_A$ and the nonet symmetry breaking parameter $\lambda$ (see Subsection
4.4 in \cite{ExtMod3}). The  experimental picture having dramatically changed since  \cite{WZWChPT},
 this assumption deserves  certainly to be revisited as will be done in the present work.
 Moreover,  we also imposed \cite{leutw96}~: 
\be
\begin{array}{lll} 
\displaystyle \hvar \hvar^\prime =-\hvar_0^2 \sin{2 \theta_P},~~~{\rm with}~~~
\displaystyle 
\hvar_0 =\frac{\sqrt{3}}{4} ~\frac{m_d-m_u}{m_s-\hat{m}}
~~~{\rm and}~~\hat{m}=\frac{1}{2} (m_u+m_d)~.
\end{array} 
\label{xxxyy}
\ee
As a whole, this lessens the number of free parameters by 2 units without any
degradation of the fit quality or any change in the HVP values.

However, for the present purpose, it has been found worthwhile to release
these constraints and let  $\theta_P$, $\epsilon$ and 
$\epsilon^\prime$ vary freely. When analyzing below the $[\pi^0,~\eta,~\etp]$  mixing
properties, this assumption will be revisited in a wider context. 

\section{Revisiting the $\tau$ Dipion Spectra~: A Puzzle?}
\label{tau_spectra}
\indentB Section 17 of \cite{ExtMod7} reported the  properties 
of our set of --  more than 50 -- data samples  when submitted to
global fits based on either of
the Reference Solution (RS) and Basic Solution (BS) variants
of the BHLS$_2$ model. Table 3 therein displays a detailed account
of the  information returned by the fits for the various physics channels.
More precisely, this Table shows that   the reported $\chi^2/N_{points}$ 
averages for the displayed groups of the held data samples
are generally of the order 1 -- with the sole exception of the
 $K^+ K^-$ data sample from \cite{BaBarKK}.  

The $\tau$ channel $\chi^2/N_{points}$ overall piece of information displayed in
 this Table 3, covers 
a data group merging the samples provided by the Aleph \cite{Aleph}, 
Cleo \cite{Cleo}  and Belle \cite{Belle} Collaborations. One can 
read\footnote{The corresponding fits have been performed 
fixing $\theta_P$ through the condition $\theta_0\equiv 0$ and imposing the
condition  $\hvar \hvar^\prime =-\hvar_0^2 \sin{2 \theta_P}$, as reminded
in Section \ref{param_free};  one  yields now more favorable  $\chi^2$
by having released these constraints as will be seen shortly. }  
therein~:  $\chi^2/N_{points} = 92/85$ (RS variant) and 
$\chi^2/N_{points} = 98/85$ (BS variant). In the following, one may
refer to these data samples as, resp. A, C and B.

However, this fair behavior of the $\tau$ channel data actually hides  contrasted  
behaviors  among the three samples gathered inside  this group.  This issue deserves
reexamination\footnote{This issue was already addressed 
in \cite{ExtMod2} in the context of an oversimplified version the broken HLS model.}
within the BHLS$_2$ \cite{ExtMod7} context.

It was noted in Section 11 of \cite{ExtMod7}   that
the subtraction poynomials resp. $P^\tau_\pi(s)$ and $P^e_\pi(s)$
of resp. the $\pi^\pm \pi^0$ and $\pi^+ \pi^-$ pion loops involved in the pion 
form factors are different, allowing this way
for relative Isospin Symmetry breaking  (IB) effects; more precisely, they are related by~:
\be
P^\tau_\pi(s) = P^e_\pi(s) +\delta P^\tau_\pi(s) ~,
\label{tau_1}
\ee
and the polynomial  $\delta P^\tau_\pi(s)$ is also determined by fit.

Within the BHLS$_2$ context, $P^e_\pi(s)$, as any  of the other loops involved, is a
second degree polynomial with floating parameters.
However, in order to get good global fits when
including  the $\tau$ data -- especially the Belle spectrum \cite{Belle} --
 the degree of $\delta P^\tau_\pi(s)$  
has been increased to the third degree in  the BS  
variant\footnote{ In the present paper, one focuses solely on the BS variant -- 
which carries 3 floating 
parameters less than the RS variant --  as RS and BS return similar pictures
with the collection of existing samples and,
especially, in the $\tau$ sector.} of BHLS$_2$. Actually,
this   $\delta P^\tau_\pi(s)$ degree assumption is not  harmless as
it corresponds   to introducing a non-renormalizable counter term in the renormalized 
BHLS$_2$ Lagrangian. As A and C well manage  within the BHLS framework with a
second degree  $\delta P^\tau_\pi(s)$, the issue raised by the Belle spectrum 
is thus
worthwhile to be cautiously examined; this is the matter of the present Section. 

\vspace{0.5cm}

For the series of (BHLS$_2$) global fits presented in the present Section, we have chosen
to discard the data covering the $e^+ e^- \ra \pi^+ \pi^- \pi^0$ annihilation channel
to lessen
the fit code execution times. The lowest energy data point of the  Cleo spectrum 
\cite{Cleo} is discarded as outlier;  with this proviso,  the three
$\tau$ spectra  \cite{Aleph,Cleo,Belle}  are fully addressed within our fits
from threshold up to 1 GeV.

Formally, the differences between the dipion spectra in the $\tau$ decay and in the $e^+ e^-$
annihilation should solely follow from Isospin Symmetry Breaking (IB) effects.
Therefore, a real  understanding of these  supposes {\it a minima}
 a   simultaneous  dealing with the
 $e^+ e^- \ra \pi^+ \pi^-$ annihilation channel and with the dipion spectra collected 
in the $\tau^\pm \ra \pi^\pm \pi^0 \nu_\tau$ decay. 
The annihilation data  addressed
in our fitting codes --
 CMD-2 \cite{CMD3pion-1989,CMD2-1998-1,CMD2-1998-2}, SND \cite{SND-1998}
KLOE \cite{KLOE10,KLOE12}, BESIII \cite{BES-III}, Cleo-c \cite{CESR}
-- have been presented\footnote{See also Section \ref{SND20txt} below.} 
in detail in Section 13 of \cite{ExtMod7}. 

Actually, the BESIII Collaboration has recently published an erratum  \cite{BESIII-cor}
to their  \cite{BES-III}  which essentially confirms the original spectrum but 
drastically reduces the statistical uncertainties.This will not be discussed at length 
and one only quotes the $\chi^2/N_{points}$ evolution: Running our standard BHLS$_2$  
code with  the {\it uncorrected}  BESIII dipion spectrum  \cite{BES-III}, the
various fits  return $\chi^2/N_{points} \simeq 35/60$, whereas running it with
the {\it corrected} data  \cite{BESIII-cor} one yields 
$\chi^2/N_{points} \simeq 50/60$; this more realistic goodness of fit 
clearly indicates that 
the errors are indeed better understood, allowing
the BESIII spectrum to really influence the physics results derived from fits.

One should also remind that the 2 dipion spectra  from KLOE08 \cite{KLOE08} and BaBar 
\cite{BaBar,BaBar2}, exhibiting  a poor consistency with all the ($> 50$) 
others, are  discarded  since the very beginning
of the HLS modelling program \cite{ExtMod3,ExtMod4}. Finally, the SND dipion spectrum \cite{SND2020}
measured over the $0.525 <\sqrt{s}<0.883$ GeV energy interval  will be analyzed separately 
below. 
\subsection{Fitting the $\tau$ Dipion Spectra}
\label{tau-fit}
\indentB The $\tau$ spectra submitted to global fits are defined by~:
\be 
\displaystyle \frac{1}{\Gamma_\tau} \frac{d \Gamma_{\pi \pi}}{ds}=
{\cal B}_{\pi \pi} \frac{1}{N}\frac{d N(s)}{ds}
\label{tau_a}
\ee
using the event distributions and  branching fractions  ${\cal B}_{\pi \pi}$
provided by each of the  Aleph \cite{Aleph}, Belle \cite{Belle} and Cleo \cite{Cleo}
Collaborations. The full $\tau$  width is derived from its lifetime taken
from \cite{RPP2016}. The relation with the pion form factor is~:
\be 
\displaystyle \frac{d \Gamma_{\pi \pi}}{ds}=
  \frac{G_F^2}{m_\tau^3} [S_{EW} G_{EM}(s)] |F_\pi^\tau(s)|^2
\label{tau_b}
\ee
where  $F_\pi^\tau(s)$ is derived from the BHLS$_2$ Lagrangian \cite{ExtMod7} and  $S_{EW}$
collects the short range radiative corrections \cite{Marciano};   
 the long range radiative corrections are collected in $G_{EM}(s)$ and
 evaluated on the basis of \cite{Cirigliano1,Cirigliano2,Cirigliano3}.   
 The normalization of the full form factor at the origin is, thus,  given
 by the product $[S_{EW} G_{EM}(s)]$ in the standard BHLS$_2$ 
 which fulfills\footnote{All breaking parameters occuring in  BHLS$_2$, in
 particular those associated with the breaking of the Isospin Symmetry
 or of the Nonet Symmetry are considered being of order $\delta$, a generic
 perturbation parameter. All expressions derived from our Lagrangian 
 are  understood truncated at this order and, therefore, terms of order
 ${\cal O}(\delta^2)$ or higher are always discarded.}  automatically
 $F_\pi^\tau(0)=1 +{\cal O}(\delta^2)$. 
 \vspace{0.5cm}
 
 Anticipating on the following Sections, let us state that one will define
 an extension of standard BHLS$_2$ \cite{ExtMod7} to allow for a violation
 of CVC in the $\tau$ sector;  it will be named
 EBHLS$_2$. The main difference --  not the single one --
 between  EBHLS$_2$ and the  standard BHLS$_2$ \cite{ExtMod7} is that it fulfills
 $F_\pi^\tau(0)=1-\lambda_3^2/2$ where $\lambda_3$ is a floating parameter
 of order ${\cal O}(\sqrt{\delta})$ reflecting a symmetry breaking.  Moreover,
 setting $\lambda_3=0$ therein allows to recover exactly the standard BHLS$_2$
 \cite{ExtMod7}. When  $\lambda_3 \ne 0$,  the rescaling generated by
  EBHLS$_2$ is numerically modulated by   accompanying changes in the 
 internal structure of the $\rho$ term in $F_\pi^\tau(s)$. On the other hand, as
 the rest of the non--anomalous Lagrangian pieces are unchanged, the properties
of $F_\pi^e(s)$ remain unchanged; in particular, the condition
 $F_\pi^e(0)=1 +{\cal O}(\delta^2)$ is still valid as in BHLS$_2$ because
 the term of ${\cal O}(\delta)$ vanishes by having stated \cite{ExtMod7} $\xi_0=\xi_8$.
 
 As the $\tau$ data analysis is  the main motivation for the forthcoming
 EBHLS$_2$  extension, it was found worthwhile to display, beside
 the standard BHLS$_2$ fit results, the corresponding  EBHLS$_2$ information,
 prior to dealing with its derivation\footnote{The derivation of EBHLS$_2$ is mainly
 addressed  in  Sections \ref{thooft-kin}  and \ref{tau_sector}.}. 

\subsection{BHLS$_2$ Global Fits Excluding the Spacelike Data}
\label{tau-nospace}
\indentB 
Table \ref{Table:T0} reports on a series of fits which aims at coping with the 
$\tau$ topic; the global fits reported in this Subsection discard the spacelike data 
\cite{NA7,fermilab2} and the discussion emphasizes solely the behavior of the dipion data
from the annihilation channel and from  the $\tau$ decay. 

As stated just above, the reference therein to the parameter $\lambda_3$  
anticipates about the EBHLS$_2$  extension proposed below, and
stating the condition $ \lambda_3=0$ is strictly identical to
having BHLS$_2$ running, in particular its BS variant  
\cite{ExtMod7} solely used all along the present work, except otherwise stated.

 \begin{itemize}
\item  
The first data column displays the global fit performed using the published A, B and C spectra
imposing the polynomial $\delta P^\tau_\pi(s)$ (see Equation (\ref{tau_1})) to be second degree.
Obviously, the $\chi^2/N_{points}$ averages are reasonable for all the displayed data samples
or groups shown (as well as  those not shown) except for Belle which yields  the unacceptable 
average $\chi^2/N_{points} =2.52$. 
\item
The simplest ({\it ad hoc}) choice to better accomodate the Belle spectrum turns out 
to allow $\delta P^\tau_\pi(s)$ to be third  degree.  Doing this, the
second data column shows that the fit returns a fair probability, as 
already known since \cite{ExtMod7}. Indeed, beside a  quite marginal improvement
 of the $e^+ e^- \ra \pi^+\pi^-$ 
account, one observes a sensible improvement in the description of the Aleph  
($\chi^2:41 \ra 22$) and Belle ($\chi^2:48 \ra 36$) spectra, whereas the Cleo
spectrum $\chi^2$ remains unchanged and satisfactory. The improvement for the Belle spectrum
 is significant  ($\chi^2/N_{points}: 2.52 \ra 1.90$) but not fully  satisfactory. 
    Nevertheless, the top panel in Figure~\ref{Fig:tauRes} shows that the normalized residuals
  for the $\tau$ spectra exhibit a reasonably flat behavior,
  thanks to having a third degree $\delta P^\tau_\pi(s)$.

So, once the degree for $\delta P^\tau_\pi(s)$ is appropriate, BHLS$_2$ gets
a fair account for the A and C data and an acceptable  one 
for the Belle spectrum.  

\end{itemize}

However, inspired by the fit summary Table VII
in the Belle paper \cite{Belle}, one has addressed two  other similar
strategies~: 

 \begin{itemize}
\item  Instead of fitting the A, B and C spectra {\it as such}, one chose 
using each of them normalized to its integral over the fitting energy 
range ($<1.$ GeV), {\it i.e} one rather fits the A, B, C {\it lineshapes} within the 
global BHLS$_2$ framework. In this case,  a second degree 
$\delta P^\tau_\pi(s)$  is already sufficient and yields a fair global fit 
(95 \% probability). The corresponding results  are displayed in the
third data column; they are clearly  satisfactory in both the annihilation 
 channel and   the $\tau$ sector. 
 In this configuration, the Belle spectrum
 undergoes an individual $\chi^2$ improvement by 10 units and 
 comes out  with a  more reasonable $\chi^2/N_{points}= 1.37$. 
\end{itemize}
\begin{table}[!phtb!]
\begin{center}
\hspace{-1.4cm}
\begin{minipage}{0.9\textwidth}
\begin{tabular}{|| c  || c  | c || c | c ||| c ||}
\hline
\hline
\hhhv  BHLS$_2$ Fit	& 
\multicolumn{4}{|c|||}{ \hhhv BHLS$_2$ $\equiv$ ($ \lambda_3=0$)} &   \hhhv $~~$ $\lambda_3$ free $~~$\\
\hhhw   \hhhw (Excl. spacelike) & \hhhw  spectra   &  \hhhw spectra (3$^{rd}$ deg.)  &  \hhhw  lineshapes & \hhhw  rescaled & \hhhw spectra\\
\hline
\hline
 \hhhv NSK  $\pi^+\pi^-$ (127)&    $138$    &  $136$      & $135$   & $138$    & $138$ \\
\hline
 \hhhv KLOE $\pi^+\pi^-$ (135)&    $146$    &  $143$      & $145$   & $144$   & $140$\\
\hline
\hhhv BESIII $\pi^+\pi^-$ (60)&    $47$     &  $47$       & $48$    & $48$    & $50$\\
\hline
\hline
 \hhhv	$\tau$ (ABC) (84)    &      $122$   &  $92$       & $79$    & $79$     & $78$ \\
\hline
\hline 
 \hhhv	$\tau$ (ALEPH) (37) &       $41$    &  $22$       & $23$    & $23$     & $21$\\
\hline
 \hhhv	$\tau$ (CLEO) (28)  &       $33$    &  $34$       & $30$    & $31$     & $32$\\
\hline
 \hhhv	$\tau$ (Belle) (19) &       $48$    &  $36$       & $26$    & $25$     & $25$\\
\hline
 \hhhv	Fit Prob.           &        66\%   &   93\%      &  95\%   &  94\%    &  96\% \\
 \hline
\hline
\end{tabular}
\end{minipage}
\end{center}
\caption {
\label{Table:T0}
Global fit properties (spacelike data excluded)~:
$\chi^2$ values  for the various sample groups; the numbers of 
data points are given between parentheses. The subtraction
polynomial $\delta P^\tau(s)$ 
is always second degree except when explictely stated (second data column). The tag
"spectra" stands for fitting with the reported A, B and C dipion spectra;
the tag  "lineshapes" stands for the case when these spectra are normalized to their 
 integral over the fitted energy range; the tag "rescaled" covers the case when
a common rescaling factor is applied to the three dipion $\tau$  spectra.  The last data column
displays the results obtained fitting within EBHLS$_2$.
}
\end{table}

Stated otherwise, once  the $\tau$ spectra are normalized,  BHLS$_2$  
provides a fairly good
simultaneous account of the A, B and C spectra and proves
that the A, B and C {\it lineshapes} are quite consistent with each other
without any need for  some {\it ad hoc} trick. Moreover, there is no point in
going beyond the second degree for $\delta P^\tau_\pi(s)$.

An obviously  similar approach to the {\it lineshape} method just emphasized is to let the  pion form factor
$|F_\pi^\tau(s)|^2$ be such  that\footnote{One should   remind that, in contrast to the neutral vector current,
the charged vector current is conserved only in the isospin symmetry limit.}
$|F_\pi^\tau(s=0)|^2 \ne 1$. This is inspired by the
standalone fit performed by Belle and  reported in  Table VII of their \cite{Belle}.

In this study, a first fit has been performed across the full energy range
of the Belle dipion spectrum   
using a Gounaris--Sakurai (GS)  pion form factor  $|F_\pi^\tau(s)|^2$ 
-- which fulfills $|F_\pi^\tau(s=0)|^2 =1$; this fit is the matter of the leftmost data
column of their Table VII and  reports $\chi^2/N_{points}=80/62$.
 Belle also reports therein  
 a second fit, having allowed for a mere rescaling~:
 $$|F_\pi^\tau(s)|^2 \ra (1 + \lambda_\tau)|F_\pi^\tau(s)|^2 $$
of their GS parametrization; the corresponding results are reported
in the rightmost data column of their Table VII with
$\chi^2/N_{points}=65/62$. The noticeable 15  unit 
gain for the $\chi^2$ value (a 4 $\sigma$ effect), resulting from a single additional
 floating parameter, stresses  the  relevance of what was just named
$\lambda_\tau$. Let us perform likewise within the global BHLS$_2$ context.

 \begin{itemize}
\item 
As the standard BHLS$_2$ \cite{ExtMod7} provides 
$|F_\pi^\tau(s=0)|^2 = 1$, one  performs as Belle 
by using  rescaling factors of the form $1+\lambda_\tau$.
We have first performed global fits using  a single
$\tau$ spectrum (A, B, C in turn) and derived the following results~:
\be
%\hspace{-0.5cm}
\left \{
\begin{array}{llll}
{\rm Aleph~:} & \lambda_\tau= (-7.63 \pm 0.66) \% ~, & \chi^2/N_{points}=15/37~,& {\rm Prob=} ~98 \%\\
{\rm Cleo~ ~~:}  & \lambda_\tau= (-3.12 \pm 0.66) \% ~, & \chi^2/N_{points}=27/28~,& {\rm Prob=} ~95 \%\\
{\rm Belle~~:} & \lambda_\tau= (-5.96 \pm 0.52) \% ~, & \chi^2/N_{points}=23/19~,& {\rm Prob=} ~92 \%
\end{array}
\right .
\label{tau_2}
\ee
which, unexpectedly,  indicate that B, as well as A and C, nicely accomodate a rescaling factor 
without degrading the description of the annihilation data. One has also performed a global fit 
merging the three $\tau$ spectra each renormalized by a common (single) scale factor. 
One gets~:
\be
\hspace{-0.4cm}
\begin{array}{llll}
{\rm A+B+C~:} & \lambda_\tau= (-6.95 \pm 0.37) \% ~, & \chi^2/N_{points}=79/84~,
& {\rm Prob=} ~94\%~~;
\end{array}
\label{tau_3}
\ee
more details can be found in the fourth data column of  Table \ref{Table:T0};
the normalized residuals derived from this fit are displayed in the middle panel
of Figure \ref{Fig:tauRes}. Comparing the fit results reported here,
one observes a 13 unit gain compared to using a third degree $\delta P^\tau(s)$,
in line with the Belle own fits -- with exactly the same model parameter freedom
as $\lambda_\tau$  comes replacing the coefficient dropped out
by reducing the degree of  $\delta P^\tau(s)$ by 1 unit.
\end{itemize}

Such a {\it common} rescaling is certainly beyond experimental biases as,
moreover, A, B and C have been collected with different detectors
by independent teams. This is also much beyond the reported uncertainties
on their respective $\tau^- \ra \pi^- \pi^0 \nu$ branching fractions which govern
the absolute scale of the fitted $\tau$ spectra.

So, one reaches an outstanding fit quality  by
 rescaling  the 3 spectra {\it by the same amount}; 
this improvement of the $\tau$ sector is {\it not} obtained at the expense of 
degrading, even marginally, the account of the $e^+e^-$ annihilation data.
Within
the BHLS$_2$ context -- and discarding the spacelike spectra -- it is found 
that  $\lambda_\tau = (-6.95 \pm 0.37) \% $, a non-negligible value.
This amount is certainly related to intrinsic details of the Lagrangian model,
noticeably the mass and width differences of the neutral and charged
$\rho$ mesons which, also, contribute to the absolute scale.

Finally, the last data column in Table \ref{Table:T0} shows that the forthcoming
EBHLS$_2$ succeeds to produce a nice fit,  very close to 
BHLS$_2$ ones corresponding to the information
displayed by Equations (\ref{tau_3}) and reported in the fourth data column of Table
\ref{Table:T0}. The normalized residuals are displayed in the
bottom panel of Figure \ref{Fig:tauRes}. 

One can observe that the three sets of normalized residuals shown in
 Figure \ref{Fig:tauRes} are almost identical and each is quite acceptable.
 Finally, one should stress that in all configurations,
 Table \ref{Table:T0} exhibits
 a fair account of the A and C spectra; it is therefore noticeable, and even amazing, that a
 remarkable simultaneous fit of A, C and B can {\it also} be derived.
  Moreover, in  both kinds of configurations (rescaling or not)
 a same and fair account of all the annihilation data is obtained. 

\subsection{BHLS$_2$ Global Fits Including the Spacelike Data}
\label{tau-space}
\indentB  The fit properties and parameter values reported just above have been derived
using the dipion spectra only in the timelike region for both the $\pi^+ \pi^-$ and
$\pi^\pm \pi^0$ pairs. Moreover, it has been shown \cite{ExtMod7} that,
within the BHLS$_2$ framework, the {\it same} analytic 
function fairly well describes the pion form factor in the spacelike {\it and} timelike energy regions. 
It is, therefore, desirable
to enforce the impact of the Analyticity requirement within the BHLS framework 
by requiring a simultaneously account of both energy regions.
\begin{table}[!phtb!]
\begin{center}
\hspace{-2.0cm}
\begin{minipage}{0.9\textwidth}
\begin{tabular}{|| c | c || c || c || c | c ||}
\hline
\hline
\hhhv Fitting Framework	& 
\multicolumn{3}{|c||}{ \hhhv BHLS$_2$ $\equiv$ ($ \lambda_3=0$)} & 
 \multicolumn{2}{|c||}{ \hhhv EBHLS$_2$ $\equiv$  ($ \lambda_3$ free)}\\
\hhhw   \hhhw (Incl. Spacelike) & spectra &  \hhhw spectra (3$^{rd}$ deg.)  &  \hhhw  rescaled  (2$^{nd}$ deg.) & \hhhw   spectra & \hhhw rescaled\\
\hline
\hline
 \hhhv NSK  $\pi^+\pi^-$ (127)&  138   &  $134$     & $137$	& $138$     & $136$ \\
\hline
 \hhhv KLOE $\pi^+\pi^-$ (135)&  139  &$146$      & $154$	& $140$     & $141$\\
\hline
\hhhv BESIII $\pi^+\pi^-$ (60)& 48 &  $47$       & $47$	& $48$      & $48$\\
\hline
\hline
\hhhv Spacelike  $\pi^+\pi^-$ (59)& 62& $65$      & $77$	& $62$    & $61$ \\
\hline
\hline
 \hhhv	$\tau$ (ABC) (84)   &     ${ \times}$ & 92      & $88$	& $81$     & $77$ \\
\hline
\hline 
 \hhhv	$\tau$ (ALEPH) (37) & 24  &   $23$       & $28$	& $25$     & $21$\\
\hline
 \hhhv	$\tau$ (CLEO) (28)  &  30 &    $33$       & $30$	& $31$     & $32$\\
\hline
 \hhhv	$\tau$ (Belle) (19) &      ${ \times}$ & 37      & $30$	& $25$     & $24$\\
\hline
 \hhhv	Fit Prob.           &  93\%&    89\%       &  79\%	&  91\%    &  94\% \\
 \hline
\hline
\end{tabular}
\end{minipage}
\end{center}
\caption {
\label{Table:T0-b}
Global fit properties (including the spacelike data)~:
$\chi^2$ of the various sample groups; their numbers of 
data points are shown between parentheses. The subtraction
polynomial $\delta P^\tau(s)$
is always second degree except when explictely stated (second data column). The tag
"spectra" stands for fitting the A, B and C $\tau$ dipion spectra as such;
the tag "rescaled" covers the case when a common
 rescaling factor is applied to the three dipion $\tau$  spectra.
}
\end{table}

So, including the spacelike data \cite{NA7,fermilab2} into the set of samples 
submitted to the global fit looks a natural step. {\it A priori}, this
should mostly affect $F_\pi^e(s)$; however, as $F_\pi^e(s)$ and $F_\pi^\tau(s)$ are deeply
intricated within the BHLS$_2$ framework, extending the fit to the spacelike region
 can be of consequence for both form factors.
This Subsection reports on the global fit results derived when including also the spacelike data. 

\vspace{0.5cm}

The fit reported in the first data column of Table \ref{Table:T0-b}, displays the fit
information when fitting with BHLS$_2$ using a second degree $\delta P^\tau(s)$
polynomial, {\it having discarded } the Belle spectrum.  As evidenced by 
its  reported probability (93\%), the fit exhibits a nice account of each group of
data samples as one always observe $\chi^2/N_{points} \simeq 1$. This proves that
the need for a third degree $\delta P^\tau(s)$ is solely caused  by the Belle (B) spectrum.

The second data column in this Table reports for the same fit including also the Belle
spectrum but with a third degree $\delta P^\tau(s)$;
it exactly corresponds 
to those already reported in the second data column of Table \ref{Table:T0}.
The average $\chi^2$ of the various sample groups are observed quite similar,
including for the Belle data sample ($\chi^2/N_{points}=1.95$); for the spacelike
data one yields a favorable $\chi^2/N_{points}=1.10$. This case corresponds to the fit
configurations previously used in \cite{ExtMod7}; with a 89\% probability, 
it is clearly a satisfactory solution and one does not observe any  degradation
of the goodness of fit by having included the spacelike data.

The fit reported in the third data column of Table \ref{Table:T0-b} is the exact analog
of the one displayed in the fourth data column of Table \ref{Table:T0}, taking also into 
account the spacelike data. Here also $\delta P^\tau(s)$ carries the
second degree. The best fit returns a global rescaling factor $1+\lambda_\tau$ with~:
\be
%\hspace{-0.5cm}
\begin{array}{llll}
{\rm A+B+C~:} & \lambda_\tau= (-4.42 \pm 0.40) \% ~, & \chi^2/N_{points}=88/84~,& {\rm Prob=} ~79\%~~.
\end{array}
\label{tau_4}
\ee

Comparing with Equations (\ref{tau_3}), one observes  a drop in 
probability produced by the inclusion of the spacelike data within the global fit procedure
($94\% \ra 79\%$). Compared to having excluded the spacelike data, the effect is noticeable on 
the  KLOE data ($\chi^2/N_{points}~: 1.07 \ra 1.14 $) and for the
$\tau$ data ($\chi^2/N_{points}~: 0.94 \ra 1.05$). 
On the other hand, having reduced the $\delta P^\tau(s)$ degree, one 
observes a clear degradation
of the spacelike data account as  $\chi^2/N_{points}~: 1.10 \ra 1.31$. 
Nevertheless, even if not the best reachable fit as will be seen shortly,
 this configuration provides  a quite  reasonable picture.

The last two data columns in Table \ref{Table:T0-b} refer to the Extended BHLS$_2$ model
fit results. In this case, one has examined the EBHLS$_2$ solution
 (fourth data column) and, for completeness, also performed  the analysis
with an additional rescaling of the $\tau$ spectra (fifth data column).

One clearly observes that the original spectra exhibit uniformly good 
 properties in all channels, and the fit yields  a  
91\% probability,  as displayed in the fourth data column.
Complemented with an additional common rescaling of the $\tau$ spectra, some marginal
improvement is observed in the description of these
as shown in the last data column of Table \ref{Table:T0-b}. 

\vspace{0.5cm}

\begin{figure}[!phtb!]
\hspace{0.cm}
\begin{minipage}{1\textwidth}
%\begin{center}
\hspace{1.cm}
\resizebox{0.8\textwidth}{!}
{\includegraphics*{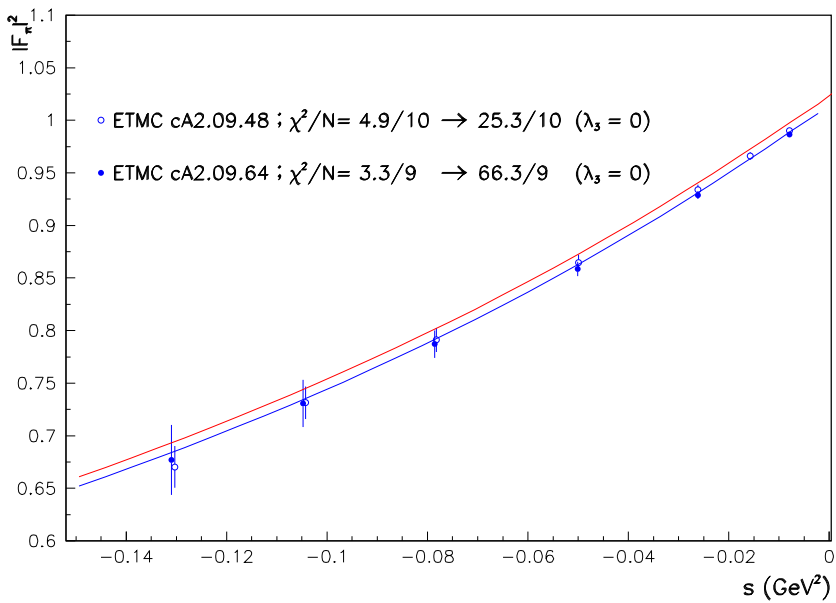}} 
%\end{center}
\end{minipage}
\begin{center}
\vspace{-0.5cm}
\caption{\label{Fig:lambda3Pred}  $|F_\pi(s)|^2$ derived from the fits with $\lambda_3=0$
(red curve) and  with $\lambda_3 \ne 0$ (blue curve). The LQCD data points from \cite{ETMC-FF},
not fitted, 
are superimposed. The values for average $\chi^2$ distances $\chi^2/N_{points}$ are shown 
for $\lambda_3 \ne 0$ and $\lambda_3 =0)$.
 }
\end{center}
\end{figure}

\begin{figure}[!phtb!]
\hspace{-1.8cm}
\begin{minipage}{\textwidth}
%\begin{center}
%\hspace{-0.3cm}
\resizebox{1.\textwidth}{!}
{\includegraphics*{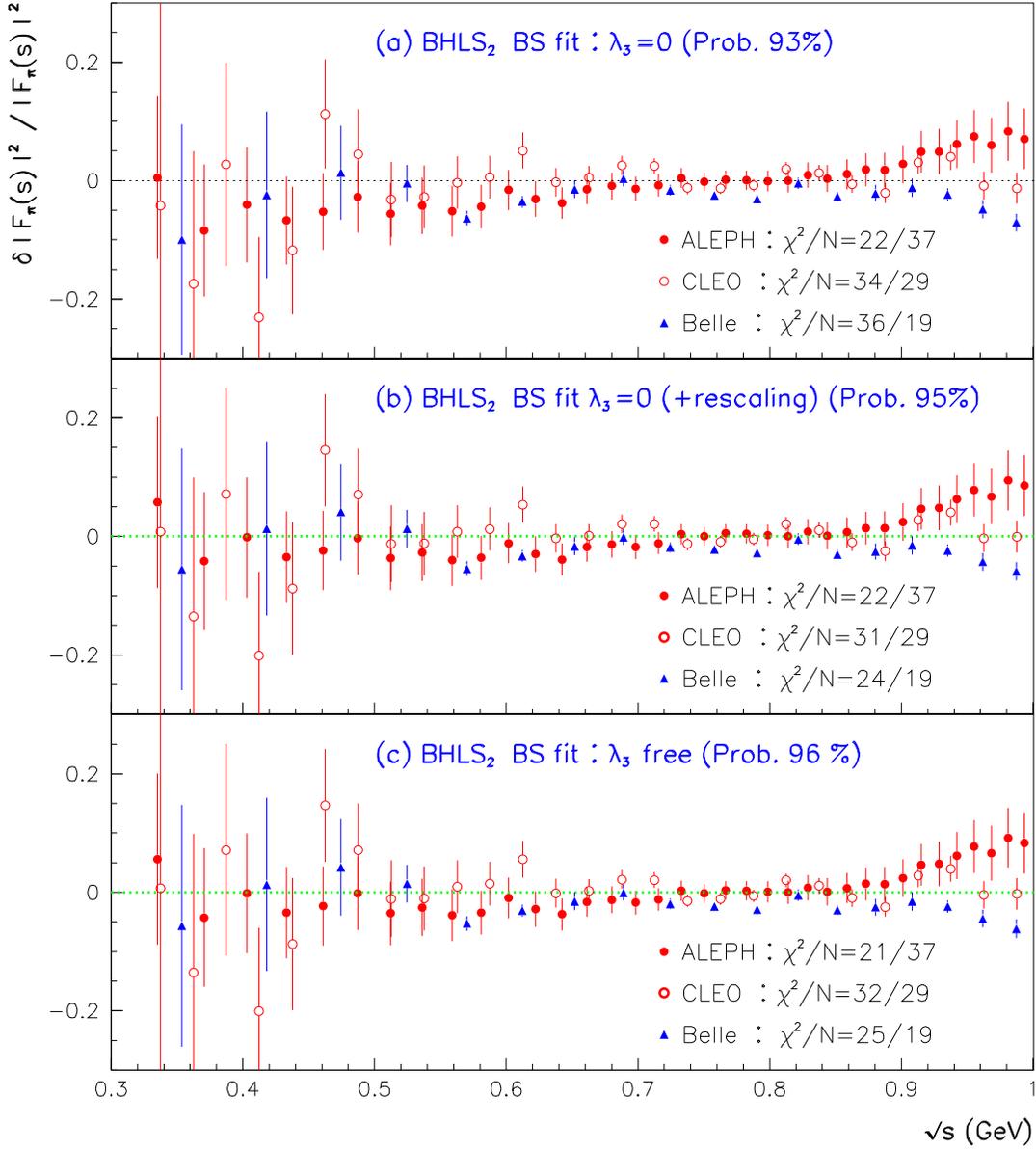}}
%\end{center}
\end{minipage}
\begin{center}
\caption{\label{Fig:tauRes} Normalized residuals derived in the global fits excluding 
the spacelike data.
Top panel displays the case when A, B and C {\it as such} are simultaneously fitted;
the middle panel displays the case where a common rescaling factor is applied to A, B and C.
The bottom panel reports the corresponding results derived by fitting A, B and C {\it as such} 
within the EBHLS$_2$ framework. }
\end{center}
\end{figure}

In contrast with the preceeding Subsection (no dealing with the spacelike data),
using a second degree  $\delta P^\tau(s)$ and rescaling the $\tau$ spectra, 
while improving the $\tau$ sector,  leads to a degraded account of the spacelike 
data ($\chi^2/N_{points}~: 1.10 \ra 1.31 $) and  a loss of the remarkable
{\it prediction} of the LQCD pion form factor data  \cite{ETMC-FF}
 reported in \cite{ExtMod7}; this is illustrated by Figure \ref{Fig:lambda3Pred}.
Comparing this Figure with  Figure 8 of \cite{ExtMod7}, derived by excluding the
$\tau$ data from the fit, one observes a good agreement with EBHLS$_2(\lambda_3 \ne 0)$ 
solution\footnote{In the broken HLS frameworks, the pion form factor
fulfills $F^e_\pi(0)= 1+ {\cal O} (\delta^2)$;
when $\lambda_3$ is left free, the fit returns a departure from 1  by 2 permil 
whereas it becomes 1.3 percent  when $\lambda_3$ is fixed to zero.
This is the origin of the "shift" exhibited by the two curves shown in Figure 
\ref{Fig:lambda3Pred}.}. 
Finally, the last two data columns in Table \ref{Table:T0-b} show that EBHLS$_2$
perfectly succeeds in recovering uniformly good $\chi^2$'s with  fit probabilities 
exceeding 90\% and a fair account of all channels. 

\subsection{Summary}
\label{mini_conclude}
\indentB
Therefore, once including the $\tau$ data within the BHLS$_2$ minimization procedure,
a fit using a third degree  $\delta P^\tau(s)$ succeeds, 
reminding the conclusions already reached in \cite{ExtMod7}; however, one has
to accept an average $\chi^2$  of 1.95 for Belle, whereas those for Aleph and Cleo
are resp. 0.62 and 1.18.

The fact that A, B and C carry a common lineshape may look hardly 
accidental. However, in contrast with Table \ref{Table:T0}, 
where an evidence for solely a rescaling looks reasonable,
Table \ref{Table:T0-b} indicates that a mere rescaling is insufficient
--  not even mentioning that the EBHLS$_2$ prediction
for the LQCD pion form factor data  \cite{ETMC-FF} severely
degrades compared to \cite{ExtMod7}. In this case, EBHLS$_2$ -- complemented or
not with a common rescaling of the $\tau$ spectra -- performs nicely and
fulfills all desirable Analyticity requirements using a second degree $\delta P^\tau(s)$
only, as should be preferred. The fits reported in  Subsections \ref{tau-nospace} and 
\ref{tau-space} convincingly illustrates that the
third degree looks somewhat {\it ad hoc} and can be avoided.

Therefore, one is led to complement the Covariant Derivative (CD) breaking
introduced in \cite{ExtMod7} by an additional term (see Section 
\ref{thooft-kin}) which breaks the kinetic sector of the HLS Lagrangian.
This points
toward a violation of CVC in the $\tau$ sector (at a $\simeq -2.5\%$ level for 
$|F_\pi^\tau(s)|$). For completeness, one has also examined the effect of
a possible rescaling  complementing the CVC breaking (generated
by a non-zero $\lambda_3$); this is reported in the last data column
of Table \ref{Table:T0-b}. This rescaling may correspond to a
(higher order?) correction to the product $S_{EW} G_{EM}(s)$ 
(at a $\simeq -2\%$ level).
However, this additional freedom does not produce a significant effect
and is discarded from now on.

\vspace{0.5cm}

To conclude, the present analysis leads to stress the importance of having at disposal
a new high statistics $\tau$ dipion spectrum;  it is, indeed, of importance to conclude about 
the specific behavior exhibited by the Belle spectrum within global fits.
The challenging properties exhibited by  fitting the $\tau$ dipion spectra 
or, alternatively, their lineshapes may look amazing enough to call for
a confirmation of the Belle spectrum properties. Indeed, it is not unlikely that the 
much higher statistics of the Belle spectrum (50 times those of Cleo \cite{Belle})
allows a finer structure to show up, calling for a more refined description.

\section{Extending the BHLS$_2$ Breaking Scheme} 
\label{thooft-kin} 
\indentB
The Hidden Local Symmetry Model \cite{HLSRef} has been supplied with specific symmetry
breaking mechanisms to provide the BHLS$_2$ framework \cite{ExtMod7} within which almost 
all $e^+e^-$ annihilation channels occuring up to the $\phi$ mass are encompassed. This
allowed for a simultaneous fit of almost all collected data samples covering 
the non--anomalous decay channels ($\pi^+ \pi^-$, $K^+ K^-$, $K_L K_S$) and
anomalous ones \cite{FKTUY} ($\pi^+ \pi^- \pi^0$, $\pi^0 \gam$, $\eta \gam$).
As clear from  \cite{ExtMod7}, one gets a fair description with superb goodness of fit.

As just emphasized in Section \ref{tau_spectra},  the decay mode 
$\tau^- \ra \pi^- \pi^0 \nu_\tau$ is also a natural 
part of the successive HLS  frameworks
\cite{taupaper,ExtMod1,ExtMod2,ExtMod3,ExtMod7}.
However, as illustrated by Table~\ref{Table:T0-b}, the issue is whether, one
requires  (or not)  to consider, besides the ALEPH \cite{Aleph} and Cleo 
\cite{Cleo} spectra, those from Belle \cite{Belle}. 
Our present goal is to define an 
extension of BHLS$_2$ which can naturally encompass simultaneously the A, C 
and B spectra and continuously recover the BHLS$_2$ framework in some
smooth limit. Anticipatively, this extension has been named EBHLS$_2$.
 
BHLS$_2$ has been constructed by considering, beside the BKY mechanism
\cite{BKY,Heath}, the Covariant Derivative Breaking and the Primordial Mixing
procedures -- see Sections 4 and 8 of \cite{ExtMod7}. These essentially 
address the vector sector of the HLS Model and rotations allow to render 
BHLS$_2$ Lagrangian canonical. This lets also the vector  meson kinetic energy supplied
by the Yang--Mills Lagrangian be canonical.

Regarding the  pseudoscalar (PS) sector, the BKY mechanism \cite{BKY,Heath,ExtMod3} also 
contributes to break the symmetries of the HLS Model.  One should also emphasize that
the mass breaking in the kaon sector is at the origin of the Dynamical Mixing of the vector
mesons \cite{taupaper}, which is the central piece of the various broken versions 
of the HLS model. Indeed, thanks to having different charged and neutral kaon loops,
the ($\rho^0$, $\omg$, $\phi$) mass matrix at one loop becomes non--diagonal 
and, thus,  imposes  another
step in the vector field redefinition \cite{ExtMod3,ExtMod7}. This forth and back play between vector
field redefinition and   Isospin Symmetry breaking in the PS sector should be noted. 

Beside the two mechanisms just listed, in order to account 
for the physics of the anomalous processes,  the 't~Hooft determinant terms 
\cite{tHooft}, more precisely its kinetic part, provide the needed nonet symmetry 
breaking in the PS  sector.
Moreover, higher order and loop terms in Chiral Perturbation Theory and QED
corrections are expected to extend the breaking of the PS kinetic energy term
beyond the singlet component. It is the purpose of the present Section  
to extend the kinetic breaking\footnote{The recent \cite{Masjuan} develops
such a kinetic breaking focusing on a treatment of the $\eta-\etp$ system.
Former studies, like  \cite{Schechter:1992iz,Kisselev:1993} for instance, already 
considered  such a breaking mechanism.}
the full $(\pi^0-\eta-\etp)$ system; one already knows from the preceeding
Section that it provides a consistent   picture of the 
$\tau$ sector as it renders consistent the account of the A, C and B spectra.

 Once these symmetries have been broken, the PS kinetic energy term of the HLS Lagrangian
 is no longer diagonal and a field redefinition is mandatory to restore its
canonical form\footnote{Regarding the vector meson sector, the layout of the full 
renormalization procedure 
has been already defined and thoroughly described in \cite{ExtMod7};
it applies here without any modification and will not be rediscussed.}.
This is performed in the two steps addressed right now.

\subsection{Diagonalization of the ${\cal L}_{A}$ PS Kinetic Energy Piece} 
\label{diag_la}
\indentB
In the BHLS/BHLS$_2$ model,  
the pseudoscalar (PS) kinetic energy term  writes \cite{ExtMod3,ExtMod7}~:
\be
 \displaystyle {\cal L}_{A,kin} = {\rm Tr} \left [ \pa P_{bare}  X_A \pa P_{bare}  X_A\right ] \,,
   \label{kin1}
\ee
where $X_A$ is the so--called BKY breaking matrix at work in 
the ${\cal L}_{A}$ sector of the BHLS$_2$ non-anomalous Lagrangian
\cite{ExtMod7} (${\cal L}= {\cal L}_A+ a {\cal L}_V$);
combining  the new breaking scheme defined in
\cite{Heath}  and the extension proposed in \cite{Hashimoto},
it writes \cite{ExtMod3}~:
\be
\left \{
\begin{array}{ll}
 \displaystyle X_A={\rm Diag} [q_A, y_A,z_A] \\[0.5cm]
 \displaystyle 
  q_A=1+\frac{\Sigma_A+\Delta_A}{2}~,&  \displaystyle 
y_A=1+\frac{\Sigma_A-\Delta_A}{2}~~.
\end{array}
\right.
\label{kin2}
\ee
The departure from unity of the $(u,\overline{u})$ and $(d,\overline{d})$ 
entries 
($q_A$ and $y_A$)  of $X_A$, numerically small \cite{ExtMod3},  are treated 
as ${\cal O} (\delta)$ perturbations\footnote{ \label{delta} We have already
defined  heuristically the first
non--leading order in perturbative expansions by some generic (perturbation)
$\delta$ parameter; this notation is prefered to the previous naming 
$\epsilon$ used in  our \cite{ExtMod7} to avoid confusion with breaking parameters
to be introduced below.} 
in amplitude calculations whereas $z_A$ occuring as the $X_A$ $(s,\overline{s})$  entry
is expected and treated as ${\cal O} (1)$; this entry can be also referred to as
flavor breaking \cite{feldmann_3,feldmann_1,feldmann_2}. 
Assuming the pion decay constant $f_\pi$ 
occuring in the HLS based Lagrangian models is the observed one, its renormalization
is unnecessary and has been shown to imply \cite{ExtMod3}  $\Sigma_A=0$. 
Therefore, phenomenologically, one is left with only two free parameters, $\Delta_A$ and 
$z_A$.

To restore the PS kinetic energy of the ${\cal L}_{A}$ piece of the BKY broken HLS
Lagrangians to canonical  form, 
 a first field transform \cite{WZWChPT} is performed~:
\be
 P_{R1}  =     X_A^{1/2} P_{bare} X_A^{1/2}~~;
\label{kin3}
\ee
$P_{R1}$ is the (first step) renormalized PS field matrix which brings
${\cal L}_{A,kin}$ back into canonical form. One has~:
 \be
\left (
\begin{array}{l}
 \displaystyle\pi^3_{bare} \\[0.5cm]
 \displaystyle\eta^0_{bare}  \\[0.5cm]
 \displaystyle\eta^8_{bare} 
\end{array}
\right )
=
\left (
\begin{array}{ccc}
1 	& \displaystyle -\frac{\Delta_A}{\sqrt{6}} & \displaystyle  -\frac{\Delta_A}{2\sqrt{3}}   \\[0.5cm]
 \displaystyle - \frac{\Delta_A}{\sqrt{6}}   & \displaystyle B  & \displaystyle A \\[0.5cm]
\displaystyle  -\frac{\Delta_A}{2\sqrt{3}}     & \displaystyle A& \displaystyle C 
\end{array}   
\right )
\left (
\begin{array}{l}
 \displaystyle \pi^3_{R1} \\[0.5cm]
 \displaystyle \eta^0_{R1}  \\[0.5cm]
 \displaystyle \eta^8_{R1}  
\end{array}
\right )
\displaystyle = W
\left (
\begin{array}{l}
 \displaystyle \pi^3_{R1} \\[0.5cm]
 \displaystyle \eta^0_{R1}  \\[0.5cm]
 \displaystyle \eta^8_{R1}  
\end{array}
\right )
\label{kin4}
\ee
which defines the matrix $W$.
We use the notation $\pi^3$ to remind the specific Gell-Mann matrix to which 
the neutral pion is associated, and devote the notation $\pi^0$ to the 
corresponding mass eigenstate. The $(\eta_0,\eta_8)$  entries of $W$
in Equation (\ref{kin4}) are given by~:
 \be
\begin{array}{llll}
\displaystyle A=\sqrt{2} \frac{z_A-1}{3z_A} ~,& B= \displaystyle \frac{2 z_A+1}{3z_A} ~,& C=\displaystyle \frac{z_A+2}{3z_A}
~,& \displaystyle  (BC-A^2 = \frac{1}{z_A})
\end{array}
\label{kin5}
\ee
which are used all along this study.
For further use, Equation (\ref{kin4}) is re-expressed~:
 \be
 \begin{array}{ llll}
\displaystyle  {\cal V}_{bare}= W {\cal V}_{R1}~,
& {\rm with~~}\displaystyle {\cal V}_{any}^t=( \pi^3_{any},~\eta^0_{any},~~\eta^8_{any})
& %\displaystyle 
(any =bare,R_1)~.
\end{array}
\label{kin4b}
\ee
 In terms of 
the $R1$ renormalized fields, ${\cal L}_{A,kin} $ is thus canonical~:
\be
{\cal L}_{A,kin} =\displaystyle \frac{1}{2} \left \{
\left [\pa_\mu  \pi^3_{R1}\right ]^2 +\left [\pa_\mu  \eta^0_{R1}\right ]^2 + \left  [\pa_\mu  \eta^8_{R1}\right ]^2  
\right \}~.
\label{kin6}
\ee

\vspace{0.7cm}

The following expression  of  $X_A$  in  the $U(3)$ algebra canonical basis 
clearly exhibits the precise structure of the BKY breaking procedure~:
\be
X_A= I + \left \{ \Delta_A T_3 +
\sqrt{\frac{2}{3}} \left [z_A-1 \right ] \left [ T_0 -\sqrt{2}~ T_8 \right ] \right \} \,,
\ee
where $I$ is the unit matrix.  $T_0=I/\sqrt{6} $ complements the usual Gell--Mann
matrices normalized by Tr$[T_a T_b] =\delta_{a b} /2$. 

Displayed in this way,
the departure from unity of the $X_A$ breaking matrix exhibits its 3 expected 
contributions. $\Delta_A$, a purely  Isospin Symmetry breaking (ISB) 
parameter,  is associated with  $T_3 ={\rm Diag} (1,-1,0)/2$ as it should. 
In contrast, the effect of the flavor breaking amount $z_A - 1$ will be met several times below. 
Its origin is naturally shared between $T_0 ={\rm Diag} (1,1,1)/\sqrt{6}$ and
$T_8 ={\rm Diag} (1,1,-2)/2\sqrt{3}$;  both simultaneously vanish in the "no--BKY
breaking" limit  $z_A=1$. So, as expected, the BKY matrix  $(s,\overline{s})$ entry,
combines correlatedly $SU(3)$ and  Nonet Symmetry (NSB) breakings in the PS sector. Let us
also note that \cite{ExtMod7} $z_A =[f_K/f_\pi]^2 + {\cal O}(\delta)$ as will be 
reminded below in the EBHLS$_2$ modified context -- see Equations (\ref{other4}).

\subsection{The Kinetic Breaking~: A Generalization of the 't~Hooft Term}
\label{XA-BRK}
\indent \indent A more direct breaking of the $U(3)$ symmetric PS field matrix
to $SU(3)\times U(1)$ has also been found phenomenologically requested to succesfully
deal with the whole BHLS realm of experimental data \cite{ExtMod3,ExtMod7}.
These are  the so--called  't~Hooft determinant terms 
\cite{tHooft,Kaiser_2000,leutw,WZWChPT};
 limiting oneself here to the kinetic energy term, one has been led 
 to supplement the HLS kinetic energy piece by~:
 \be
\begin{array}{llll}
 \displaystyle  {\cal L}_{'tHooft}= \lambda \frac{f_\pi^2}{12} 
 {\rm Tr} \ln{\pa_\mu U} \times   {\rm Tr} \ln{\pa^\mu U^\dagger}~,&
 U=\xi_L^\dagger \xi_R= exp{[2i P /f_\pi]} ~~,
\end{array}
\label{kin7}
\ee
where $P$ is the  usual $U(3)$ symmetric pseudoscalar bare field matrix 
\cite{ExtMod3,ExtMod7} and $f_\pi$ the (measured) charged pion decay
constant. This relation is connected with $\det{\pa U}$ by the identity~:

$$ \ln{{\rm det}\pa_\mu U}= \ {\rm Tr} \ln{\pa_\mu U}~~.$$
Expanding $\ln{\pa_\mu U}$ in Equation (\ref{kin7}),  the leading order
term is\footnote{In the literature, $\lambda$ is named $\Lambda_1$
\cite{Kaiser_2000,leutw,feldmann_3}. 
Removing the derivative symbols in Equation (\ref{kin7})  generates 
 a singlet mass term -- the topological susceptibility  --  to account for 
the $\etp$ mass.}~:
\be
\displaystyle  {\cal L}_{'tHooft}= \frac{\lambda}{2} 
\pa_\mu \eta^0_{bare} \pa^\mu \eta^0_{bare}
\label{kin8}
\ee
only involving the singlet PS bare field $\eta^0_{bare}$.

The 't~Hooft term tool, already used in the previous broken HLS versions,
can be fruitfully generalized. Indeed, Equation (\ref{kin7}) can be interpreted as~: 
\be
\displaystyle  
{\cal L}_{'tHooft}= \frac{f_\pi^2}{2}
 {\rm Tr} \ln{ X_H \pa_\mu U} \times   {\rm Tr} \ln{X_H \pa^\mu U^\dagger}~
\label{kin9}
\ee
where\footnote{Assuming $\lambda$ be positive, which is supported
by our  former fit results \cite{ExtMod3,ExtMod7}.} $X_H=\sqrt{\lambda} T_0$. 

Equation (\ref{kin9}) gives a hint that other well-chosen forms
 of the $X_H$ matrix  may exhibit interesting properties.
Indeed, it clearly permits  to define mechanisms not limited to only
nonet symmetry breaking.  
This leads us to propose the following choice for $X_H$~:
\be
\displaystyle  
 X_H = \lambda_0 T_0 +\lambda_3 T_3+\lambda_8 T_8
\label{kin10}
\ee
as it manifestly allows for a breaking of Isospin Symmetry and 
enrich  the HLS Model ability  to cover the
 $(\pi^0, ~\eta, ~\eta^\prime)$ mixing properties. As will be seen shortly,
it also leads to differentiate the pion pair couplings to
$\gamma$ and $W^\pm$.
With this choice, Equation (\ref{kin9}) becomes at leading order~:
\be
\displaystyle  
{\cal L}_{'tHooft}= \frac{1}{2} \left [
\lambda_3 \pa_\mu \pi^3 +\lambda_0  \pa_\mu \eta^0+\lambda_8  \pa_\mu \eta^8
\right ]
\left [
\lambda_3 \pa^\mu \pi^3 +\lambda_0  \pa^\mu \eta^0+\lambda_8  \pa^\mu \eta^8
\right  ]~~.
\label{kin11}
\ee
This form for $X_H$ is certainly not the unique way to generalize the usual
't~Hooft term. For instance, among other possible choices, one could quote~:
\be
\begin{array}{ll}
\displaystyle  
{\cal L}_{'tHooft}= \frac{f_\pi^2}{2} \sum_{a=0,3,8}
 {\rm Tr} \ln{ X_{H_a} \pa_\mu U} \times   
 {\rm Tr} \ln{X_{H_a} \pa^\mu U^\dagger}~~,&
\displaystyle  
 X_{H_a} = \lambda_a T_a~~,
 \end{array}
 \label{kin11b}
\ee
which turns out to drop the crossed terms in Equation (\ref{kin11}) -- and in all
expressions reported below. 
On the other hand, as will be seen in
Section \ref{kroll_brk1},   it happens that a generalization such as Equation
(\ref{kin10}) is necessary to allow BHLS$_2$  to fulfill expected properties
of the axial currents in a non-trivial way.

In order to deal with kaons or charged pions, one could also define appropriate 
projectors $X_H$; however,  this does not look necessary as
 the BKY breaking already 
produces the needed  breaking effects \cite{ExtMod3,ExtMod7}.

In the broken HLS frameworks previously defined, the (single) 't~Hooft breaking parameter
$\lambda~(=\lambda_0^2)$ was counted as  ${\cal O}(\delta)$ when truncating 
the Lagrangian to its leading  order terms in all the previously defined BHLS$_2$
breaking parameters \cite{ExtMod7}. Consistency,
thus, implies to count all the $\lambda_i$ just introduced  as  ${\cal O}(\delta^{1/2})$.

\subsection{The PS Kinetic Energy of the Extended BHLS$_2$ Lagrangian}
\indent \indent
The full PS kinetic energy term of the broken HLS Lagrangians is provided by their
 ${\cal L}_{A}^\prime \equiv {\cal L}_{A} +{\cal L}_{'t~Hooft}$ Lagrangian piece~:
\be
 \displaystyle {\cal L}_{kin}^\prime = {\rm Tr} \left [ \pa P_{bare}
   X_A \pa P_{bare}  X_A\right ]
 + 2 ~ \{ {\rm Tr} \left [X_H \pa  P_{bare} \right ] \}^2  ~~.
   \label{kin12}
\ee
Performing the change of fields of Equations (\ref{kin4}) 
which diagonalizes  ${\cal L}_{A,kin} $
and using  $X_H$ as given in Equation (\ref{kin10}), the full 
kinetic energy term ${\cal L}_{kin}^\prime$  can be written~:
 \be
\begin{array}{ll}
 \displaystyle {\cal L}_{kin}^\prime =& \displaystyle  
 \frac{1}{2} \left [ (1+\lambda_3^2) \pa_\mu \pi^3_{R1} \pa^\mu \pi^3_{R1} 
+(1+\tlambda_0^2) \pa_\mu \eta^0_{R1} \pa^\mu \eta^0_{R1} 
+(1+\tlambda_8^2) \pa_\mu \eta^8_{R1} \pa^\mu \eta^8_{R1} \right .\\[0.5cm]
~~& + \displaystyle \left . 
2 \tlambda_0 \tlambda_8~ \eta^0_{R1} \eta^8_{R1}
+2\lambda_3\tlambda_0~\pi^3_{R1}\eta^0_{R1} + 2\lambda_3\tlambda_8~\pi^3_{R1}\eta^8_{R1} 
\right ]  ~~,
\end{array}
\label{kin13}
\ee
omitting the kaon and (charged)  pion terms which are standard and displayed 
elsewhere \cite{ExtMod3,ExtMod7}.  We have defined~:
 \be
\begin{array}{ll}
 \displaystyle \tlambda_0= \lambda_0 B + \lambda_8 A~,
&  \displaystyle \tlambda_8= \lambda_0 A + \lambda_8 C
 \end{array}
 \label{kin14}
\ee
where $A$, $B$, $C$ are given by Equations (\ref{kin5}). One should note the intricate
combination of the 't~Hooft breaking parameters with the BKY parameter $z_A$.

Defining the (co-)vector
${\cal V}^t_{R1} = (\pi^3_{R1},~\eta^0_{R1},~\eta^8_{R1})$, 
${\cal L}_{kin}^\prime$ can be written~:

 \be
  \displaystyle {\cal L}_{kin}^\prime= 
  \frac{1}{2} ~{\cal V}_{R1}^t \! \cdot \! M  \! \cdot \! {\cal V}_{R1}\,,
\label{kin15}
\ee
$M$ being the sum of the unit matrix and of a rank 1  one writes~:
 \be
\begin{array}{ll} 
\displaystyle
 M = 1+ a \cdot a^t ~,& {\rm where~~~}  a^t=(\lambda_3,~\tlambda_0,~\tlambda_8)~~.
 \end{array}
\label{kin16}
\ee
The second step renormalized fields 
${\cal V}^t_{R} = (\pi^3_{R},~\eta^0_{R},~\eta^8_{R})$
are defined by~:
 \be
\displaystyle
 {\cal V}_{R} = \left [1+ \frac{1}{2}a \cdot a^t \right ] \cdot {\cal V}_{R1}
   \label{kin17}
\ee
which brings the kinetic energy into canonical form~:
 \be
  \displaystyle {\cal L}_{kin}^\prime = 
  \frac{1}{2} ~{\cal V}_{R}^t \cdot  {\cal V}_{R} + {\cal O}(\delta^2)~~.
\label{kin18}
\ee

At the same order, one has~:
 \be
\displaystyle
 {\cal V}_{R1} = \left [1- \frac{1}{2}a \cdot a^t \right ] \cdot {\cal V}_{R}
 + {\cal O}(\delta^2)
   \label{kin19}
\ee
and, finally, using Equations (\ref{kin5}) and (\ref{kin4b})~:
 \be
\displaystyle
 {\cal V}_{bare} = W \cdot \left [1- \frac{1}{2}a \cdot a^t \right ] \cdot {\cal V}_{R}+
 {\cal O}(\delta^2)~~,
   \label{kin20}
\ee
where $W$ is defined in Equation (\ref{kin4}).
 
\section{PS Meson Mass Eigenstates~: The Physical PS field Basis}
\label{PhysicalPSfields} 
\indentB The PS field $R$ basis renders the kinetic energy term
canonical; nevertheless, this $R$ basis is not expected to diagonalize the PS
mass term into its  mass eigenstates ($\pi^0,~\eta,~\etp$). Indeed, for instance, 
up to small perturbations, 
the $\eta^0_R$ and $\eta^8_R$ are almost pure singlet and octet field combinations,
while the physically observed $\eta$ and $\etp$ mass eigenstate fields are  
  mixtures of these. In order to preserve the canonical structure
of the PS kinetic energy one should consider the transformation from $R$ fields 
to the physically observed mass eigenstates; as  the PS mass term 
(not shown) is certainly a positive definite quadratic form, this transformation
should be a pure rotation.

In the traditional approach,  the physical $\eta$ and $\etp$ fields are related
to the singlet-octet states by the so--called  one angle transform~:
\be
\left (
\begin{array}{l} 
\displaystyle \eta \\[0.5cm]
\displaystyle  \eta^\prime
\end{array}
\right )
=
\left (
\begin{array}{cc} 
\displaystyle  \cos{\theta_P}&\displaystyle  -\sin{\theta_P} \\[0.5cm]
\displaystyle  \sin{\theta_P}&\displaystyle  \cos{\theta_P} 
\end{array}
\right )
\left (
\begin{array}{l} 
\displaystyle \eta^8_R \\[0.5cm]
\displaystyle  \eta^0_R
\end{array}
\right ) ~.
 \label{current12}
\ee
However, extending to the  mass eigenstate ($\pi^0,~\eta,~\etp$) triplet, one expects a 
 3--dimensional rotation and thus 3 angles. Adopting 
 the Leutwyler parametrization \cite{leutw96}, one has~:
\be
\left (
\begin{array}{l} 
\displaystyle \pi^3_R \\[0.5cm]
\displaystyle \eta^8_R \\[0.5cm]
\displaystyle  \eta^0_R
\end{array}
\right )
=
\left (
\begin{array}{ccc} 
\displaystyle 1 & \displaystyle -\hvar & \displaystyle -\hvar^\prime \\[0.5cm]
\displaystyle \hvar \cos{\theta_P}+\hvar^\prime \sin{\theta_P} &
\displaystyle \cos{\theta_P} &\displaystyle \sin{\theta_P} \\[0.5cm]
\displaystyle -\hvar \sin{\theta_P}+\hvar^\prime \cos{\theta_P} &
\displaystyle -\sin{\theta_P} &\displaystyle \cos{\theta_P}  
\end{array}
\right )
\left (
\begin{array}{l} 
\displaystyle \pi^0 \\[0.5cm]
\displaystyle \eta \\[0.5cm]
\displaystyle  \etp
\end{array}
\right )
 \label{current12-1}  
\ee
to relate the $R$ fields which diagonalize the kinetic energy to the physical
({\it i.e. } mass eigenstates) neutral PS fields. The three angles occuring there
($\hvar$, $\hvar^\prime$ and even $\theta_P$)
are assumed  ${\cal O}(\delta)$ perturbations; nevertheless, 
it looks better to stick close to the one--angle picture by keeping the 
trigonometric functions of $\theta_P$,
the so--called third mixing angle \cite{WZWChPT};
for clarity and for the sake of comparison with other works,
$\theta_P$ is not treated as manifestly small.
The Leutwyler rotation matrix can be factored out into
a product of 2 rotation matrices~:
\be
\left (
\begin{array}{ccc} 
\displaystyle 1 & \displaystyle -\hvar & \displaystyle -\hvar^\prime \\[0.5cm]
\displaystyle \hvar \cos{\theta_P}+\hvar^\prime \sin{\theta_P} &
\displaystyle \cos{\theta_P} &\displaystyle \sin{\theta_P} \\[0.5cm]
\displaystyle -\hvar \sin{\theta_P}+\hvar^\prime \cos{\theta_P} &
\displaystyle -\sin{\theta_P} &\displaystyle \cos{\theta_P}  
\end{array}
\right )
=
\left (
\begin{array}{ccc} 
\displaystyle 1 & \displaystyle 0 & \displaystyle 0 \\[0.5cm]
\displaystyle 0 & \displaystyle \cos{\theta_P} &\displaystyle \sin{\theta_P} \\[0.5cm]
\displaystyle 0 & \displaystyle -\sin{\theta_P} &\displaystyle \cos{\theta_P}  
\end{array}
\right )
\left (
\begin{array}{lll} 
\displaystyle 1 & \displaystyle -\hvar & \displaystyle -\hvar^\prime \\[0.5cm]
\displaystyle \hvar  & \displaystyle 1 &\displaystyle 0 \\[0.5cm]
\displaystyle \hvar^\prime  & \displaystyle 0  &\displaystyle 1 
\end{array}
\right ).
 \label{current12-2}
\ee
Substantially, the second matrix in the right-hand side of  Equation (\ref{current12-2})  
reflects Isospin breaking effects.  In the following, one names 
 $M(\theta_P)$ and $M(\hvar)$ the two matrices showing up there; they fulfill~:
\be
\displaystyle  \left [ M(\theta_P) \cdot M(\hvar) \right ]^{-1}=
\widetilde{M}(\hvar)\widetilde{M}(\theta_P)
\label{current12-3}
\ee 
up to terms of degree higher than 1 in $\delta$. This implies \cite{leutw96}~:
\be
\left (
\begin{array}{l} 
\displaystyle \pi^0\\[0.5cm]
\displaystyle \eta \\[0.5cm]
\displaystyle  \etp
\end{array}
\right )
=
\left (
\begin{array}{ccc} 
\displaystyle 1 &\displaystyle \hvar \cos{\theta_P}+\hvar^\prime \sin{\theta_P} & 
\displaystyle -\hvar \sin{\theta_P}+\hvar^\prime \cos{\theta_P} \\[0.5cm]
\displaystyle -\hvar & \displaystyle \cos{\theta_P} &\displaystyle -\sin{\theta_P} \\[0.5cm]
\displaystyle -\hvar^\prime &\sin{\theta_P} & \cos{\theta_P}
\end{array}
\right )
\left (
\begin{array}{l} 
\displaystyle \pi^3_R \\[0.5cm]
\displaystyle \eta^8_R \\[0.5cm]
\displaystyle  \eta^0_R
\end{array}
\right )~.
 \label{current12-4} 
\ee

 As for their perturbative order,  $\hvar$ and $\hvar^\prime$ are treated as ${\cal O}(\delta)$.
  Equations (\ref{current12-4}) and  (\ref{kin20}) allow to define the (linear) relationship between the
  physical $\pi^0$, $\eta$ and $\etp$ states and their bare partners occuring in the original HLS Lagrangians.

\section{Extended BHLS$_2$~: The Non--Anomalous Lagrangian}
\label{tau_sector}
\indentB
The non--anomalous EBHLS$_2$ Lagrangian in the present approach can also be written~:
\be
\displaystyle
{\cal L}_{HLS} = {\cal L}_{A}+a {\cal L}_{V}  +{\cal L}_{p^4}\,,
\label{AA1}
\ee
as in \cite{ExtMod7}. As in this Reference, one   splits up the first two terms
in a more appropriate way~:
\be
\displaystyle
{\cal L}_{A}+a {\cal L}_{V}  = {\cal L}_{VMD}+{\cal L}_{\tau}~~.
\label{AA2}
\ee
${\cal L}_{VMD}$ essentially adresses the physics of the $e^+e^-$
annihilations to charged pions and to kaons pairs; within the present breaking scheme,
it remains strictly identical to those displayed in Appendix A.1 of \cite{ExtMod7}.
The ${\cal L}_{p^4}$ is also unchanged as it does not address PS mesons interactions; it
is identical to those displayed in Appendix A.3  of \cite{ExtMod7}. Both pieces have not
to be discussed any further and their expressions will not be reproduced here
to avoid lengthy repetitions.
\vspace{0.4cm}

All modifications induced by the generalized 't~Hooft kinetic breaking mechanism 
are concentrated in the ${\cal L}_{\tau}$ piece and are displayed
right now. Using\footnote{As stated in Section \ref{param_free}, the 
breaking parameter $\Sigma_V$ is phenomenologically out of reach and one imposes
$\Sigma_V \equiv 0$ within our fits; it is, nevertheless, kept in the model expressions
for information.} ($m^2=a g^2 f_\pi^2$)~:
\be
\displaystyle 
m_{\rho^\pm}^2=m^2\left[ 1+ \Sigma_V \right]~~~,~~~
f_{\rho W} =  a g  f_\pi^2 \left[ 1+ \Sigma_V \right]\,,
\label{AA4}
\ee
the expression of ${\cal L}_{\tau}$  in terms of bare PS fields 
is given,  at lowest order in the breaking 
parameters, by~:
\be
\hspace{-0.5cm}
\begin{array}{ll}
{\cal L}_{\tau}& = \displaystyle - \frac{i V_{ud}g_2}{2} W^+ \cdot 
\left [ (1 -\frac{a (1+ \Sigma_V )}{2} ) \pi^- \parsym \pi^3_b
+\frac{1}{\sqrt{2}} [1 -\frac{a}{2z_A}(1+ \Sigma_V )] K^0 \parsym K^-
\right]
\\[0.5cm]
\hspace{-1.cm} ~& \displaystyle  + m_{\rho^\pm}^2 \rho^+ \cdot \rho^- 
-\frac{g_2V_{ud}}{2}  f_{\rho W} W^+ \cdot \rho^-
+ \frac{i a g }{2} (1+ \Sigma_V)
\rho^- \cdot \left [\pi^+ \parsym \pi^3_b +\frac{1}{z_A\sqrt{2}}\overline{K}^0 \parsym K^+  
\right ]  \\[0.5cm]
\displaystyle 
\hspace{-1.cm} ~& \displaystyle + \frac{f_\pi^2 g_2^2}{4} \left \{ 
\left[\displaystyle [(1+\frac{\Delta_A}{2}) z_A
+a  z_V (1+ \frac{\Sigma_V }{2})] |V_{us}|^2 + 
[1+a(1+ \Sigma_V )] |V_{ud}|^2\right]
\right \}W^+ \cdot W^-  \\[0.5cm]
\hspace{-3.cm} ~& \displaystyle 
+\frac{1}{9} a e^2 f_\pi^2 \left (5+5\Sigma_V+z_V\right ) A^2~~,
\end{array}
\label{AA5}
\ee
where one has limited oneself to only display the terms relevant for
our purpose.  The (classical) photon and $W$ mass terms \cite{HLSRef,Heath}
are not considered  and are given only for completeness. 
However, it is worth reminding that the photon mass term does not prevent 
the photon pole to reside at $s=0$ as required \cite{Heath1998}, at leading
order.  
The interaction part of ${\cal L}_{\tau}$  can be split up into several  pieces. 
Discarding  couplings of the form $W K \eta_8$, one can write~:
\be
{\cal L}_{\tau} = \displaystyle {\cal L}_{\tau,K}
+{\cal L}_{\tau,K}^\dagger +{\cal L}_{\tau,\pi}  +{\cal L}_{\tau,\pi}^\dagger
 + m_{\rho^\pm}^2 \rho^+ \cdot \rho^- 
\label{AA6}
\ee
with~:
\be
\displaystyle {\cal L}_{\tau,K} =-\frac{i}{2\sqrt{2}}
\left \{
g_2 V_{ud} \left[ 1 -\frac{a}{2z_A}(1+ \Sigma_V ) \right] W^+
 +\frac{a g }{z_A} (1+ \Sigma_V)\rho^+
\right\} 
 \! \cdot\!~  K^0 \parsym K^-   
\label{AA7}
\ee
and~:
\be
\displaystyle {\cal L}_{\tau,\pi} =\displaystyle - \frac{i}{2}   
\left \{
g_2 V_{ud} \left[ 1 -\frac{a}{2}(1+ \Sigma_V ) \right] W^+
+ a g (1+ \Sigma_V)\rho^+
\right\} 
\cdot \! \pi^- \parsym \pi^3_b
-\frac{g_2V_{ud}}{2}  f_{\rho W} W^+  \! \cdot \!\rho^-
\label{AA8}
\ee
where the subscript $b$ indicates that the $\pi^0$ field is bare. Equations
(\ref{kin20}) provide the relationship  				
between bare and renormalized states, in particular~:
\be
\pi^3_b= \displaystyle   \left \{ 1 -\frac{\lambda_3^2}{2} \right \} \pi^3_R -
\left \{ \frac{1}{\sqrt{6}} \Delta_A + \frac{\lambda_3 \tlambda_0}{2} 
\right \} \eta^0_R -
\left \{ \frac{1}{2\sqrt{3}} \Delta_A + \frac{\lambda_3 \tlambda_8}{2} 
\right \} \eta^8_R ~~.
\label{AA9}
\ee
The occurence of the $\lambda_3$ parameter generates a decoupling 
of the $W \pi^\pm\pi^0$ and $A \pi^+\pi^-$ interaction
intensities. Using also Equation (\ref{current12-4}), 
 Equations (\ref{AA8}) and (\ref{AA9}) give at first non--leading order~:
\be
\left \{
\begin{array}{lll} 
\displaystyle {\cal L}_{\pi^0 \pi^\pm} = & \displaystyle - \frac{i}{2}   
\left \{g_2 V_{ud} \left[ 1 -\frac{a}{2}(1+ \Sigma_V ) \right] W^+ + a g (1+ \Sigma_V)\rho^+
\right\}  \left \{ 1 -\frac{\lambda_3^2}{2} \right\} \cdot \! \pi^- \parsym \pi^0 \,,\\[0.5cm]

\displaystyle {\cal L}_{\eta \pi^\pm} = & \displaystyle + \frac{i}{2}   
\left \{g_2 V_{ud} \left[ 1 -\frac{a}{2}(1+ \Sigma_V ) \right] W^+ + a g (1+ \Sigma_V)\rho^+
\right\} \\[0.5cm]
~&\displaystyle  \times \left [
\left \{ \frac{1}{\sqrt{6}} \Delta_A + \frac{\lambda_3 \tlambda_0}{2} \right \} \cos{\theta_P}
-\left \{ \frac{1}{2\sqrt{3}} \Delta_A + \frac{\lambda_3 \tlambda_8}{2} \right \} \sin{\theta_P}
+\hvar
\right ]\cdot \! \pi^- \parsym \eta    \,,\\[0.5cm]
\displaystyle {\cal L}_{\etp \pi^\pm} = & \displaystyle + \frac{i}{2}   
\left \{g_2 V_{ud} \left[ 1 -\frac{a}{2}(1+ \Sigma_V ) \right] W^+ + a g (1+ \Sigma_V)\rho^+
\right\}  \\[0.5cm]
~&\displaystyle  \times \left [
\left \{ \frac{1}{2\sqrt{3}} \Delta_A + \frac{\lambda_3 \tlambda_8}{2} \right \} \cos{\theta_P}
+\left \{ \frac{1}{\sqrt{6}} \Delta_A + \frac{\lambda_3 \tlambda_0}{2} \right \} \sin{\theta_P}
+\hvar^\prime \right ]\cdot \! \pi^- \parsym \etp  \,,
\end{array}
\right .
\label{AA10}
\ee
in terms of physical  neutral PS fields, and then~:
\be
\displaystyle {\cal L}_{\tau,\pi} ={\cal L}_{\pi^0 \pi^\pm}+{\cal L}_{\eta \pi^\pm}+{\cal L}_{\etp \pi^\pm}
-\frac{g_2V_{ud}}{2}  f_{\rho W} W^+  \! \cdot \!\rho^-
+{\mathrm h.c.}  + m_{\rho^\pm}^2 \rho^+ \cdot \rho^- ~~.
\label{AA11}
\ee

Once more, the rest of the ${\cal L}_A+ a {\cal L}_V$ is unchanged compared to their BHLS$_2$ 
expressions \cite{ExtMod7}.

Regarding the pion form factor in the $\tau$ decay, the changes versus Subsection 11.1 in \cite{ExtMod7}
and the present EBHLS$_2$ are very limited~:
\be
\left \{ 
\begin{array}{lll}
W^\mp \pi^\pm \pi^0 ~~{\rm coupling~:}& \displaystyle  \left[ 1 -\frac{a}{2}(1+ \Sigma_V ) \right]
&\Longrightarrow 
\displaystyle \left[ 1 -\frac{a}{2}(1+ \Sigma_V ) \right]\left \{ 1 -\frac{\lambda_3^2}{2} \right\} \,,\\[0.5cm]
\rho^\mp \pi^\pm \pi^0 ~~{\rm coupling~:}& ~~~~\displaystyle  ag (1+ \Sigma_V )
&\Longrightarrow 
\displaystyle ~~~ag (1+ \Sigma_V ) \left \{ 1 -\frac{\lambda_3^2}{2} \right\}~~.
\end{array}
\right .
\label{tranfo_tau}
\ee
This implies a global rescaling of the BHLS$_2$ pion form factor $F_\pi^\tau(s)$ by 
 $1-\lambda_3^2/2$; it also implies that the  $\pi^0 \pi^\pm$ loop acquires a factor of 
 $[1-\lambda_3^2]$. The BHLS$_2$ $W^\pm -\rho^\mp$ transition amplitude $F_\rho^\tau(s)$
 and the $\rho^\pm$ propagator $[D_\rho(s)]^{-1}$ are changing correspondingly
(See Subsection 11.1 in \cite{ExtMod7}). 
 On the other hand,  $F_\pi^e(s)$ remains identical to its BHLS$_2$ expression,
as well as both kaon form factors.

\section{Extended BHLS$_2$~: The Anomalous Lagrangian Pieces}
\label{anomalous_pieces}
\indentB If  only the ${\cal L}_\tau$ part of the non--anomalous BHLS$_2$ Lagrangian 
is affected by the kinetic breaking presented above, all the anomalous FKTUY
pieces  \cite{FKTUY,HLSRef} are concerned. 

The Lagrangian pieces of relevance for the phenomenology we address are, on the one hand~: 
\be
\left \{ 
\begin{array}{lll}
\displaystyle 
{\cal L}_{AAP}= -\frac{3 \alpha_{em}}{\pi f_\pi}(1-c_4) 
~\epsilon^{\mu \nu \alpha \beta} \pa_\mu A_\nu \pa_\alpha A_\beta \mathrm{Tr} \left [ Q^2 P \right ]  \,,\\[0.5cm]
\displaystyle 
{\cal L}_{VVP}= -\frac{3 g^2}{4 \pi^2 f_\pi}~c_3 
~\epsilon^{\mu \nu \alpha \beta} \mathrm{Tr} \left [ \pa_\mu V_\nu \pa_\alpha V_\beta   P \right ]  \,,
\end{array}
\right .
\label{anomLags_1}
\ee
where $Q$ is the usual quark charge matrix and 
$A$, $V$ and $P$ denote resp. the electromagnetic field, the vector field matrix and the
U(3) symmetric bare pseudoscalar field matrix as defined in \cite{ExtMod3} regarding their normalization.
As one did not find any important improvement by assuming $ c_3 \ne c_4 $, the difference of these has been 
set to zero; consequently, the ${\cal L}_{AVP}$ Lagrangian piece \cite{HLSRef} drops out.

On the other hand, the following pieces should also be considered~:
\be
\left \{ 
\begin{array}{lll}
\displaystyle 
{\cal L}_{APPP}= -i\frac{3 e}{3 \pi^2 f_\pi^3} \left[1-\frac{3}{4}(c_1-c_2+c_4)\right]
~\epsilon^{\mu \nu \alpha \beta} A_\mu
\mathrm{Tr} \left [Q ~\pa_\nu P \pa_\alpha   P \pa_\beta   P\right ]
 \,, \\[0.5cm]
\displaystyle 
{\cal L}_{VPPP}= -i\frac{3 g}{4 \pi^2 f_\pi^3} \left[c_1-c_2-c_3\right]
\displaystyle 
~\epsilon^{\mu \nu \alpha \beta} 
\mathrm{Tr} \left [V_\mu \pa_\nu P \pa_\alpha   P \pa_\beta   P\right ]~~.
\end{array}
\right .
\label{anomLags_2}
\ee
These involve, beside $c_3$ and $c_4$, a third parameter $c_1-c_2$ which is also not fixed within the
HLS framework \cite{HLSRef} and should be derived from the minimization procedure. In order to ease  the
reading of the text we have found it worth pushing them in Appendices \ref{AAP-VVP} 
and \ref{APPP-VPPP}.

Regarding the
pseudoscalar fields, the Lagrangian pieces listed in  Appendices \ref{AAP-VVP} and \ref{APPP-VPPP}, 
are expressed in terms of the physically observed $\pi^0,~\eta,~\etp$ whereas, for simplicity,
the vector mesons  are expressed in terms of their ideal combinations~: $\rho^0_I,~\omg_I$ 
and $\phi_I$.  The procedures to derive the couplings to the physically observed $\rho^0,~\omg$ and
$\phi$ and construct the cross-sections for the $e^+e^- \ra (\pi^0/\eta) \gam$ annihilations
are given in full details in  Section 12 of \cite{ExtMod7}.
Nevertheless, one has found worthwhile to construct the amplitude and the cross-section for the
$e^+e^- \ra \pi^0 \pi^+ \pi^-$ annihilations in the Extended BHLS$_2$ framework;
this information\footnote{One may note that $F_2(s)$ in Equation (\ref {eq2-30})
corrects for an error in \cite{ExtMod7} missed in the Erratum ($3\xi_3/2 \ra 2 \xi_3$). }
 is provided in Appendix \ref{3pionMod}.

\section{Update of the 3$\pi$ Annihilation Channel}
\label{BESIII} 
\indentB  The BESIII Collaboration has recently published the Born cross section
spectrum \cite{BESIII_3pi} for the $e^+e^- \ra \pi^+ \pi^- \pi^0$ annihilation
over the $0.7 \div 3.0$ GeV energy range collected  in the ISR mode. This new data
sample comes complementing the spectra collected at the VEPP-2M Collider by CMD2
\cite{CMD2KKb-1,CMD2-1998-1,CMD2-1998,CMD2-2006} and SND \cite{SND3pionLow,SND3pionHigh}
covering the $\omg$ and $\phi$ peak regions. Besides, the only data on the 3$\pi$ cross section 
stretching over the intermediate region was collected much earlier \cite{ND3pion-1991} by the 
former Neutral Dectector (ND). As the  measurement by BaBar \cite{BaBar_3pi} only covers
the $\sqrt{s} > 1.05$ GeV  region, it is of no concern for physics studies up to the $\phi$ signal.
On the other hand, previous independent analyses \cite{Kubis_3pi,ExtMod7} indicate that the CMD2 spectrum
\cite{CMD2-2006} returns an  average $\chi^2$ per point much above 2 units which led  to discard it
from global approaches. 

The BESIII sample \cite{BESIII_3pi} is the first 3 pion sample to encompass the whole range 
of validity of the HLS Model, providing a doubling of the number of candidate data 
points and, additionally, the first cross-check of the  cross section behavior
in the energy region in between the  narrow $\omg$ and $\phi$ signals.  

We first examine  how it fits within the global HLS framework in isolation ({\it i.e.}
as single representative of the 3$\pi$ annihilation channel) and conclude about its consistency
with the already analyzed data samples covering the {\it other} annihilation channels,
namely  $\pi \pi$, 
$(\pi^0/\eta) \gam$  and both $K \overline{K}$ modes. A second step is devoted to 
consistency studies between the BESIII spectrum and those previously  collected 
 in the same 3$\pi$ channel by 
the ND   \cite{ND3pion-1991}, CMD2 \cite{CMD2KKb-1,CMD2-1998-1,CMD2-1998} and SND 
\cite{SND3pionLow,SND3pionHigh} detectors.

\subsection{The BESIII 3$\pi$ Data Sample in Isolation within EBHLS$_2$}
\label{BESIII_isolated}
\indentB
The fit procedure already developped within the previous BHLS frameworks \cite{ExtMod3,ExtMod7} 
 -- and closely followed here -- relies on a global $\chi^2$ minimization.    In order
to include the BESIII sample within the EBHLS$_2$ framework\footnote{Actually,
the 3$\pi$ channel is  marginally sensitive to the differences between BHLS$_2$
 and its extension to EBHLS$_2$.}, one should first define its 
contribution to the global $\chi^2$. This requires to  define the error covariance
matrix merging appropriately the statistical and systematic uncertainties provided
 by the BESIII Collaboration together with its spectrum and paying special care 
to the normalization uncertainty treatment. This should be done by closely
following the information provided together with its spectrum by the 
Collaboration\footnote{In the process of sample combination frameworks, additional 
issues may arise; for instance, the consistency of the absolute energy calibration
of the various experiments with each other should be addressed as done in
\cite{ExtMod7} with the dikaon data samples collected
by   CMD3 \cite{CMD3_K0K0b,CMD3_KpKm} and BaBar \cite{BaBarKK}.}.

For definiteness, the data point of the BESIII sample \cite{BESIII_3pi} at 
the energy squared $s_i$ is~:
$$ m_i \pm \sigma_{stat,i} \pm \sigma_{syst,i} \,,$$
using obvious notations; it is useful to 
define $\sigma(s_i)=\sigma_{syst,i}/m_i$, the $i^{\rm th}$ experimental fractional systematic 
error. Then,
the elements of the full covariance matrix $W$ associated with the BESIII spectrum
write~:
\be
W_{ij}=V_{ij} + \sigma(s_i) \sigma(s_j) A_i A_j  \,,
\label{bes3_1}
\ee
where the indices run over the number of data points ($i,j= 1, \cdots N$).
$V$ is the (diagonal) matrix of the squared statistical errors ($\sigma_{stat,i}^2$),
and  $\sigma(s_i)$ is the reported fractional systematic error 
at the data point of energy (squared) $s_i$, defined as just above.
The systematic errors are considered point-to-point correlated and reflecting
a (global) normalization uncertainty.

At start of the fit iterative procedure, the natural choice for $A$ 
is the vector of measurements itself ($A_i=m_i$); in the iterations afterwards,
it is highly recommended \cite{D'Agostini,Blobel_2006,ExtMod5} to 
replace the measurements by the fitting model function $M$ ($A_i=M(s_i) \equiv M_i$) 
derived at the previous iteration step  for the concern of avoiding the
occurence of biases. Then, the experiment contribution to the global $\chi^2$ writes~:
\be
\chi^2 = (m - M)_i W^{-1}_{ij} (m - M)_j ~~.
\label{bes3_2}
\ee

Moreover, a normalization correction naturally follows from the global scale uncertainty. It is
a {\it derived} quantity of the minimization procedure. Defining the vector $B$
($B_i=\sigma(s_i) M_i$), this correction is given by \cite{ExtMod5,ExtMod7}~:
\be
\displaystyle \mu = \frac{B_i V^{-1}_{ij} (m - M)_j }{1+B_i V^{-1}_{ij} B_j} ~~.
\label{bes3_3}
\ee
For the purpose of graphical representation, one could either perform the 
replacement $m_i \ra m_i - \mu \sigma(s_i) M_i$, {\it or}  apply the correction to the
model function $M_i \ra [1+ \mu \sigma(s_i)] M_i$. When comparing graphically
several spectra,  the former option should clearly be prefered as, indeed, even if
not submitted to the fit, other parent data samples can be fruitfully represented in the 
same plot by performing the change\footnote{Numerically, $\mu$ is derived by
using the function values $M(s)$  at the central values of the fit parameters.}~: 
$$m^\prime_i \ra m^\prime_i - \mu \sigma^\prime_i(s^\prime_i) M(s^\prime_i)~,$$
using the BESIII fit function $M(s)$. 
Such a plot is obviously a relevant visual piece of information.

A global fit involving all data covering the $\pi \pi$, 
$(\pi^0/\eta) \gam$  and both $K \overline{K}$ channels and {\it only} the
BESIII spectrum to cover the 3$\pi$ annihilation final state
has been performed and returns, for the BESIII sample\footnote{We 
have prefered skipping the first few data points  more subject to 
non--negligible background; the spectrum is thus fitted in the energy 
range $\sqrt{s} \in [0.73 \div 1.05]$ GeV. }, $\chi^2/N= 170/128$.

\begin{figure}[!phtb!]
\hspace{-1.8cm}
\begin{minipage}{\textwidth}
%\begin{center}
%\hspace{-0.3cm}
\resizebox{1.2\textwidth}{!}
%{\includegraphics*{fig_2020/residuals_bes3pi_isolation_all.eps}}
{\includegraphics*{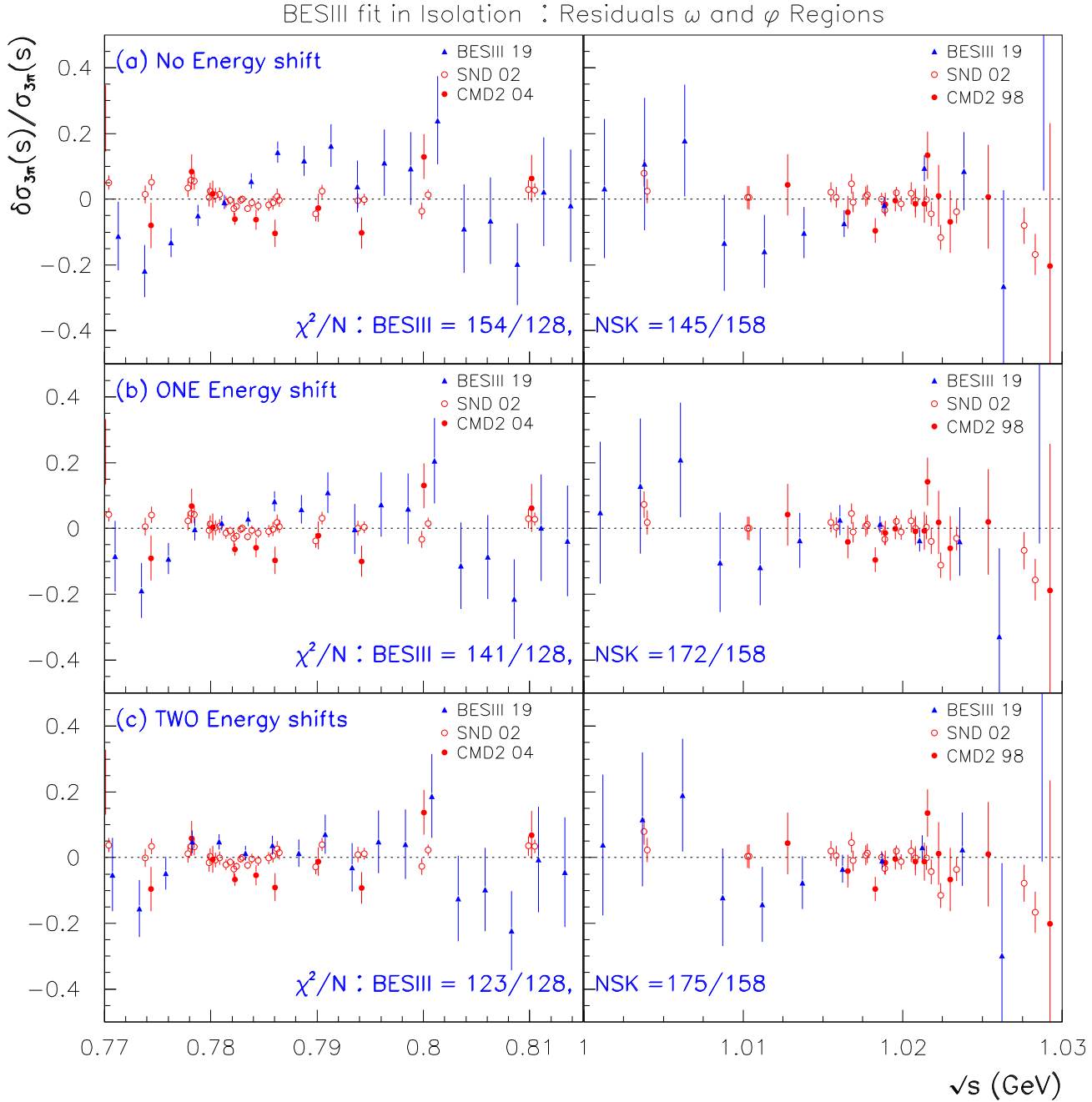}}
%\end{center}
\end{minipage}
\begin{center}
\caption{\label{Fig:ResBesIII} Normalized residuals of EBHLS$_2$ fits to the  BESIII 
$3 \pi$ data in isolation under 3 different configurations~:
No energy shift (top panels), one global
energy shift (middle panels) and two energy shifts (bottom panels).
The normalized residuals are defined as $\delta \sigma(s)/\sigma(s)$
where $\delta \sigma(s_i) = m^\prime_i- M(s_i)$ -- see the text for the definitions.
The partial
$\chi^2/N$'s are displayed. Also shown, the NSK $\chi^2/N$'s  distance
of the CMD2 and SND data to the best fit solutions derived from fitting BESIII
data in isolation in each  configuration.}
\end{center}
\end{figure}

The top panels in Figure \ref{Fig:ResBesIII} display the distribution of
the BESIII normalized residuals $\delta \sigma(s)/\sigma(s)$ corrected as reminded 
just above. In the $\omg$ region, at least,  the normalized residual distribution
 is clearly energy dependent. The normalized (pseudo-)residuals of the {\it unfitted} data
  samples displayed\footnote{For short, these will be referred to below
as NSK.}, namely those from \cite{CMD3pion-1989,SND3pionLow} in the 
$\omg$ region and from \cite{CMD2KKb-1,SND3pionHigh} in the $\phi$ region, 
likewise corrected for the normalization uncertainty, are, instead, 
satisfactory\footnote{Thus, the normalization correction applied to each of the NSK 3$\pi$ data samples
is determined by the fit of solely the BESIII data \cite{BESIII_3pi} within the global 
framework.}, despite being unfitted. The fit process allows to compute the (global) $\chi^2$ 
distance of the NSK samples to the (BESIII) fit function and returns
$\chi^2/N=180/158$, a reasonable value for unfitted data. 

However, the behavior of the BESIII residuals may indicate a mismatch between the
$\omg$ and $\phi$ pole positions in the BESIII sample compared to the other 
($\simeq 50$) data samples involved in the (global) fit; indeed, the narrow 
$\omg$ and $\phi$ signals are already present in the $(\pi^0/\eta) \gam$ and
$K \overline{K}$ channels, and therefore, as when dealing with the CMD3 and BaBar 
dikaon samples in \cite{ExtMod7}, a mass recalibration (shift) could be 
necessary to avoid mismatches with the pole positions for $\omg$ and $\phi$ 
in the other data samples.
We thus have refitted the BESIII data, by allowing for such a mass shift to 
recalibrate the BESIII energies and match our reference energy scale\footnote{Our 
reference energy is actually defined consistently by
more than 50 data samples. As an important part of these  has been 
collected at the VEPP-2M Collider in Novosibirsk, we denote, when 
needed, our reference energy by  $E_{NSK}$.}.  So, we define~: 
$$E_{BESIII} = E_{NSK} + \delta  E_{BESIII}$$ 
and let $\delta  E_{BESIII}$ 
vary within the fit procedure. The fit returns 
$\delta  E_{BESIII}= (-286.09 \pm 44.19)$ keV with  $\chi^2/N_{BESIII}= 141/128$
and, thus, the (noticeable)
gain of 29 units should be attributed to  only allowing for a non-zero
 $\delta  E_{BESIII}$.  The corresponding normalized 
residuals, displayed in the middle row
of Figure \ref{Fig:ResBesIII}, are clearly much improved, whereas the $\chi^2$
distance of the NSK 3$\pi$ data sample to the BESIII fit function stays alike. 

Owing to the sharp improvement produced by this mass shift, it was tempting to
check whether the energy (re-)calibration could be somewhat different at the
$\omg$ and the $\phi$ masses. For this purpose, it is appropriate to redefine
the fitting algorithm by stating~;

\be
\displaystyle 
E_{BESIII} = E_{NSK} + 
\left \{
\begin{array}{lll}
\delta  E_{BESIII}^\omg , & E_{BESIII} < E_{mid} \\[0.5cm]
\delta  E_{BESIII}^\phi, & E_{BESIII} > E_{mid}
\end{array}
\right .
\label{bes3_4}
\ee
where  $E_{mid}$  should be chosen appropriately, {\it i.e.} significantly 
outside the $\omg$  and $\phi$ peak  energy intervals.
As obvious from the bottom panel of Figure \ref{PIPIPI}, the 3$\pi$ cross section in the
intermediate energy region is almost flat and indicates that the choice for  $E_{mid}$ is 
far from critical; one chose   $E_{mid}=0.93$ GeV.

The corresponding global fit has been performed and returns $\chi^2/N_{BESIII}= 123/128$,
an additional gain of 18 $\chi^2$ units comes, to be cumulated with the previous 29 unit
gain. The recalibration constants versus $E_{NSK}$ are~:
$$\left \{\delta  E_{BESIII}^\omg =(-518.92 \pm 72.04)~{\rm keV}~,~
\delta  E_{BESIII}^\phi=(-118.58 \pm 58.72)~{\rm keV} \right \}~~.
$$ 
After having applied this recalibration, the BESIII
normalized residuals, shown in the panels of the  bottom row in Figure 
\ref{Fig:ResBesIII}, are observed flat, as well as their NSK partners also displayed.

One should remark that $\delta  E_{BESIII}^\omg$ is in striking  correspondance
with the central value for the energy shift reported by BESIII \cite{BESIII_3pi} 
compared to their Monte Carlo ($-0.53 \pm 0.25$ MeV)
and is found highly significant (about $7.5 \sigma$). $\delta  E_{BESIII}^\phi$
is also consistent with this number but quite significantly different from 
$\delta  E_{BESIII}^\omg$. Actually, comparing the three rightmost panels in Figure 
\ref{Fig:ResBesIII}, one observes that the main gain of decorrelating the 
energy calibration
at the $\omg$ and $\phi$ peaks widely improves the former energy region;
the later looks almost insensitive, as reflected by the fact that the non-zero
$\delta  E_{BESIII}^\phi$ is only a $2\sigma$ effect. 

So, once two energy recalibrations have been performed, the description of the BESIII sample
is quite satisfactory and, fitted with the other annihilation channels, the  
$\chi^2$ probability is comfortable (91.6\%).

At first sight, the differing energy shifts just reported may look surprising as, for 
ISR spectra, the energy calibration is very precisely
fixed by the energy at which the accelerator is running at
meson factories. However, such energy shifts could be related
with unaccounted effects of the secondary photon emission 
expected to affect the resonances showing up at lower energies. In
the case of the BESIII spectrum, this concerns the $\phi$ and $\omega$
regions, where photon radiation effects get enhanced by the
resonances, causing shifts between the physical resonance parameters
and their observed partners\footnote{ Note that, for the NSK scan experiments, 
photon emission on resonances  is corrected locally resonance by resonance. }. 
This topic is specifically addressed in Appendix \ref{fred_isr}, where it is 
shown -- and illustrated by Table~\ref{Table:T_shifts} therein -- 
that the expected shifts produced   by   secondary ISR photons
are in striking correspondence with the fitted
$\delta  E_{BESIII}^\omg$ and $\delta  E_{BESIII}^\phi$. 

\subsection{Exploratory EBHLS$_2$ global Fits including the BESIII 3$\pi$ Sample}
\begin{figure}[!phtb!]
\vspace{-1.cm}
\hspace{0.9cm}
\begin{minipage}{0.8\textwidth}
%\begin{center}
%\hspace{-0.3cm}
\resizebox{1.2\textwidth}{!}
%{\includegraphics*{fig_2020/residuals_bes3pi_nsk_all.eps}}
{\includegraphics*{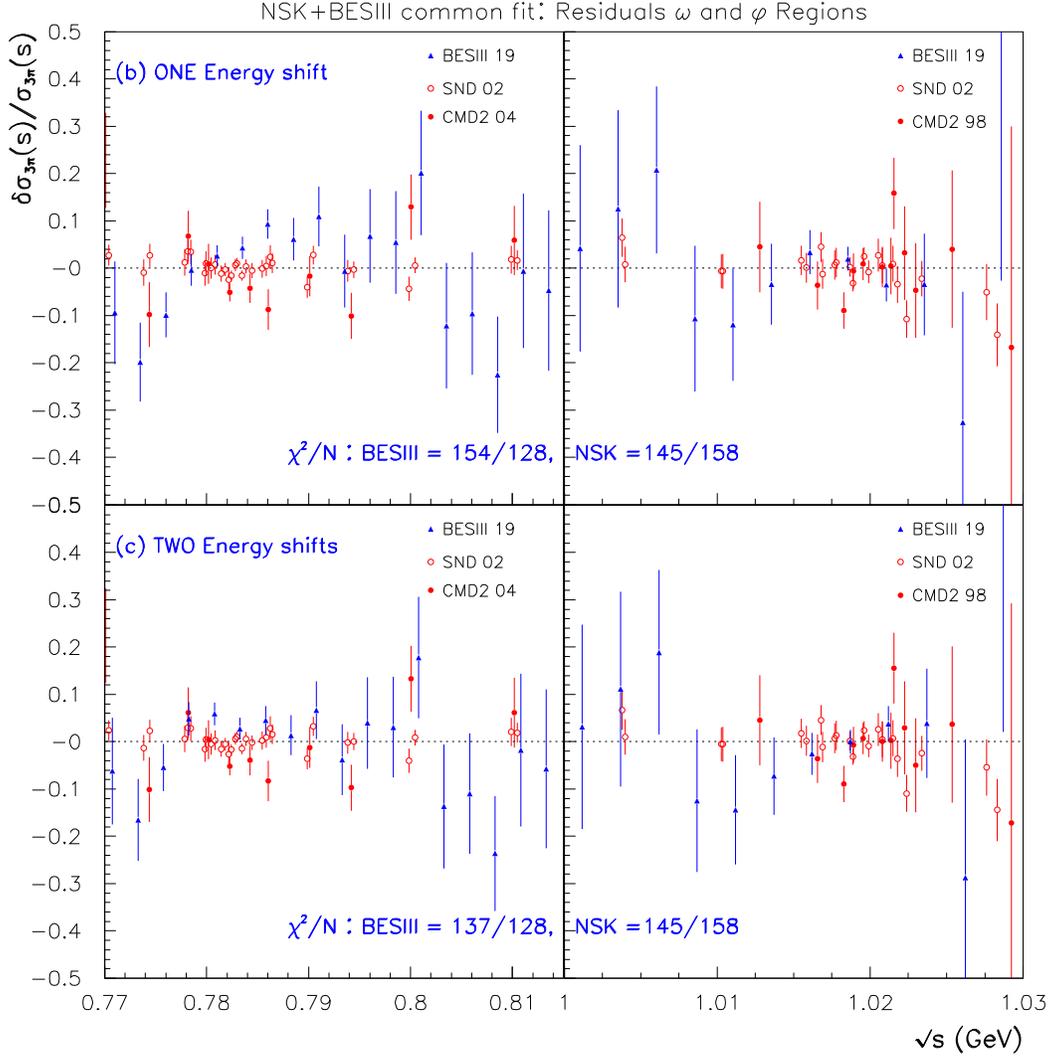}}
%\end{center}
\end{minipage}
\begin{center}
\caption{\label{Fig:Res_NSK_BesIII} Normalized residuals of EBHLS$_2$ fits to the  BESIII, CMD2 and
SND $3 \pi$ data under 2 different configurations~: Top panels correspond to a
(global) fit with only one energy shift for the BESIII spectrum, the bottom ones
are derived allowing different $\delta  E_{BESIII}^\omg$ and $\delta  E_{BESIII}^\phi$.
The partial $\chi^2/N$ are displayed for the BESIII sample on the one hand, and for
the CMD2 and SND ones (NSK) on the other hand. }
\end{center}
\end{figure}

\indentB Having proved that the BESIII 3$\pi$ data sample suitably fits the global EBHLS$_2$
framework, we perform the analysis by merging the BESIII and the parent CMD2 and SND data 
samples within a common fit procedure.  
For completeness, we have first performed a global fit allowing for a single
energy calibration constant. The fit returns $\chi^2/N=1284/1365$ and 84.7 \% probability.  
The $\chi^2/N $ values for BESIII (154/128) and NSK (145/158) are also reasonable, however, 
the normalized residuals for BESIII shown in the top panels of Figure 
\ref{Fig:Res_NSK_BesIII} -- especially the leftmost panel --
still exhibit a structured  behavior.

Therefore we have allowed for two independent energy shifts
 $\delta  E_{BESIII}^\omg$ and 
$\delta  E_{BESIII}^\phi$ within the iterative fit procedure. Convergence is reached 
with $\chi^2/N=1267/1365$ and 91.2\% probability.  The $\chi^2/N $ values for BESIII 
(137/128) is improved by 17 units whereas it is unchanged for NSK (147/158); so,
 the improvement of the total $\chi^2$ only proceeds from
the 17 unit reduction of the BESIII partial $\chi^2$. The bottom panels in Figure
\ref{Fig:Res_NSK_BesIII} are indeed observed flat in the $\omg$ and $\phi$
regions (this last distribution is still less sensitive to the fit quality improvement).
The energy recalibration constants of the BESIII data with regard to
the NSK energy scale are~:
$$\{\delta  E_{BESIII}^\omg =(-486.11 \pm 71.51)~{\rm keV}~,~
\delta  E_{BESIII}^\phi=(-135.31 \pm 59.16)~{\rm keV} \}~~,
$$ 
in fair accord with those derived in the global fit performed discarding 
the NSK 3$\pi$ data. Compared to its fit in isolation, the BESIII data $\chi^2$
is degraded by 137-123=14 units, while (see Table~\ref{Table:T3} third data column)
the NSK data $\chi^2$ is degraded by 147-135=12 units compared to the fit performed
discarding the BESIII data sample, the rest being unchanged. Regarding the 
average per data point, the degradation is
of the order 0.1  $\chi^2$--unit for both the NSK and BESIII samples, a quite insignificant 
change. So, one can conclude
that the full set of consistent data samples can welcome the  BESIII sample \cite{BESIII_3pi},
once the energy shifts $\delta  E_{BESIII}^\omg$ and $\delta  E_{BESIII}^\phi$ are
applied.

\begin{figure}[!phtb!]
\begin{minipage}{\textwidth}
\begin{center}
\resizebox{0.9\textwidth}{!}
%{\includegraphics*{fig_2020/3pion_BESIII.eps}}
{\includegraphics*{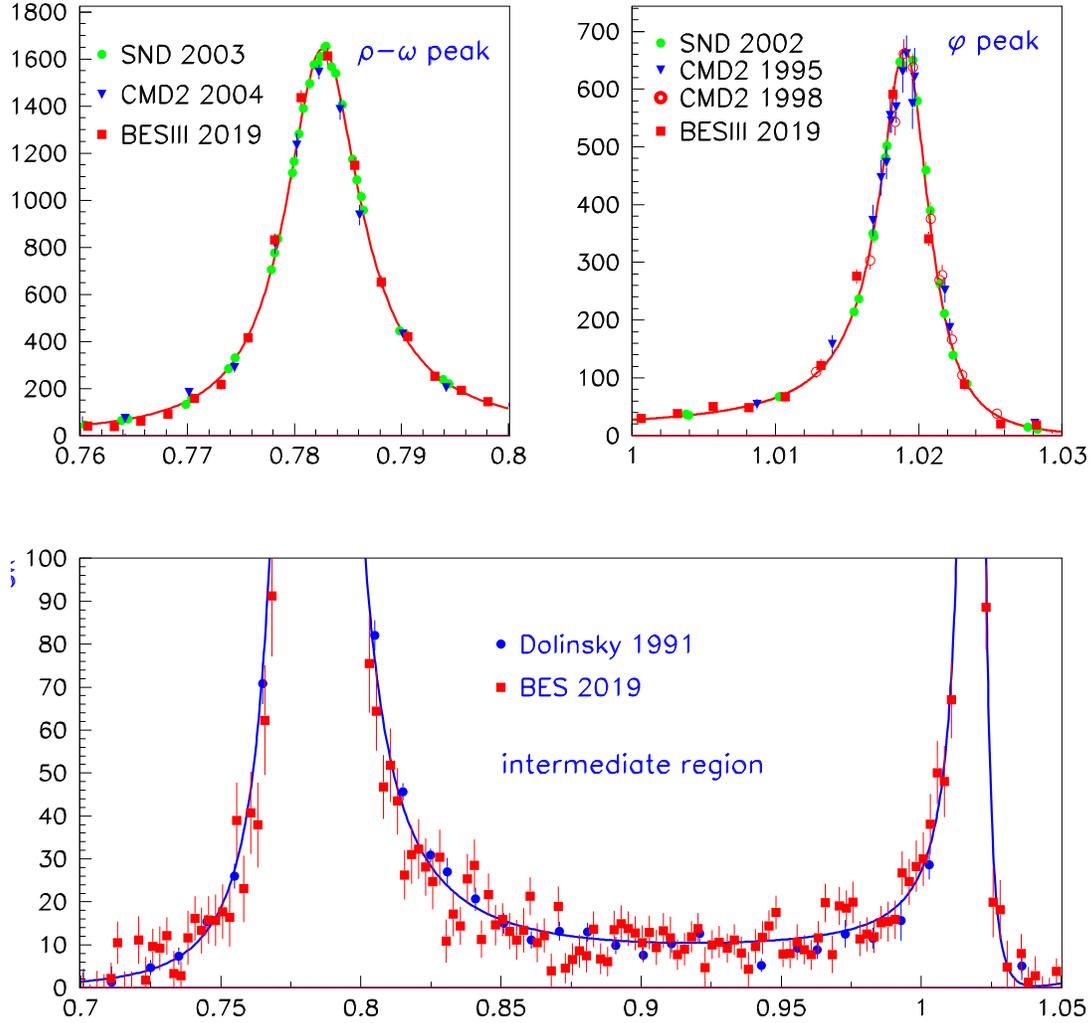}}
\end{center}
\end{minipage}
\begin{center}
\vspace{-0.3cm}
\caption{\label{PIPIPI} The global EBHSL$_2$ fit with the $\pi^+ \pi^-\pi^0$ spectra,
corrected for their normalization uncertainty; only statistical errors are shown.
The top panels display the data and fit in the $\omg$ and $\phi$ mass intervals, the
bottom panel focuses on the intermediate energy region. The energy recalibration has
been applied to the BESIII data. 
} 
\end{center}
\end{figure}

Figure \ref{PIPIPI} displays the fit function and data in the 3$\pi$ channel. All data
are normalization corrected  as emphasized above and, additionally, the energy
shifts induced by having different $\delta  E_{BESIII}^\omg$ and $\delta  E_{BESIII}^\phi$
calibration constants are applied to the BESIII data sample; one should also note
the nice matching of the ND and BESIII data in the intermediate region. 
Additional fit details of this new global fit are given in 
the third data column of Table~\ref{Table:T3}.

\section{Revisiting the $2 \pi$ Annihilation Channel}
%\section{Analysis of the $2 \pi$ Annihilation Channel : SND20}
\label{SND20txt}
\indentB  A fair understanding of
the dipion annihilation channel, which provides by far the largest contribution ($\simeq
75\%$)  to the muon HVP, is an important issue.
Fortunately, the  $e^+ e^- \ra \pi^+ \pi^-$  cross-section
is also the most important channel encompassed within the BHLS \cite{ExtMod3,ExtMod5}
and BHLS$_2$ \cite{ExtMod7} frameworks previously developped. All  the existing dipion
data samples were examined within the context of these two  variants of 
the HLS Model. As some of them   exhibit strong
tensions \cite{ExtMod3} with significant consequences on the derived physics 
quantities, it looks worthwhile to revisit this issue when a new measurement
arises, at least to check if the consistency pattern previously favored
deserves reexamination.

Beside the data samples formerly collected and gathered in  \cite{Barkov}, 
an important place should be devoted to the data from CMD-2 
\cite{CMD2-1995corr,CMD2-1998-1,CMD2-1998-2} and SND  \cite{SND-1998} 
collected in scan mode on the VEPP-2M collider at Novosibirsk; these CMD2 and SND
samples are collectively referred to below as NSK.  
These were followed  by higher statistics samples, namely, the KLOE08 spectrum \cite{KLOE08}
collected at Da$\Phi$ne and those collected by BaBar \cite{Davier2009} at PEP-II, both using the 
Initial State Radiation (ISR) method \cite{Benayoun_isr}.  Slightly later, the KLOE Collaboration
produced two more ISR data samples, KLOE10  \cite{KLOE10} and KLOE12  \cite{KLOE12}, the latter
being tightly related with KLOE08 (see Figure 1 in \cite{KLOEComb}). In this reference,
the KLOE-2 Collaboration has also  published a dipion spectrum derived by combining the KLOE08, 
KLOE10 and KLOE12 spectra; this combined spectrum is referred to below as KLOE85, thus named according
to its number of data points.

Two more data samples, also collected in the ISR mode, were appended to this list by BESIII 
\cite{BES-III} -- with recently improved statistical errors  \cite{BESIII-cor} -- 
and a CLEO-c group \cite{CESR}.  
Finally, the SND collaboration has just published a data sample \cite{SND20} 
collected in scan mode on the new VEPP-2000 Facility at Novosibirsk; this spectrum,
seemingly still preliminary, 
is referred to below as SND20. Another high statistics data sample, also collected in 
scan mode, is expected  from the CMD3 Collaboration \cite{CMD3prelim19}.

Important tension between some of these samples -- namely KLOE08 and BaBar --
and all others have already been  identified \cite{ExtMod5,WhitePaper_2020}; 
the occurence of the new data sample from SND  (and its comparison with NSK, KLOE
and BaBar \cite{SND20}) allows to reexamine this consistency issue 
and gives the opportunity to remind how it is dealt within global frameworks.

\subsection{The Sample Analysis Method~: A Brief Reminder}
 \label{remind}
 \indentB
The broken HLS modelings previously developped, especially BHLS$_2$ as well as its present 
extension, aim at providing    frameworks which encompass a large part of the low 
energy physics, the realm of 
the non-perturbative regime of QCD, and extend up to the $\phi$ mass region; 
they have rendered possible fair  simultaneous accounts of the six major $e^+ e^-$ 
annihilation channels ($\pi^+ \pi^-$, 
$\pi^+ \pi^- \pi^0$, $K^+ K^-$, $K_L K_S$, $(\pi^0/\eta) \gam$)   up to 1.05 GeV/c;
slightly modified (EBHLS$_2$), this framework also provides now a satisfactory
understanding  of the A, B and C dipion spectra from the decay of the $\tau$ lepton. 

As already noted several times, 
the Lagrangians which substantiate the various broken HLS models emphasize a property
expected from QCD~: The different annihilation channels should be correlated
via their  common underlying QCD background; this is reflected within our Effective Lagrangians
by the fact that all their model parameters are simultaneously involved in the amplitudes for any
of the accessible physics processes they encompass. 
A straightforward  example is represented by  $g$, the universal 
vector coupling and this property is, more generally, exhibited by  
the expressions for the various amplitudes derived from within the various
broken HLS Lagrangians.

Most of the Lagrangian parameters are not known {\it ab initio} and  are derived from
the data via a global fit involving all channels and, possibly, all available data samples. For this
purpose, the provided data samples and associated information (data points, statistical errors, systematics,
correlated or not) 
are supposed reasonably\footnote{Of course, it is unrealistic to expect that 
the spectrum and uncertainties defining any data sample have been {\it perfectly} determined- see below. } 
well estimated. With this at hand, one can construct a motivated
global $\chi^2$ and derive the Lagrangian parameters through a minimization procedure
 like {\sc minuit}.

\vspace{0.5cm}

Among the various kinds of uncertainties reported by the different experiments, a special
care should be devoted to the global normalization uncertainties -- which can be energy
dependent as already dealt with in Subsection \ref{BESIII_isolated}. Actually, as for 
energy scale recalibrations (see also Subsection \ref{BESIII_isolated}), it looks obvious 
that  the most appropriate normalization of a given sample
can only be determined by comparing with several other independent spectra covering 
the same physics channel. Even more, a global treatment of these provides the 
best normalization of each sample versus all the others by a kind of bootstrap mechanism.

Actually a global fit, when possible,  looks the best tool to determine the most
appropriate normalization of each spectrum in accord with its reported uncertainties, 
including its normalization uncertainties; this is reminded in details in  
\cite{ExtMod5,ExtMod7} and above in Section  \ref{BESIII}. The goodness of the corresponding
fit tells the confidence one can devote to the normalization corrections.

 The  probability of the best fit reflects the quality of the experimental information and the 
 relative consistency of the various data samples involved in the procedure within the model framework;
 we have  now 3 significantly different HLS frameworks at hand which have been shown in  
 \cite{ExtMod7} and just above  to lead to a consistent picture.  
 
\subsection{Samples Covering the $\pi^+ \pi^-$ Channel~: A Few Properties}
\label{select}
\indentB
Let us  first consider the data samples already identified as not exhibiting significant 
tension among them  within the BHLS$_2$ frameworks, the previous one \cite{ExtMod7}
or the present one; this defines a Reference set of data samples, named ${\cal H}_R$. 
This  already covers the $3\pi $ data samples already considered 
in Section \ref{BESIII} and all the existing data samples covering the  $\pi^0 \gam$
and  $\eta \gam$ decay channels. Regarding the dikaon spectra, 
we refer the reader to our analysis in
\cite{ExtMod7} where the tensions between the  CMD3 spectra \cite{CMD3_KpKm,CMD3_K0K0b}
and the others from SND, BaBar and (corrected \cite{ExtMod7}) CMD2  has led to discard
them from the analysis\footnote{The CMD3 data have, nevertheless, been dealt 
with to estimate systematics in the muon HVP evaluation \cite{ExtMod7}.
}. 
As for $\tau$ dipion spectra, it has been shown in Section \ref{thooft-kin} 
that the residual tension observed in the account for Belle compared to Aleph and Cleo 
can be absorbed. The reference set  ${\cal H}_R$ also includes  the spacelike pion 
 \cite{NA7,fermilab2} and kaon form factor spectra  \cite{NA7_Kc,fermilab2_Kc} which
are satisfactorily understood within the BHLS$_2$ frameworks \cite{ExtMod7}.

For the purpose of reexaming sample tensions, it looks appropriate to also
include in  ${\cal H}_R$, the pion form factor spectra collected by 
BESIII \cite{BES-III,BESIII-cor} and Cleo-c \cite{CESR}. Indeed, anticipating somewhat 
on our fit results, it has been observed that, alone or together with either of the
NSK, KLOE, BaBar samples, or with any combination of these, they get the same
individual sample $\chi^2$'s, with fluctuations  not exceeding 1.5 units for each of them.
Fitting the  ${\cal H}_R$ sample set thus defined within the present framework
returns $\chi^2/N = 926/1021$ and a 94\% probability;  in this fit, the BESIII and Cleo-c
samples yield
$$[\chi^2/N]_{BESIII} = 49/60,~~~~ [\chi^2/N]_{Cleo-c}=27/35~~.$$ 

One can consider the probability of the global fit (here 94\%) as a faithful
tag of mutual consistency of the (more than 50) samples included  in ${\cal H}_R$ 
which fairly fit the broken HLS framework.

%\begin{table}[!pt!]
\begin{table}[!phtb!]
\begin{center}
%\hspace{-1.6cm}
\begin{minipage}{0.9\textwidth}
\begin{tabular}{|| c  || c  | c | c ||}
\hline
\hline
\hhhv $\chi^2/N_{\rm pts}$ [Prob]	  &  \hhhv + NSK &  \hhhv + KLOE &  \hhhv + BaBar\\
%\hhhw   \hhhw ~~~ & ~~    & ~~ & ~~ 	\\
\hline
\hline
 \hhhv NSK  $\pi^+\pi^-$ (127)     &  $129/127$ [95.3\%]   & $142/127$ [91.2\%]     & $138/127$ [51.7\%] \\
\hline
 \hhhv KLOE $\pi^+\pi^-$ (135)     & $136/135$ [91.2\%]    & $132/135$ [95.2\%]     & $148/135$ [31.9\%] \\
\hline
 \hhhv	BaBar $\pi^+\pi^-$ (270)   & $328/270$ [51.7\%]    &  $354/270$ [31.9\%]    & $326/270$ [62.9\%] \\
\hline
\hline
\end{tabular}
\end{minipage}
\end{center}
\caption {
\label{Table:T5}
Properties of the global fits performed  with the present upgraded BHLS$_2$ model using
the  ${\cal H}_R$ sample collection with one among the NSK, KLOE 
and BaBar samples and with pairs of these. The Table is organized in such a way
that the first line displays the value for $\chi^2/N_{NSK}$ returned by fitting
the 3 configurations (${\cal H}_R$+NSK), (${\cal H}_R$+NSK+KLOE),
(${\cal H}_R$+NSK+BaBar); the corresponding fit probabilities are shown
within squared brackets. The second and third lines display the similar
information for KLOE and BaBar. The number of data points in each of 
NSK, KLOE and BaBar is reminded in the first column for convenience.}
\end{table}

We have made two kinds of global fits~: 
\begin{itemize}
\item
{\bf i)} Fits involving  ${\cal H}_R$ and each of NSK, KLOE\footnote{ KLOE indicates
the simultaneous use of KLOE10 and KLOE12} and BaBar in turn; the diagonal in Table~\ref{Table:T5}
reports the main results, namely the value returned for $\chi^2/N$ of resp. NSK, KLOE, BaBar and
the probability of the global fit at the corresponding Table entry. 

\item
{\bf ii)} Fits involving  ${\cal H}_R$  and  the pairwise combinations (NSK, KLOE), (NSK, BaBar)
and (KLOE, BaBar) in turn; the main fit results are reported in the non-diagonal entries of 
Table~\ref{Table:T5}. In order to make easier the comparison of each of the NSK, KLOE, BaBar 
accounts provided by these "pairwise" fits, we have organized the non-diagonal entries in a specific
manner~: The entry (NSK, KLOE) provides $[\chi^2/N]_{NSK}$ and the global fit probability whereas
the entry (KLOE, NSK) provides $[\chi^2/N]_{KLOE}$ and the (same) fit probability.
The same rule applies {\it mutatis mutandis} to the other pairwise fits~: (NSK, BaBar) and
(KLOE, BaBar) together with ${\cal H}_R$.
\end{itemize}

Relying on Table~\ref{Table:T5}, one clearly observes that the single mode fits for NSK and
KLOE are fairly good and in nice accord with the results returned by the corresponding pairwise 
fit. The pattern is somewhat different when BaBar is involved. 

In order to be complete, let us briefly summarize  the fit results obtained within
the present framework concerning KLOE08 and
the KLOE85 sample derived by the KLOE2 Collaboration from their
combination of the KLOE08, KLOE10 and KLOE12 spectra\cite{KLOEComb}.

\begin{itemize}
\item Regarding  KLOE08~: The global fit for  ${\cal H}_R$ + KLOE08 returns $\chi^2/N_{KLOE08}=95/60$
and a 74.7\% global fit probability. With an average $<\chi^2> \simeq 1.5$, one does not 
consider  confidently the results derived from fit to this combination compared to KLOE10+KLOE12.

\item Regarding KLOE85~: The fit for ${\cal H}_R$ + KLOE85 returns $\chi^2/N_{KLOE85}=83/85$
(global fit probability 94.7\%) which clearly indicates that the KLOE08 issue is reasonably 
well dealt within the KLOE85 combination \cite{KLOEComb}.

One has also performed the pairwise fit ${\cal H}_R$ + KLOE85 + NSK. In this case, one gets~:
$$[\chi^2/N]_{NSK} = 160/127,~~~~ [\chi^2/N]_{KLOE85}=93/85~~,$$ 
with a 80.7\% probability. This fit is obviously reasonable\footnote{One should note that
the estimates for $a_\mu(\pi^+\pi^-, \sqrt{s}\le 1.05)~ {\rm GeV}$ marginally differ~:
$493.18 \pm 0.90$ (${\cal H}_R$ + NSK + KLOE) and  $493.75 \pm 0.79$ (${\cal H}_R$ + NSK + KLOE85)
in units of $10^{-10}$.}
 but less satisfactory than
 ${\cal H}_R$ + NSK + KLOE, as the tension between KLOE85 and NSK is  large, much larger
 than when using  ${\cal H}_R$ +NSK + KLOE as displayed in Table~\ref{Table:T5}.
\end{itemize}
\subsection{The case for the 2020 SND Dipion Sample~: Fits in Isolation}
\label{SND20-1}
\begin{figure}[!phtb] 
\begin{minipage}{\textwidth}
\begin{center}
\resizebox{\textwidth}{!}
%{\includegraphics*{fig_2020/bs_snd_alone_std.eps}}
%{\includegraphics*{fig_2020/bs_snd_alone_std_cor_uncor.eps}}
{\includegraphics*{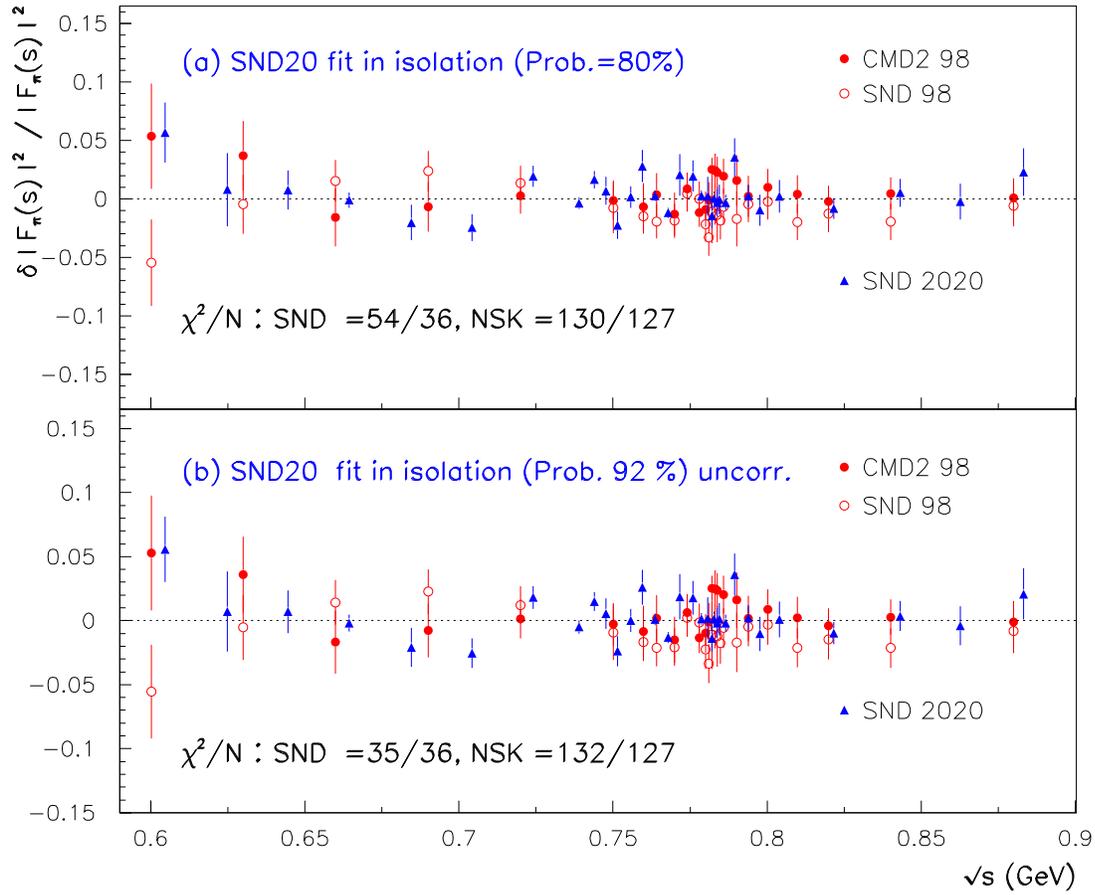}}
\end{center}
\end{minipage}
\begin{center}
\vspace{-0.5cm}
\caption{\label{SND20}  Fit of the SND20 data \cite{SND20} in isolation within the BHLS$_2$
framework. 
The top panel displays the results corresponding to a fit where SND20 systematics
are fully point-to-point correlated, whereas the bottom panel is obtained by
treating the SND20 systematics as fully uncorrelated. The NSK spectra are displayed
but not fitted.
See text for more explanations.} 
\end{center} 
\end{figure}

\indentB In order to analyze  the new data sample recently provided by the SND 
Collaboration  \cite{SND20}, the treatment of the reported systematic errors  has been
performed  as emphasized above for the $3\pi$ data from BESIII
(see Subsection \ref{BESIII_isolated}) as the systematics are expected to be
fully point-to-point correlated\footnote{There is no explicit statement in
 \cite{SND20} about how the systematics should be understood; however, this 	
 assumption corresponds to what is usually understood with the data collected at
 the Novosibirsk facilities. This  topic is further discussed just below. }. 
 For the present analysis, we have first performed 
global fits\footnote{For convenience, here,  the $3\pi$  annihilation 
channel and data are discarded from the fit procedure. } 
where the single representative for the $e^+ e^- \ra \pi^+ \pi^-$
annihilation channel is SND20, the new SND data sample  \cite{SND20};
the spacelike pion form factor data \cite{NA7,fermilab2} have also been discarded from 
the fits in isolation.
Figure \ref{SND20} summarizes our results.
 
The top panel in  Figure \ref{SND20} indicates that, in single mode, the best
fit returns a reasonable probability. However, this comes together with a large average 
$<\chi^2>_{SND20}=54/36=1.5$ (to be compared with the diagonal in Table~\ref{Table:T5}).
Nevertheless, amazingly, the SND20 form factor derived from this global fit provides 
a fairly good account of the NSK (CMD2 and SND) data {\it not submitted to the fit} as
one yields $<\chi^2>_{NSK}=130/127=1.02$, much better than SND20 itself. 
The NSK (pseudo-)residual spectrum is consistent with flatness and, addtionally, 
the ratio $130/127$ indicates that there no significant energy calibration 
mismatch between the NSK samples and SND20 -- this may have shown up in the 
$\rho^0-\omg$ drop-off region.

The bottom panel in Figure \ref{SND20} displays  results derived by assuming the SND20
systematics fully uncorrelated ({\it i.e.} the non-diagonal elements of the error
covariance matrix are dropped out). The global fit is successful and returns 
a 92\% probabability. The gain for SND20 is noticeable as 
$<\chi^2>_{SND20}=35/36=0.97$ and, clearly, the (alternative) 
pion form factor derived by fitting only the SND20 data (in this manner) within the 
global framework is almost unchanged; this is the way the $\chi^2$ distance of the NSK samples to this
alternative fit form factor can be understood~: $<\chi^2>_{NSK}=132/127=1.04$. Moreover, once again,
the NSK (pseudo)residual distributions are as flat as (and almost identical  to) those displayed in the top
panel of Figure \ref{SND20}. 
We got substantially the same results and conclusions by enlarging the
SND20 pion form factor statistical 
errors\footnote{The pion form factor squared
Table in \cite{SND20} gives numbers with only one decimal digit, so adding 0.04 to the 
statistical uncertainty rounds down to 0.}  
by 0.04 and keeping unchanged the
systematics -- {\it i.e.} treated as point-to-point correlated; this also points towards the
interest  to have, beside information on correlations, information on the accuracy of the
uncertainties \cite{WhitePaper_2020}. So, the way the SND20 uncertainties should be 
understood deserves clarification.
\begin{figure}[!phtb] 
\begin{minipage}{\textwidth}
\begin{center}
\resizebox{\textwidth}{!}
%{\includegraphics*{fig_2020/3pion_BESIII.eps}}
%{\includegraphics*{fig_2020/residuals_snd_vs_all.eps}}
{\includegraphics*{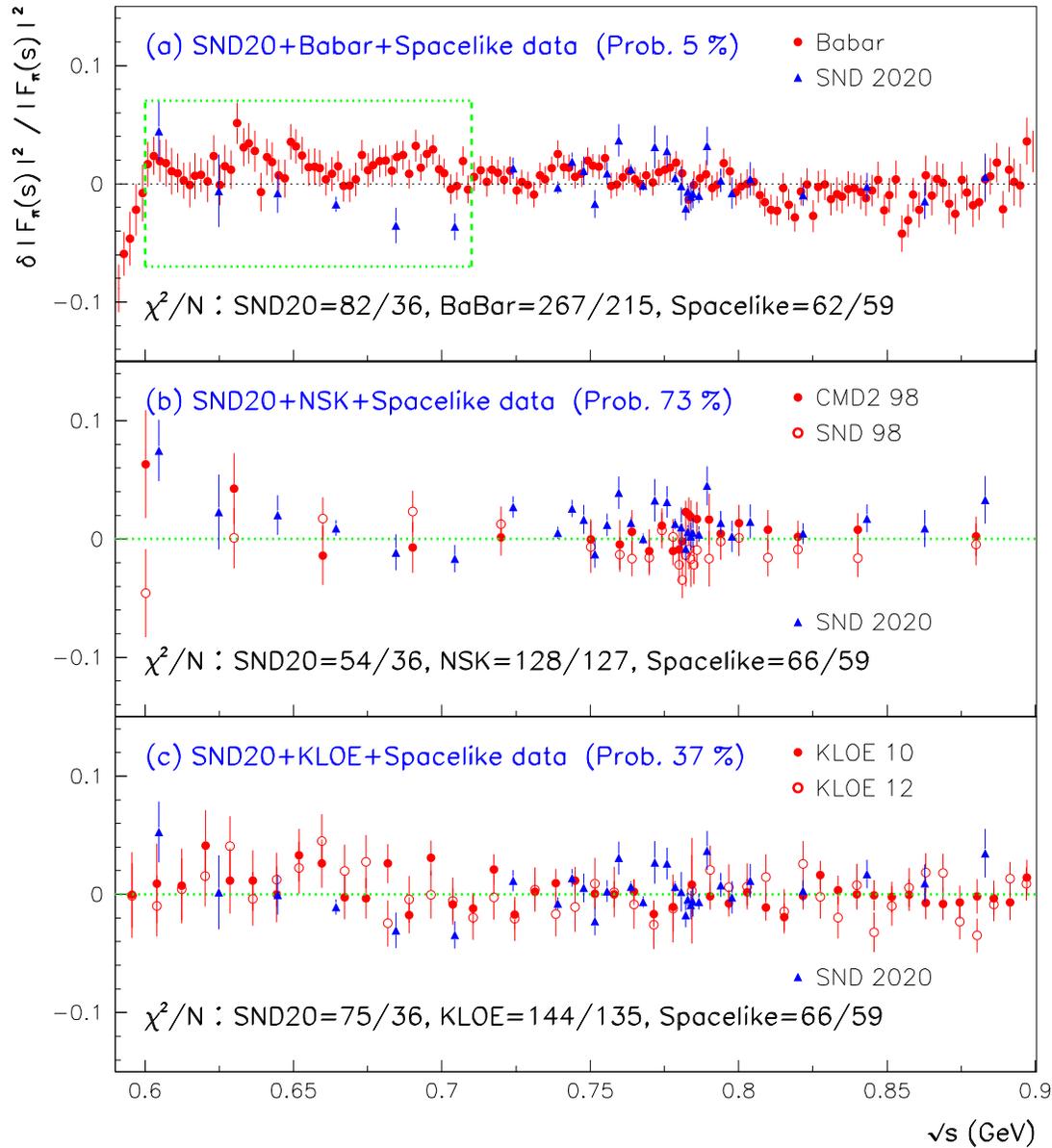}}
\end{center}
\end{minipage}
\begin{center}
\vspace{-0.3cm}
\caption{\label{SND20_vs_all} Fits of the SND20 dipion data \cite{SND20} together with
the spacelike data \cite{NA7,fermilab2}. The top panel shows the fit residuals when
the timelike dipion channel is covered by  the SND20 and BaBar \cite{Davier2009} samples;
similarly, the middle panel displays the fit residuals when covering the  
 timelike dipion channel by the SND20  \cite{SND20} and 
NSK \cite{CMD2-1995corr,CMD2-1998-1,CMD2-1998-2,SND-1998} spectra;  the bottom panel reports
likewise the case when the timelike dipion channel is covered by  the SND20
and KLOE\cite{KLOE10,KLOE12} samples. All reported systematics are treated as point-to-point 
correlated.} 
\end{center}
\end{figure}
\subsection{The case for the 2020 SND Dipion Sample~: Pairwise Fits}
\label{SND20-2}
\indentB
An interesting topic addressed in \cite{SND20} is the consistency of SND20
with respectively NSK (e.g. CMD2 \cite{CMD2-1995corr,CMD2-1998-1,CMD2-1998-2} and
SND-98  \cite{SND-1998}), KLOE ( KLOE10  \cite{KLOE10} and KLOE12  \cite{KLOE12})
and BaBar \cite{Davier2009}. 
For this purpose, it looks worthwhile performing global fits by including pairwise combination
to cover the $\pi^+ \pi^-$ annihilation channel\footnote{It has been found appropriate to re-introduce
the spacelike data from \cite{NA7,fermilab2} within the minimization procedure. As the pion form factor in
the spacelike and timelike regions is the same analytic function, this is a constraint.}.
 This allows to observe
the tension between the partners in the pair and to get a probability which emphasizes
 their global consistency. Our main fit results are collected in Figure \ref{SND20_vs_all}.

The middle panel in Figure \ref{SND20_vs_all}
 shows the case for the  global fit with the (SND20+NSK) combination.
As could be expected, this confirms the fit of SND20 in isolation reported in the top panel of Figure
\ref{SND20}~:  $<\chi_{NSK}>$ is negligibly improved whereas $<\chi_{SND20}>$ is unchanged;
the large value for $<\chi_{SND20}>=1.5$ is responsible for the global fit probability reduction
compared to fits with NSK alone (or combined with KLOE), as can be seen in 
Table~\ref{Table:T5}. With this proviso, BHLS$_2$ confirms the statement that SND20 and NSK are consistent
 \cite{SND20}  with a 73\% probability.

\begin{figure}[!phtb]
%\hspace{-1.5cm}
\begin{minipage}{0.50\textwidth}
{\includegraphics[width=\textwidth]{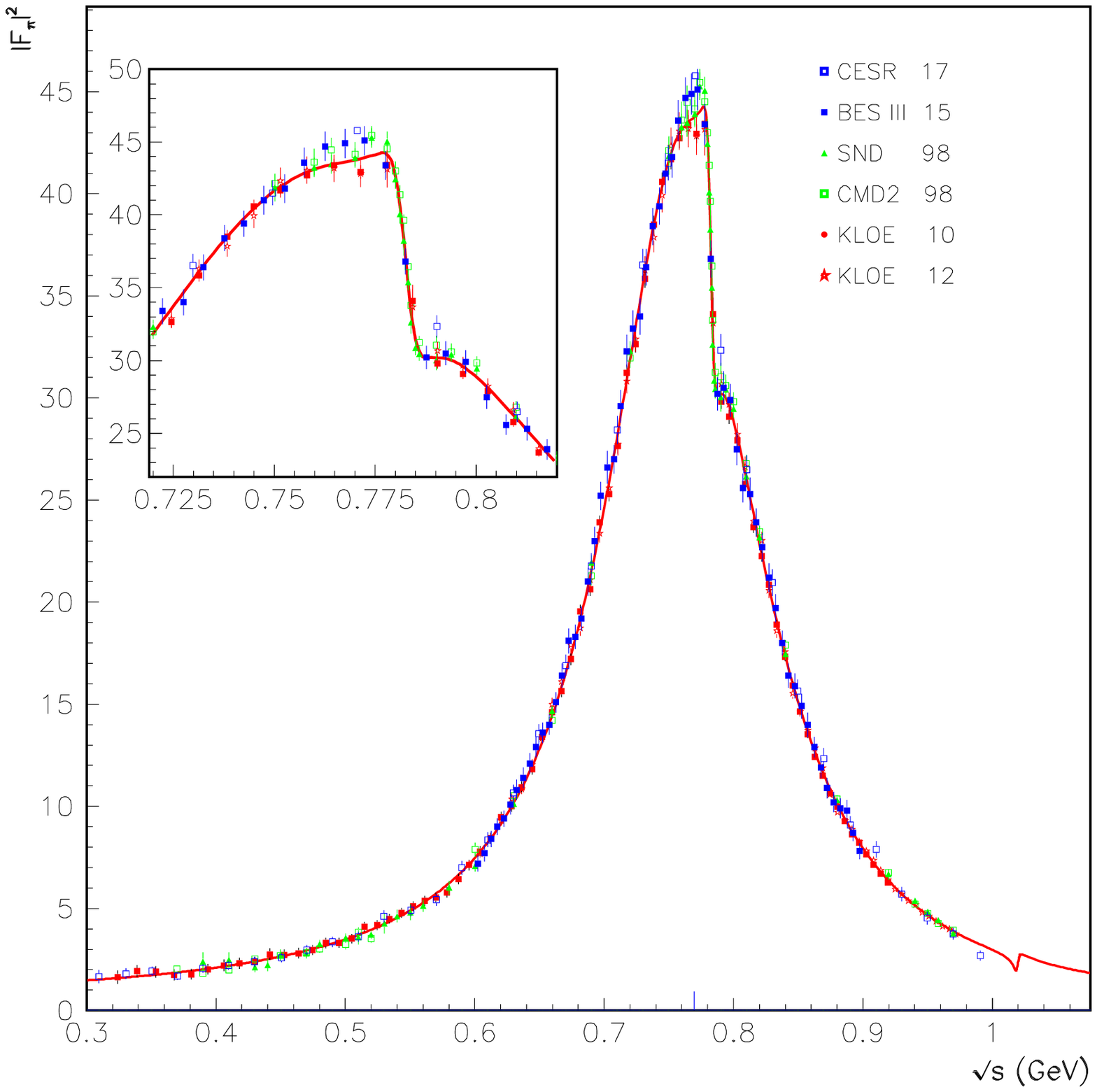}}
\end{minipage}
\hspace {0.01\textwidth}
\begin{minipage}{0.50\textwidth}
\includegraphics[width=\textwidth]{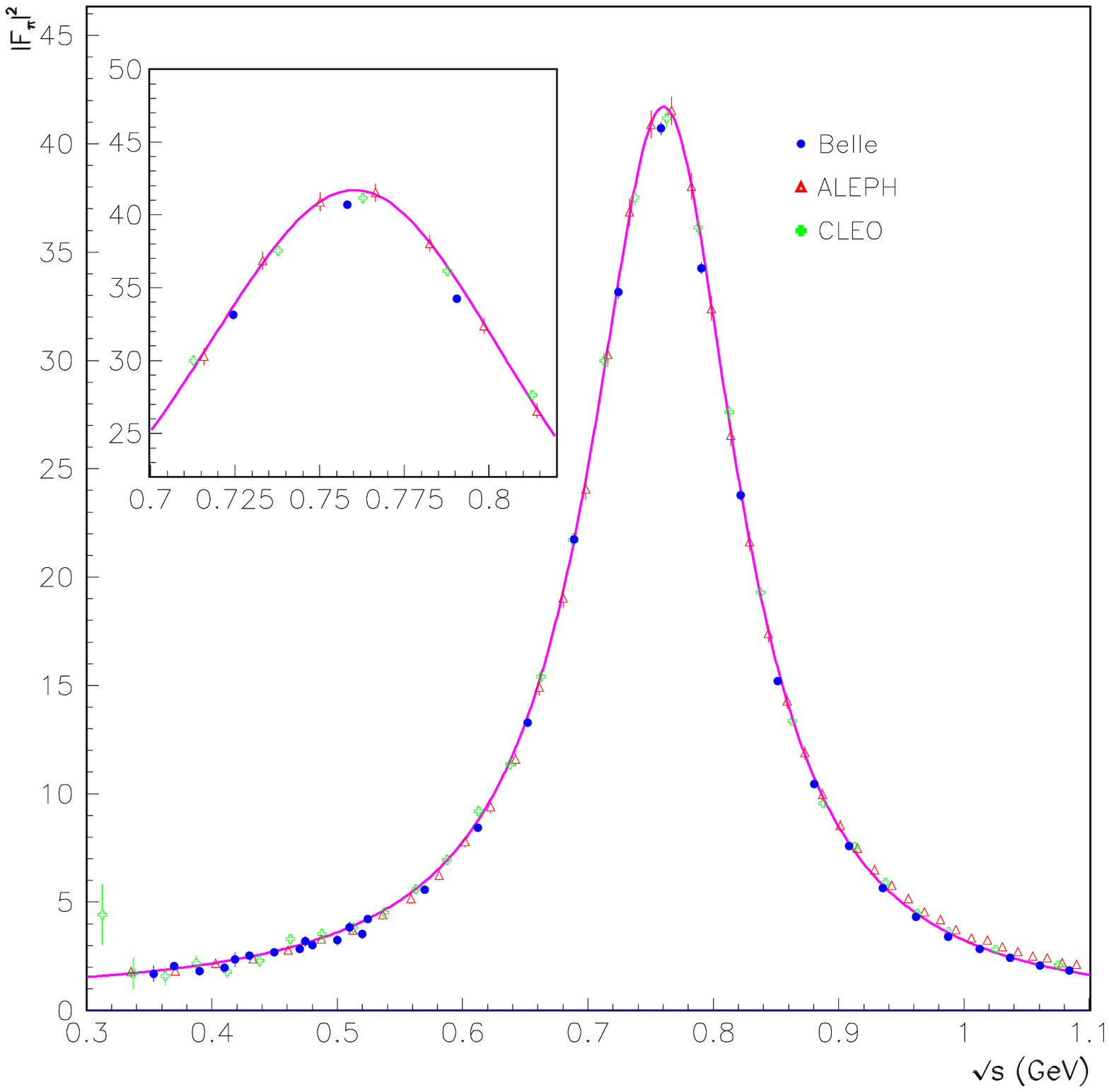}
\end{minipage}
\begin{center}
\vspace{-0.3cm}
\caption{\label{Fig:PIPI} BHLS$_2$ fit to the  $\pi \pi $ data, the 
upgraded BS solution~:
The left-hand panel shows the pion form factor squared in the
$e^+e^-$ annihilation and the right-hand one displays the same
spectrum in the $\tau$ decay. The fitted regions extend up to $s=1.0$ GeV$^2$.}
\end{center}
\end{figure}
In order to address the consistency topic about SND20 and BaBar already studied in \cite{SND20},
we have run our global fit procedure with  the (SND20+BaBar) combination. To stick 
the closest possible to the study
 reported in \cite{SND20}, we have found it worth to exclude 
from the fit the part of the BaBar spectrum with $\sqrt{s} \in [0.60,~0.71]$. The results are
displayed in the top panel of Figure \ref{SND20_vs_all} and show some resemblance between
SND20 and BaBar (normalized) residuals outside the BaBar excluded region (delimited by the green
rectangle). Nevertheless, the fit probability is  poor and its $<\chi_{SND20}>=82/36=2.3$
indicates a significant tension compared to the fit in isolation ($<\chi_{SND20}>=1.5$)

Finally, the bottom panel in Figure \ref{SND20_vs_all}, reports the main fit results obtained
by fitting the (SND20+KLOE) combination. One  can note the results compared to the fits in isolation~:
$<\,\chi_{SND20}>=2.1$ (versus $<\chi_{SND20}>=1.5$) while $<\chi_{KLOE}>=144/135=1.1$ 
(versus $<\chi_{KLOE}>=0.98$) and a 37\% probability. Here also, one can observe that the residual
distributions for SND20 are not really at variance with those for KLOE along the whole energy range. 

So, on the  whole, the SND20 spectrum  \cite{SND20} does not help in clarifying the consistency issue raised by
the existing dipion spectra and, presently,  SND20 does not bring in more support, or less support, to the choice performed
following our previous analyses  (see Tables~\ref{Table:T3} and \ref{Table:T5}) 
and illustrated by Figure \ref{Fig:PIPI}.

\section{Overview of the EBHLS$_2$ Fits}
\label{global-fits}
\begin{table}[!phtb!]
\begin{center}
\begin{minipage}{1.0\textwidth}
\begin{tabular}{|| c  || c  | c || c || c ||}
\hline
\hline
\hhhv  EBHLS$_2$ 	&\multicolumn{2}{|c||}{ \hhhv BS  ($ \lambda_3=0$)} & \multicolumn{2}{|c||}{ \hhhv  ($ \lambda_3 \ne 0$)} \\
\hline
\hhhv  \hhhw ~~	                           & \hhhv Excl. $\tau$    &  \hhhv Incl. $\tau$  & \hhhv BS  & \hhhv RS \\
\hline
 \hhhv NSK $\pi^+ \pi^-$  (127)            &   136 &  134 &   138  &  136     \\
\hline
 \hhhv KLOE $\pi^+ \pi^-$ (135)            &   141 &  146 &    139  &   139     \\
\hline
 \hhhv  BESIII $\pi^+ \pi^-$ (60)          &  48   & 47  &   49  &  	48	\\
\hline 
 \hhhv	Spacelike $\pi^+ \pi^-$ (59)       &  62   & 67&  62  &	 60		\\
\hline
\hline
 \hhhv	$\tau$ (ABC) (84)                 &   $\times$& 93 & 82  & 	80	 \\
\hline
\hline
 \hhhv	$\pi^0 \gam$ (112)                &  89  &  88 &  88     &   87		\\
\hline
 \hhhv	$\eta \gam$  (182)               &  120  & 120 & 124      &  124		 \\
\hline
 \hhhv	NSK $\pi^+ \pi^-\pi^0$ (158)      &  142   & 146  & 147   & 147		\\
\hline
 \hhhv	BESIII $\pi^+ \pi^-\pi^0$ (128)    & 138 &  138 &  137    & 	137	\\
\hline
\hline
 \hhhv	NSK $K_L K_S$ (92)                 &  103  & 104 & 103 &  104		\\
\hline
 \hhhv	NSK $K^+ K^-$ (49)                 &  41   & 42 & 39  &  39		 \\
\hline
 \hhhv	BaBar $K^+ K^-$ (27)               &  41   & 42 & 41 &  41		\\
\hline 
 \hhhv	Spacelike $K^+ K^-$ (25)           &  18  &  19 & 17 &  18		\\
\hline
 \hhhv	Decays (8)                        &  5  &  4 & 9 & 	9		\\
\hline
\hline
 \hhhv $\chi^2/N_{\rm pts}$ &    $1179/1280$  &  $1285/1365$      & 1269/1365   & 1262/1365 \\
 \hhhw Probability      &         93.3\%      &  83.1  \%          & 90.0 \%    &  91.5 \%\\
\hline
\hline
\end{tabular}
\end{minipage}
\end{center}
\caption {
\label{Table:T3}
Global fit properties  of the EBHLS$_2$ fits~; second line in the Table title
 indicates the running conditions regarding the data samples submitted to fit
 or the running of the BS or RS variants when $\lambda_3 \ne 0$. The number of data points
 involved is given between parentheses in the first column.
The last lines display  the  global $\chi^2/N_{\rm pts}$ and probability of each fit.}
\end{table}
\indentB
Some of the general  properties of the global fits performed in the EBHLS$_2$ framework
have already been emphasized in Sections \ref{tau_spectra}, \ref{BESIII} and \ref{SND20txt}
with special emphasis on resp. the $\tau$ decay dipion spectra, the $e^+ e^- \ra \pi^+ \pi^- \pi^0$
annihilation channel and the crucial $e^+ e^- \ra \pi^+ \pi^- $ one. The general features of our
fitting algorithm concept are detailed in Section 15 of our \cite{ExtMod7}, for instance. 
 Let us, for convenience,  remind the gross features of our global fit method~:
\begin{itemize}
\item The contribution of each data sample to the global $\chi^2$ to be minimized 
is constructed using solely the uncertainties {\it exactly as they are provided by 
each experiment  without any external input}. Additionally,
this kind of input may influence the numerical outcome of the fits in an uncontrolled way.
On the other hand, if not already performed by the 
relevant experiments, the reported uncorrelated systematics are merged appropriately
with the statistical error covariance matrix, 

\item The correlated systematic errors -- possibly $s$-dependent -- are treated with
special care  \cite{ExtMod5} as emphasized in Subsection \ref{BESIII_isolated} in
 order to avoid the
so-called Peelle Pertinent Puzzle \cite{Peelle} which generally results in biasing
the evaluation of physics quantities  based on  $\chi^2$ minimization procedures
\cite{ Chiba,Fruehwirth}. Moreover, an iterative procedure is used which has been proved
 to avoid biases  \cite{ExtMod5}. 

In the case of a global $\chi^2$ minimization, it should be stressed that the
absolute scale of each experiment is  derived in full consistency with those of all
the other experiments (or data samples),
especially -- but not only -- with those  collected by other groups in the same physics 
channel\footnote{Actually, this looks the natural way, if not the single one,
 to derive reliably and consistently the correction to the normalization of the
 various spectra.}. 
Several examples  can be found in \cite{ExtMod7}
where it is shown that the derived scale corrections  quite well compare with 
the corresponding experimental expectations.

\item When merging the different data samples which cover the same energy range,
their different energy calibrations may exhibit some mismatch; this issue
was previously encountered in our \cite{ExtMod7} with the energy calibration of the dikaon 
spectra from CMD-3 \cite{CMD3_K0K0b,CMD3_KpKm} and BaBar \cite{BaBarKK}
versus those of the corresponding samples from CMD2 and SND; 
this issue happened again herein when dealing with
the BESIII 3$\pi$ sample \cite{BESIII_3pi} and is solved accordingly
(see Section \ref{BESIII} above). It should be noted that, for signals as narrow
as the $\omg$ and $\phi$ mesons, global fit technics are certainly the best suited way to
match  the energy scales of various spectra otherwise poorly consistent.

\item In order to {\it confidently} rely on global fit outputs to evaluate physics quantities,
one should discard data samples which exhibit noticeable inconsistencies with the rest 
of the benchmark samples. Our requirement to identify such samples has generally been 
to get an average $\chi^2$ per point  smaller than $\simeq 1.5$. 
\end{itemize} 
\vspace{0.3cm}

Compared to the fits reported in \cite{ExtMod7} and as already noted in Section 
\ref{param_free}, we have here released the constraint (Equation(\ref{xxxyy}))  on the 
product $\epsilon \epspr$ and also let the mixing angle $\theta_P$ float freely. 
Moreover, as several preliminary fits typically return~:
\be
\begin{array}{lll} 
\displaystyle \Delta_A=[0.55 \pm 4.59] \times 10^{-2}~,~~~{\rm and}~~~
\displaystyle  \lambda_8=[2.18 \pm 4.18] \times 10^{-2}~,
\end{array} 
\label{fitres_1}
\ee
\noindent imposing  $\Delta_A=\lambda_8=0$ looks worthwhile; indeed, Equations (\ref{fitres_1}) 
clearly shows that the physics presently addressed in the EBHLS$_2$ framework does not 
exhibit a significant sensitivity to these parameters when left free. These
constraints will be reexamined\footnote{See below the Sections
 devoted to the  $[\pi^0,\eta,\etp]$ mixing.}
 in the context where the $[\pi^0,\eta,\etp]$ 
mixing is also addressed.  

Table \ref{Table:T3} reports the fit results under 4 configurations; the energy scale
corrections for the BaBar dikaons \cite{BaBarKK} and the BESIII samples  \cite{BESIII_3pi}
are floating parameters. The first 2 data columns, actually, update the BHLS$_2$ fit results derived
for the BS variant given in \cite{ExtMod7}; for the fit performed including the $\tau$ spectra,
the polynomial $\delta P^\tau(s)$ here is third degree.
 
For the EBHLS$_2$ fits reported in the last two data columns, 
$\delta P^\tau(s)$ is second degree\footnote{This means that BHLS$_2$ and EBHLS$_2$
actually carry an identical parameter freedom.}.
The third data column displays the $\chi^2$ contributions of various groups oyf data samples to
the global $\chi^2$ using the BS configuration. For completeness, the last data column reports 
the EBHLS$_2$ fit results obtained under its RS configuration \cite{ExtMod7}.

One should note that, substantially, these pure EBHLS$_2$ fits and the 
BHLS$_2$ fit excluding the $\tau$ spectra (first data column) 
exhibit  similar and favorable $\chi^2$  averages per point for all groups of 
data samples 
with the sole exception of the individual decay modes which is doubled. In the
EBHLS$_2$ BS or RS configurations, the single mode which significantly departs from 
$\chi^2 \le (1.0 \div 1.2)$, is $\eta \ra \gam \gam$, which returns 4, e.g. a $2 \sigma$
difference with the Review of Particle Properties \cite{RPP2016}. On the other 
hand, one may consider, in view of Table \ref{Table:T3}, that the BS variant 
of EBHLS$_2$ does not need to be improved by the Primordial Mixing mechanism 
introduced in \cite{ExtMod7} to construct the RS variant of BHLS$_2$. 

The numerical values of the  model parameters of the EBHLS$_2$/BHLS$_2$ framework
will be examined and commented in a wider context involving also the treatment of 
the  $[\pi^0,\eta,\etp]$ mixing properties in Section \ref{BHLS-mixing}.

\section{Evaluation of $a_\mu$, the Muon HVP}
\label{muon-hvp}
\indentB
As the previous BHLS releases  \cite{ExtMod3,ExtMod7},
 EBHLS$_2$  encompasses the bulk of the low energy 
$e^+ e^- \ra {\rm hadrons}$
annihilations up to, and including, the $\phi$ mass region.  Therefore, 
taking into account suitably 
the various kinds of uncertainties reported by the different experiments
affecting the spectra they collected, a {\it fully} global fit  
is expected to lead to precise evaluations  of the contributions 
 to $a_\mu^{HVP-LO}$ from the energy region $\sqrt{s} \le 1.05$ GeV.
Within this approach, the specific contribution 
of the hadronic channel $H_i$ is obtained by means of the
cross section $\sigma(H_i,s) \equiv \sigma(e^+ e^- \ra H_i, s)$ with
parameter values derived from 
the global fit performed within the EBHLS$_2$ framework~:
\be
\begin{array}{lll}
\displaystyle a_\mu(H_i) =\frac{1}{4 \pi^3} \int_{s_{H_i}}^{s_{cut}} ds ~K(s) 
~\sigma(H_i,s)~,&
H_i=\{ \pi^+\pi^-, \pi^0 \gam, \eta \gam, \pi^+\pi^- \pi^0, K^+ K^-, K_L K_S \}~~.
\end{array}
\label{hvp-1}
\ee
$K(s)$  \cite{FredBook,Fred09} is the usual kernel which enhances 
the weight of the 
threshold regions compared to the higher energy regions  of the $H_i$ spectrum; 
$s_{H_i}$ is the threshold of the $ H_i$ hadronic  channel  
and $\sqrt{s_{cut}}=1.05$ GeV is the validity limit, common to the different 
HLS frameworks. 
\subsection{Remarks on the $\sqrt{s} \le 1.05$ GeV Contribution to $a_\mu(\pi \pi)$ }
\label{amu_pipi}
\indentB
We find it of special concern to substantiate what supports the choices performed in
 our global fit approach in connection with the muon HVP outcome.
For this purpose, the analysis of the $\pi^+ \pi^-$ channel 
properties within the (E)BHLS$_2$ framework is of special relevance. 
Global fits have been performed
involving the data samples collected for all the HLS final states except for
$\pi  \pi$ and complemented in turn by each of the various KLOE samples already 
commented in \cite{ExtMod7} to feed solely the $\pi^+ \pi^-$ channel. 
\begin{table}[!htbp] 
%\hspace{-1.cm}
%\begin{center}
\begin{tabular}{|| c  || c  || c |c ||}
\hline
\hline
\hhhd  $\pi^+ \pi^- $ Data Sample  &  \hhhv Direct integration  \cite{Anastasi:2017eio}  &\hhhv BHLS$_2$ 
 &  $\chi^2_{\pi^+ \pi^-}/N_{\pi^+ \pi^-}$ (Prob.)\\
\hline
\hline
\hhhv  KLOE10      & $376.0 \pm 3.4$    & $375.04 \pm 2.35$  	& 69/75 (78\%)	\\
\hline
\hhhv  KLOE12      & $377.4 \pm 2.6$    & $376.74 \pm 1.59$ 	& 59/60 (80\%)   \\
\hline
 \hhhv  KLOE85     & $377.5 \pm 2.2$    & $377.17 \pm 0.89$  	& 95/85 (65\%)	 \\
\hline
 \hhhv  KLOE08 	   & $378.9 \pm 3.2$    &   $373.78 \pm 1.84$	& 130/60 (14\%)  \\
\hline
\hline
\end{tabular}
%\end{center}
\caption {\label{Table:T1}
The $\pi^+ \pi^- $ contribution to the HVP-LO in the range 
$[0.35,~0.85]$ GeV$^2$ in units of $10^{-10}$.
The direct integration evaluations are read off Fig. 6 
in \cite{Anastasi:2017eio}. The  (E)BHLS$_2$ evaluations are derived by fits
as sketched in the text; the last data column displays  relevant pieces of the fit
information.}
\end{table}

Table \ref{Table:T1} -- reprinted from Table 3 in \cite{WhitePaper_2020} -- 
clearly shows that the (E)BHLS$_2$ central values are obviously in close 
correspondence with those derived by directly integrating  the 
data \cite{Anastasi:2017eio}, except for KLOE08 which, correlatedly,
exhibits a poor global fit probability; this substantiates the reason why 
one may prefer discarding poorly fitted data samples to avoid biases,
possibly large. However, one should also remark that its effect within the
KLOE85 combination is much softer as the KLOE85 fit probability (65\%) remains
comparable to those for KLOE10 and KLOE12 which are both higher 
and almost identical (78\% and 80 \%).

Table \ref{Table:T1} also shows the important reduction of the uncertainties 
induced by the  non--$\pi^+ \pi^- $ channels involved in the reported global 
fits; this reduction is,  of course,  amplified
when including the other accepted $\pi^+ \pi^- $ samples in the fit procedure
as will be seen shortly. On the other hand, when the fit probability is poor,
the values returned by the fits for the uncertainty
and the central value should be handled with care.

\vspace{0.5cm}
\indentB
Table \ref{Table:T6} shows a breakdown of the contributions  to $a_\mu(\pi \pi)$ 
from different  energy intervals.
The top lines display the results derived by other groups, namely CHS \cite{Colangelo:rpi}, 
DHMZ  \cite{Davier_2019}
and KNT \cite{Keshavarzi_2019} while the bottom lines show the
EBHLS$_2$ outcome from fits performed under
the various indicated configurations. The favored configuration, 
which corresponds to a good account of all the channels
encompassed within the EBHLS$_2$ framework, is tagged by "KLOE+X". Nevertheless, 
in order to really compare the 
global fit method with \cite{Colangelo:rpi,Davier_2019,Keshavarzi_2019} it is worth 
relying on the same set of experimental data. To this end, we have also run
our code including the BaBar data sample within the set of $\pi^+ \pi^-$ fitted 
spectra so that the sample contents are similar in all the discussed approaches;
nevertheless, in order to avoid the effects of
energy calibration mismatch  between the BaBar and KLOE 
spectra within the fit procedure, 
we have  removed the BaBar $\rho^0-\omg$ drop-off region from the fit.
The corresponding results are
given in Table \ref{Table:T6} under the tag "KLOE+BaBar+X".

\begin{table}[!ptbh!]
\hspace{-0.5cm}
{\small
\begin{tabular}{|| c  || c  | c | c || c||}
\hline
\hline
\hhhv  \hhhw $\sqrt{s}$ Interval (GeV)	  & \hhhv $\sqrt{s} \le 0.6$    &  \hhhv  $0.6 \le \sqrt{s} \le 0.9$    
& \hhhv $0.9 \le \sqrt{s} \le 1.0$   & \hhhv $\sqrt{s} \le 1.0$  \\ %[-0.3cm]
\hline							    							 
\hline
 \hhhv CHS18 \cite{Colangelo:rpi}          
			 &     $110.1 \pm 0.9 $        &  $369.6 \pm 1.7 $        & $15.3\pm 0.1$      & $495.0\pm  2.6$     \\
\hline
 \hhhv DHMZ19  \cite{Davier_2019}
                        &     $110.4\pm 0.4 \pm 0.5$  & $371.5 \pm 1.5 \pm 2.3$	 & $15.5 \pm 0.1 \pm 0.2$     & $497.4 \pm 1.8 \pm 3.1$     \\
\hline
 \hhhv KNT19   \cite{Keshavarzi_2019}          
 &     $108.7  \pm 0.9$        & $369.8 \pm 1.3$	         & $15.3 \pm 0.1$     & $493.8 \pm 1.9$              \\
\hline
\hline
\hhhv  KLOE+BaBar +X	&\multicolumn{4}{|c||}{ \hhhv  $\chi^2/N_{pts}$~: BaBar=1.45, KLOE=1.15, NSK=1.10} \\ %[-0.3cm]
\hhhv Prob=11.4\% 
                	& $108.83\pm 0.09$        &  $369.06 \pm 0.62$	 &  $15.36 \pm 0.38$     & $493.19  \pm 0.73$     \\
\hline
\hline
\hhhv  KLOE +X	&\multicolumn{4}{|c||}{ \hhhv  $\chi^2/N_{pts}$~: KLOE=1.03, NSK=1.09} \\ %[-0.3cm]
\hhhv Prob=90.0\% (Incl. $\tau$)
                        &  $107.79\pm 0.12$      &   $366.76 \pm 0.73$	 & $15.16 \pm 0.42$       & $489.70 \pm 0.84$      \\
\hline
\hhhv Prob=93.3\% (Excl. $\tau$)
                        & $107.67\pm 0.13$        &  $367.21 \pm 0.84$	 &  $15.17 \pm 0.48$     & $490.05 \pm 0.98$     \\
\hline
\hline
\end{tabular}
}
%\end{center}
\caption {Breakdown of $10^{10} \times a_\mu[\pi \pi]$ by energy intervals.
The displayed data for CHS18, DHMZ19 and KNT19 are extracted
from Table 6 in \cite{WhitePaper_2020}.
The EBHLS$_2$ fits are reported using BaBar and KLOE10/12 and the later only together with
the NSK, BESIII and Cleo-c dipion spectra, globally referred to as X. The data collected
in the eighties \cite{Barkov} are also part of X.
\label{Table:T6}
}
\end{table}

Regarding the reported central values for $ a_\mu[\pi \pi]$, it is clear 
that CHS18, DHMZ19, KNT19 and the evaluation derived from the KLOE+BaBar+X fit are similar; 
nevertheless, one should point out
the higher similarity of the KNT19 and EBHLS$_2$(KLOE+BaBar+X) evaluations. Indeed, the 
difference between their
 central values are resp. 0.1, 0.7 and 0.2  for resp. the $\sqrt{s} \le 0.6$ GeV,
$0.6$ GeV $\le \sqrt{s} \le 0.9$ GeV and $0.9$ GeV $ \le \sqrt{s} \le 1.0$ GeV energy intervals.
One may infer that this fair agreement is mostly due to having similar  treatments of the 
correlated systematics
in the BHLS approaches \cite{ExtMod5} and in the KNT dealing \cite{Keshavarzi_2019}.

As in  global approaches the data collected in the 
non--$\pi^+ \pi^-$ channels are equivalent to having at disposal an additional  statistics
in the $\pi^+ \pi^-$ channel, one expects smaller errors  for the 
 (E)BHLS$_2$ evaluations of $a_\mu[\pi \pi]$; this is indeed  what is observed for the
 $\sqrt{s} \le 0.6$ GeV and $0.6$ GeV $ \le \sqrt{s} \le 0.9$ GeV contributions
 to $a_\mu[\pi \pi]$ but, surprisingly, not for the $0.9$ GeV $\le \sqrt{s} \le 1.0$ 
 GeV interval. Nevertheless, integrated up to 1.0 GeV, the contribution to
 $ a_\mu[\pi \pi]$ exhibits an uncertainty improved by a factor of $\simeq 2.5$
 compared to the other approaches reported in Table \ref{Table:T6}.

This comparison proves that the  observed central value differences 
between BHLS$_2$ and the others -- especially KNT -- are mostly due to having 
discarded BaBar (and KLOE08) and only marginally to the global fit method. 
Finally, the last 2 lines of Table \ref{Table:T6} shows the effect of 
including the $\tau$ data. 
The use of these generates an additional (modest) improvement of the 
uncertainties as could be expected, and a marginal shift. 
{\bfEPJC The  comfortable probabilities reached by the EBHLS$_2$(KLOE+X) fits
should also be noted.  As  reminded in Section \ref{global-fits}, 
they are reached without resorting to error information beyond what is provided
by the various experiments like error inflation factors, for instance. }

As noted several times, the validity range of the
HLS approaches to $e^+ e^-$ annihilations extends up to $\simeq 1.05$ GeV,
thus including  the $\phi$ mass region. However, the $[1.0,1.05]$ GeV energy interval
of the dipion spectrum is poorly known; indeed, apart from the BaBar spectrum\footnote{Our
 \cite{ExtMod4} provided a study of $\phi$ mass region in the BaBar spectrum.} 
\cite{BaBar}, the most recent  information about this spectrum piece follows 
from the old SND results \cite{Achasov:2000pu} which underly the RPP \cite{RPP2016}
entries for the $\phi \ra \pi \pi$ decay. 

 As clear from Figure \ref{Fig:PIPI}, in this mass region  the spectrum is widely
 dominated by the tail of the $\rho$ resonance with, on top of it, a tiny effect
 due to the narrow $\phi$ signal. A direct numerical estimate derived from the scarce
 data collected around the $\phi$ mass gives 
 $a_\mu(\pi \pi, [1.0,1.05]~{\rm GeV})=[3.35 \pm 0.04] \times 10^{-10}$. 
 On the other hand,
 relying on the RPP \cite{RPP2016} information, EBHLS$_2$ returns~:
$$a_\mu(\pi \pi, [1.0,1.05]~{\rm GeV})=[3.07 \pm 0.11] \times 10^{-10}~~;$$
replacing within the data set fitted via EBHLS$_2$ the RPP
$\phi \ra \pi \pi$ datum by the BaBar  [1.0,1.05] GeV spectrum piece returns 
$a_\mu(\pi \pi, [1.0,1.05]~{\rm GeV})=[3.10 \pm 0.10] \times 10^{-10}$. Therefore, 
some (mild) systematics
affects this mass region as the cross section lineshape is not really well defined
(see Figure 1 in \cite{ExtMod4}).

\subsection{Contribution to the Muon HVP of the Energy Region $\le 1.05$ GeV}
\label{HVP-HLS}
\indentB
The sum $a_\mu(HLS) = \sum_i  a_\mu(H_i)$ of the quantities defined by Equation (\ref{hvp-1})
represents about 83\% of the total muon HVP; it can be computed with fair precision
using the EBHLS$_2$ fit information to construct the relevant cross sections; these
are derived by sampling the model parameters using  
the parameter central values and the error covariance matrix returned 
by the {\sc minuit} minimization procedure. Sampling out the model parameters allows
to compute a large number of estimates for the different $ a_\mu(H_i)$ and for  $a_\mu(HLS)$, 
the average values  of which defining our reconstructed central values and their r.m.s. 
giving their standard deviations.

The fitted cross sections are also used to estimate  the FSR  contributions for
the $\pi^+ \pi^-$, $\pi^+ \pi^-\pi^0$ and $K^+ K^-$ final states and the Coulomb interaction
 effect which is significant for the $K^+ K^-$ final state as  the kaons are slow in the $\phi$
energy region. 

The HLS model functions are describing VP amputated data, accordingly, 
all the data submitted to our global fits are amputated from their
photon VP factor. Uncertainties related to VP amputation and FSR estimates are
included below as separate systematics.

\begin{table}[!ptbh!]
%\hspace{-0.7cm}
{\small
\begin{tabular}{|| c  || c  | c || c | c|| c ||}
\hline
\hline
\hhhv  EBHLS$_2$ 	&\multicolumn{2}{|c||}{ \hhhv BS  ($ \lambda_3=0$)} & \multicolumn{2}{|c||}{ \hhhv  ($ \lambda_3 \ne 0$)}&  \hhhv Data Direct  \\ %[-0.3cm]
\hline
\hhhv  \hhhw ~~	        & \hhhv Excl. $\tau$    &  \hhhv Incl. $\tau$           & \hhhv BS 		 & \hhhv RS &  \hhhw Integration \\ %[-0.3cm]
\hline							    							 
\hline
 \hhhv $\pi^+ \pi^-$    &     $493.12 \pm 0.98 $     &  $492.77 \pm 0.85$        & $492.77\pm 0.86$      & $493.00\pm  0.90$         & $496.26\pm 3.46$     \\
\hline
 \hhhv $\pi^0 \gam$     &     ~~$4.41 \pm 0.02$      &  ~~$4.40 \pm 0.02$	 & ~~$4.41 \pm 0.02$     & ~~$4.41 \pm 0.02$      &  ~~$4.58 \pm 0.08$   \\
\hline
 \hhhv $\eta \gam$      &     ~~$0.64 \pm 0.01$      & ~~ $0.65 \pm 0.01$	 & ~~$0.65 \pm 0.01$     & ~~$0.65 \pm 0.01$      & ~~$0.55 \pm 0.06$         \\
\hline
 \hhhv $\pi^+ \pi^- \pi^0 $ & ~$44.40\pm 0.32$        &  ~$44.41 \pm 0.32$	 & ~$44.45 \pm 0.32$     & ~$44.41 \pm 0.30$    & ~$44.80 \pm 1.72$
       \\
\hline
 \hhhv $K^+ K^- $      &      ~$18.20 \pm 0.10$       &  ~$18.17 \pm 0.09$	 & ~$18.20\pm 0.09$       & ~$18.29\pm 0.11$    & ~$18.98 \pm 0.28$       \\
\hline
 \hhhv $K_L K_S$       &      ~$11.67 \pm 0.06$       &  ~$11.67 \pm 0.06$ 	 & ~$11.66 \pm 0.06 $	   & ~$11.60 \pm 0.06 $   & ~$12.61 \pm 0.27$    \\
\hline
\hline
 \hhhv HLS Sum         &      $572.44 \pm 1.08$       &  $572.06 \pm 0.95$	 & $572.14\pm 0.95$    & $573.07 \pm 1.00$       & $577.77 \pm 3.89$     \\
\hline
\hline
 \hhhv $\chi^2/N_{\rm pts}$ &     1179/1280           &  1285/1365    & 1269/1365  & 1262/1365 & $\times$   \\
 \hhhw Probability      &           93.3\%             & 83.1 \%        & 90.0 \%         & 91.5 \%    &  $\times$     \\
\hline
\hline
\end{tabular}
}
%\end{center}
\caption {EBHLS$_2$ contributions to $10^{10} \times a_\mu^{\rm HVP-LO}$
integrated up to 1.05 GeV, including FSR and Coulomb interaction among the
(slow) kaons involved in the $K^+ K^- $ final state. 
The running conditions are indicated
on top of the Table, BS and RS stand resp. for the so-called Basic and Reference
variants  defined in \cite{ExtMod7}. The last column displays the evaluation
through a direct integration of the data.
\label{Table:T7}
}
\end{table}

{\bfEPJC
% FRED CORRECTION 1
Regarding the FSR correction of the $\pi^+\pi^-\pi^0$ channel, we
assume that the FSR correction of the 2$\pi$ channel applies to the
2$\pi$ subsystem of the 3$\pi$ final state as well. Thus we take
$\sigma_{3\pi\gamma}(s)\approx
\sigma_{3\pi}(s)[\frac{\alpha}{\pi}\,\eta(s')]$ as an estimate, assuming
that the invariant mass square $s'$ of the charged $\pi^+\pi^-$
subsystem may be approximately identified as $s'\approx s$. This
 is justified because the main contribution comes from the
$\rho^0$ enhanced intermediate state  $(\gamma \rho^0 \pi^0)$, {\it i.e.}
the resonance enhancement happens at about the same $s\sim M_\rho^2$ in both
the $2\pi$ and the $3\pi$ channels (see
also~\cite{Jegerlehner:2017kke}). One then obtains a FSR contribution
$0.17 \times 10^{-10}$ to which  a 5\% error is assigned. The same
approximation is accepted by the BESIII Collaboration and their recent
3$\pi$ spectrum~\cite{BESIII_3pi}  already includes the FSR correction 
computed this way.
}

Table \ref{Table:T7} collects the results derived from EBHLS$_2$ fits
performed under various conditions. The largest difference beween
the central values for the HLS sums does not exceed $0.4\times 10^{-10}$
and reflects the effect of using or not the $\tau$ dipion spectra -- together
with slightly improved uncertainties ($\simeq$ 10\%) in the former option. The second data column
collects the results derived by assuming $\lambda_3\equiv 0$ and $\delta P^\tau(s)$
 third degree ({\it i.e.} the previous BHLS$_2$ framework);
 the third data column information is derived by letting
$\lambda_3$ free and fixing the $\delta P^\tau(s)$ degree to 2. Despite their
different probabilities, their HVP's differ
by only $6 \times 10^{-12}$ and their uncertainties too. Comparing the third 
and fourth data
columns also shows that the gain achieved by using 
the Primordial Mixing Mechanism  \cite{ExtMod7} is, by now,  negligible.

Therefore, it looks consistent to choose as final evaluation of the BHLS channel
contribution to $a^{\rm HVP-LO}_\mu$ up to $1.05$ GeV~:
$$ a_\mu^{\rm HVP-LO} (HLS) =[572.14 \pm 0.95]  \times 10^{-10}$$
up to additional systematics considered just below.
%\clearpage

On the other hand, the last data column in Table \ref{Table:T7} displays the results derived
by a direct integration of the annihilation data; in this approach, the normalization
of each of the combined spectra is the nominal one and all uncertainties
(correlated or not) are combined to provide its weight in the combined spectrum.
This brings us back to the discussion presented in the previous Subsection~:
It is not surprising to observe the data shifting compared to expectations
and their  uncertainties enlarged by the correlated contributions.
This effect is the largest for the $\pi\pi$ contribution but represents
only a $\simeq 1 \sigma_{exp}$ effect. The difference for the dikaon
contributions is rather due to taking the CMD3 data into account in the direct 
integration
whereas they are absent from the set of data samples submitted
to the EBHLS$_2$ fit procedure (see \cite{ExtMod7}); their effect is, nevertheless, 
taken into account as systematics.
\subsection{Systematics in the HLS Contribution of the Muon HVP}
\label{HVP-syst}
\indentB
Subsection \ref{amu_pipi} has
illustrated, specifically on the $\pi \pi$ channel, that a possible hint 
for a significant bias induced by the global fit method itself is tiny. Indeed, Table \ref{Table:T1}
shows that, as long as the fit probabilities are good, the values for $a_\mu(\pi\pi)$ derived from
the fit  are very close to the KLOE Collaboration own evaluations \cite{Anastasi:2017eio}.
More precisely, the EBHLS$_2$ fit estimates are distant by only resp.
 $0.28 \sigma_{exp.}$,$0.25 \sigma_{exp.}$ and $0.15 \sigma_{exp.}$ 
from the KLOE own direct integration evaluations for  resp. 
KLOE10 (78\% prob.), KLOE12 (80\% prob.) and the KLOE85 combined data set (65\% prob.) 
 The example of KLOE08 is, however, also interesting~: Indeed, even if  the fit probability is poor (14\%)
-- and for this reason excluded from our reference set of data samples -- the fit
is away from the KLOE direct integration of this spectrum by only $1.6 \sigma_{exp.}$.

On the other hand, the set of accepted data samples
being similar, Table \ref{Table:T6} also indicates that 
 the way the normalization uncertainty is dealt with accounts
for the bulk of the differences between the various approaches. The (similar)
choices made by KNT  \cite{Keshavarzi_2019} and us \cite{ExtMod5} look  
the best grounded one and leads to consistent central values. The better precision
reached within the broken HLS frameworks mostly proceeds from the global fit tool
they allow which numerically correlates the various annihilation channels
as if the statistics in each channel was larger than nominal.
One should also remind the marvellous agreement (still valid) between the
BHLS$_2$ prediction and the Lattice QCD 
 form factor spectra  \cite{ETMC-FF} emphasized in  \cite{ExtMod7}  
 (see  Figure 8 therein) -- and  in Figure \ref{Fig:lambda3Pred} above.
  
 Therefore, once a  canonical treatment of the various kinds
  of systematic  uncertainties  reported by the various groups together with their spectra
  is applied, 
  one may consider that these are already absorbed in the uncertainties derived from the
  {\sc minuit} minimization procedure.
  
\vspace{0.5cm}

However,  additional sources of uncertainty can be invoked. Until 
 EBHLS$_2$ is experimentally strengthened by new high statistics dipion spectra to be 
collected in the $\tau$
decay, one may consider that the difference between using or not the $\tau$ data contributes
a systematic uncertainty which can increase $a_\mu^{\rm HVP-LO}$
by at most $0.32 \times 10^{-10}$ (see Table \ref{Table:T7}).  
On the other hand, it is worthwhile to anticipate on the treatment of the $[\pi^0,\eta,\etp]$
mixing properties addressed in this paper from Section \ref{PS_mixing_1} onwards.
This will emphasize the relevance of the kinetic
breaking mechanism defined in Section \ref{thooft-kin} and lead to consider
 a possible shift  of $a_\mu^{\rm HVP-LO}$ by $\pm 0.3 \times 10^{-10}$ 
(see Section \ref{side-results}).

The poor knowledge of
the dipion spectrum in the $\phi$ mass region has  been emphasized. Here also,
considering the numbers given in Subsection \ref{amu_pipi},  
the central value for $a_\mu^{\rm HVP-LO}$ might undergo a shift of
$+0.28 \times 10^{-10}$.
In \cite{ExtMod7}, assuming their systematics uncorrelated, 
fits related with the CMD3 dikaon data \cite{CMD3_KpKm,CMD3_K0K0b} are reasonably good;
therefore, leaving them outside our reference sample set may result
in missing  $+0.54 \times 10^{-10}$ when evaluating $a_\mu(K\overline{K})$. 

The still preliminary SND dipion data \cite{SND20} examined above has been submitted
to our standard global fit by including it into the set of accepted dipion 
spectra. The fit returns, with a probability of 66.2\%, 
$$a_\mu({\pi \pi}) = 493.26 \pm 0.81 ~~{\rm and}~~~ 
a_\mu^{\rm HVP-LO}=572.60\pm 0.89~~,$$
in units of $10^{-10}$.
As the average $\chi^2/N_{points}$ for this SND sample is large ($\simeq 2$) we gave up
including it inside the fitted sample set and preferred 
 affecting the difference  $+0.48 \times 10^{-10}$ to the systematics. 

{\bfEPJC
% FRED Correction 2  now on page 49
Well identified other sources of systematics deserve to be addressed:
(i) The uncertainty\footnote{A part of this effect might already be
accounted for in the experimental uncertainties.}  on the total photon
VP ($\gamma V P $) has been estimated to $\pm 0.29 \times 10^{-10}$
and (ii) The FSR effect in the HLS energy range covers its
contributions to the $\pi^+\pi^-$, $\pi^+\pi^-\pi^0$ and $K^+K^-$
annihilation channels; its value amounts to $4.81 \times 10^{-10}$
over the whole non-perturbative region and may be conservatively
attributed a 2\% uncertainty\footnote{ \bfEPJC In the HLS region
($\sqrt{s} < 1.05$ GeV), we have the FSR contributions $4.26 \times
10^{-10}$ from $\pi^+\pi^-\gamma$, $0.17 \times 10^{-10}$ from
$\pi^+\pi^-\pi^0\gamma$ and $0.38 \times 10^{-10}$ from
$K^+K^-\gamma$. Altogether this amounts to $4.81 \times 10^{-10}$ 
to which  a 2\% uncertainty is attributed, {\it i.e.}
$0.10 \times 10^{-10}$.}. In the non-HLS range above 1.05 GeV the
contributions listed in Table~\ref{Table:T8} include FSR effects
estimated via the Quark Parton model. The $n_f=$ 3-, 4- and 5-flavor range
LO contributions are to be multiplied by the radiative correction
factors $\frac{3\alpha}{4\pi}\,N_c\,\sum_{i=1}^{n_f} Q^2_i$, which
yields $0.42 \times 10^{-10}$ as a total FSR effect and one may assign
a 10\% uncertainty here. The uncertainty values just given actually
affect the HVP over the whole energy range.
}

These possible additional source of systematics rather play as   shifts and, thus,
should not be combined with the uncertainty returned by the fit. Summing up all these
estimates,   our final result can be completed with the
most pessimistic systematic uncertainty\footnote{
A model uncertainty, estimated  to $\pm 0.3 \times 10^{-10}$
 in Section  \ref{side-results}, has been added linearly to the systematics.}~:
\be        
a_\mu^{\rm HVP-LO} (HLS, \sqrt{s} \le 1.05~{\rm GeV}) =572.14 \pm [0.95]_{fit} + [^{+2.31}_{-0.69}]_{syst.}  
\label{HLS_part}
\ee
in units of $10^{-10}$.  

Finally, the tiny 
contribution generated by the  "non--HLS" channels \footnote{It is provided by the low energy tails of 
channels like $e^+ e^- \ra 4 \pi $, $2\pi \eta$
or $\etp \gam \cdots$ the thresholds of  which being smaller than the $\phi$ meson mass.} 
should be considered to fully complement the
$[s_0=m_{\pi^0}^2,s_{cut}]$ energy interval contribution to the muon HVP; 
it has been  re-estimated  by direct integration
of the (sparse) existing data to $=[1.21 \pm 0.17] \times 10^{-10}$. 

\subsection{The Muon HVP and Anomalous Magnetic Moment}
\label{HVP-total}
\indentB To finalize our HLS based estimate of the muon HVP, our result Equation
(\ref{HLS_part}), complemented for the non--HLS channel contribution below
$\sqrt{s_{cut}}=1.05$ GeV already given, should be supplied by the contributions
from above this energy limit. This is displayed in the left-hand part
of Table \ref{Table:T8}; the different contributions up to  $\sqrt{s_{cut}}=5.20$ GeV
are derived by a numerical integration of the experimental data (annihilation
spectra and $R(s)$ ratio measurements) as for the $\Upsilon$ energy interval.
This part carries a significant uncertainty. The rest, evaluated using perturbative QCD,
 is reported under the tag "pQCD" and exhibits a high precision. 

\begin{table}[!htpb!] 
%\hspace{-2.2cm}
\begin{center}
%\begin{minipage}{0.5\textwidth}
{\scriptsize
\begin{tabular}{|| c  | c || c  ||||c  | c || }
\hline
\hhhv Contribution from	&  Energy Range           &  $10^{10} \times
a_\mu^{\rm HVP-LO} $ & Contribution from & $10^{10} \times a_\mu $\\
\hline
\hline
\hhhq missing channels 	&  $\sqrt{s} \le 1.05$   &  $1.21 \pm 0.17 $
& LO-HVP  &    $687.48 \pm 2.93 +\left[^{+2.31}_{-0.69} \right]_{\rm syst} $ \\
\hline
\hhhv $J/\psi$ 		& ~~~  			  &  $8.94\pm 0.59$   & NLO  HVP\cite{Keshavarzi_2019} & $-9.83  \pm 0.07 $\\
\hline
\hhhv $\Upsilon$        & ~~~  			  &  $0.11 \pm 0.01$   & NNLO HVP \cite{NNLO} &$~1.24  \pm 0.01 $\\
\hline
\hhhv hadronic		& (1.05, 2.00)		  &  $62.95 \pm 2.53$  & LBL \cite{LBL,WhitePaper_2020} &  $9.2 \pm 1.9$\\
\hline
\hhhv hadronic		& (2.00, 3.20)		  &  $21.63 \pm 0.93 $  & NLO-LBL \cite{LbLNLO} & $0.3 \pm 0.2$ \\
\hline
\hhhv hadronic		& (3.20, 3.60)		  &  $3.81 \pm 0.07 $   &  QED \cite{Passera06,Kinoshita2} &$11~658~471.8931 \pm 0.0104$\\
\hline
\hhhv hadronic		& (3.60, 5.20)		  &  $7.59 \pm 0.07$  &  EW \cite{Gnendiger:2013pva,Czarnecki:2002nt} &$15.36\pm 0.11 $\\
\hline
\hhhq  pQCD	        & (5.20, 9.46)		  &  $6.27 \pm 0.01$  &  Total Theor. &$11~659~175.33\pm 3.49 
 +\left[^{+1.62}_{-0.0} \right]_{\rm syst}$ \\
\hline
\hhhv hadronic     	& (9.46, 11.50)		  &  $0.87 \pm 0.05$  & Exper. Aver. \cite{FNAL:2021} &$11~659~206.1 \pm 4.1$ \\
\hline
\hhhq  pQCD		& (11.50,$\infty$)	  &  $1.96 \pm 0.00$ &
$10^{10} \times \Delta a_\mu$ &$30.77 \pm 5.38 -\left[^{+2.31}_{-0.69} \right]_{\rm syst} $ \\
\hline
\hhhv Total		& 1.05 $\to \infty$	  &  {$115.34 \pm 2.77$}   & Significance ($n \sigma$) &$5.72 \sigma$\\
\hhhe ~~~		& + missing chann.	  &  {~~~} &  {~~~}		&~~~~~	\\
\hline
\hline
\end{tabular}
} % \scriptsize
%\end{minipage}
\end{center}
\caption{
\label{Table:T8}
The left-hand side Table  displays the updated contributions to
$a_\mu^{\rm HVP-LO} $  from the various energy regions and
includes the contribution of the non-HLS channels in the $\sqrt{s} < 1.05$ GeV region; only total errors are shown.
The right-hand side Table provides the various contributions to $a_\mu $ in accord with Table 1 in \cite{WhitePaper_2020} 
together with our own datum for $a_\mu^{HVP-LO}$. The result for
$\Delta a_\mu=a_\mu({\rm exp})-a_\mu({\rm th})$, based on the EBHLS$_2$ fit and the average of the BNL and FNAL 
measurement \cite{BNL,FNAL:2021}, is also given; the effect of the systematic is discussed in the 
body of the text. 
 }
\end{table}
Summing up the various components, our evaluation of the muon HVP integrated over the
full energy range is~: 
\be
a_\mu^{\rm HVP-LO}  = \left \{ 687.48 \pm [2.93]_{fit} + [^{+2.31}_{-0.69}]_{syst.} \right \} \times  10^{-10}~.
\label{HVP-tot}
\ee
In order to derive the anomalous magnetic moment of the muon, its HVP should be complemented
with the contributions other than the LO--VP~: Higher order HVP effects, light-by-light, QED and electroweak
inputs. For consistency with others, we have used for these the values given in Table 1 of  \cite{WhitePaper_2020}.
This sums up to~:
\be
a_\mu^{HLS}  =11~659~175.33\pm 3.49  +\left[^{+2.31}_{-0.69} \right]_{\rm syst}
\label{amu-tot}
\ee
in units of $10^{-10}$, which exhibits  a $5.72 \sigma$ difference with the experimental average \cite{FNAL:2021}.
If taking into account the possible shift of the $a_\mu$ central value following from our 
systematics upper bound,  the significance for $\Delta a_\mu=a_\mu^{BNL}-a_\mu^{HLS}$ can decrease
down to $5.31 \sigma$.

A  final remark should be asserted~: One may  found amazing the jump in significance of $\Delta a_\mu$
compared to  \cite{ExtMod7} ; a mere comparison of the EBHLS$2$ numerical outcome 
with those of our previous work clearly shows that it is almost unchanged.
The  changes reported here  are solely due to the 30\% reduction of the uncertainty 
produced by the averaging\footnote{Using only the BNL datum leads to a significance 
for  the $\Delta a_\mu$ central value of $4.7 \sigma$.}
the FNAL  \cite{FNAL:2021} and  BNL \cite{BNL}measurements. 
\subsection{A Challenging Value for $a_\mu^{HVP-LO}$}
\label{HVP-BaBar}
\indentB
In Subsection \ref{select} , we have revisited the consistency topic
of the various available dipion spectra. The most relevant
fit properties of these are collected in Table \ref{Table:T5}.
Comparing the $\chi^2/N$  averages for NSK, KLOE and BaBar in global fits
where each of them is used as single representative for the $\pi \pi$
channel and fits using their pairwise combinations  permits
several conclusions  reflected by the fit probabilities displayed 
therein\footnote{The 3-pion data are discarded from the fits reported in this Table.}.
Namely~:
\begin{itemize}
\item The tension exhibited by the pairwise fit involving KLOE and NSK is marginal 
compared to the fits using each of them in isolation~: The fit probabilities
are quite similar.
\item In the pairwise fit involving KLOE and BaBar, one observes a strong tension
reflected by the drop in probability between the pairwise fit and those
with KLOE and BaBar in isolation.
\item The pairwise fit of the NSK and BaBar spectra also exhibits some tension
between them, but at a softer level~: If the drop in probability versus
the NSK fit in isolation is large (a factor of $\simeq 2$), the corresponding 
drop in probability versus BaBar in isolation is small ($62.9 \% \ra  51.7 \%$). 
\end{itemize}

This motivates examining a global fit involving the NSK\footnote{Including the
former data collected in \cite{Barkov}.}, BESIII \cite{BES-III,BESIII-cor}, 
Cleo-c \cite{CESR} and BaBar spectra, the KLOE data samples being excluded.
In contrast with the fits reported in Table \ref{Table:T5}, this special fit includes
the three pion spectra and involves 1500 data points; it converges at 
$\chi^2_{total}=1484$, yielding a 39.5 \% probability. This fit is not
as good   as the standard one (see the Subsection just above) which results   
in a  $\simeq 90\%$ probability, but is reported in
some details here for completeness. 

\begin{figure}[!hptb]
\hspace{0.5cm}
%\vspace{-0.5cm}
\begin{minipage}{0.9\textwidth}
\begin{center}
\vspace{-1.5cm}
\resizebox{1.0\textwidth}{!}
%{\includegraphics*{fig_2020/amu_th_2021_fnal_luigi_red.eps}}
%{\includegraphics*{fig_2020/amu_th_2021_fnal_luigi_magenta.eps}}
%{\includegraphics*{fig_2020/amu_th_2021_fnal_bmw_magenta.eps}}
{\includegraphics*{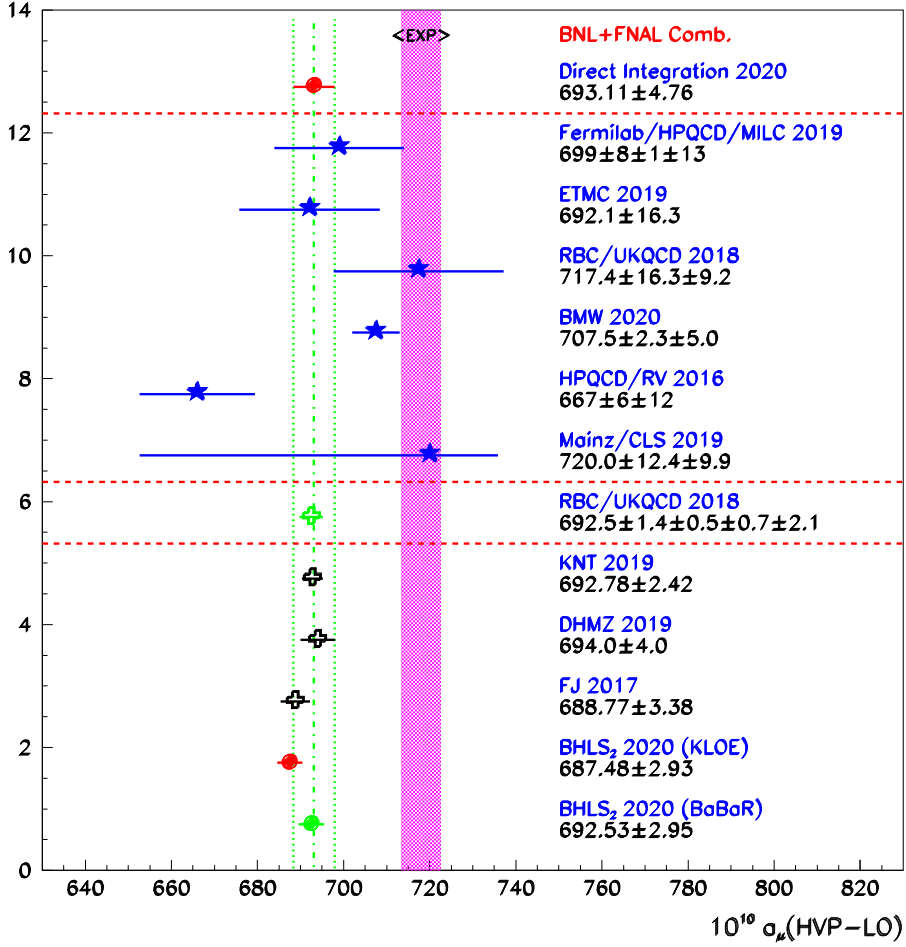}}
\end{center}
\end{minipage}
\begin{center}
\vspace{-1.cm}
\caption{\label{Fig:amu_hvp_lo}  Recent evaluations of $10^{10} \times
a_\mu^{\rm HVP-LO}$:
On top, the result  derived by a direct integration of the data combined with perturbative
QCD; the dotted vertical lines indicate the $\pm 1 \sigma$ interval. The LQCD data are
followed by the results derived using dispersive methods from
\cite{Keshavarzi_2019}, \cite{Davier_2019} and \cite{Jegerlehner_2017}. 
The two HVP-LO evaluations derived using   EBHLS$_2$ fitting codes 
are given at bottom (see text). The $\pm  1 \sigma$ interval corresponding to
the BNL+FNAL average \cite{FNAL:2021}  is shown by the shadded area. }
\end{center}
\end{figure}

For the region up to $1.05$ GeV, one gets~:
\be
a_\mu(\pi \pi)= 497.83 \pm 0.90~~~{\rm and}~~~a_\mu(HLS)=577.19\pm 1.00
\label{amu_babar}
\ee
in units of $10^{-10}$.  The corresponding standard results
can be read off Table \ref{Table:T7}, more precisely its third
data column; the increases produced by BaBar (excluding KLOE) for these quantitites
 are equal~: $\delta a_\mu(\pi \pi)=\delta a_\mu(HLS)=5.07 \times 10^{-10}$. So the 
 differences is fully carried by the $\pi \pi$ channel.
 The information displayed in the left-hand Table \ref{Table:T8}
 allows to derive the full HVP-LO~:
\be
a_\mu^{HVP-LO}({\rm BaBar})= [692.53 \pm 2.95]  \times 10^{-10}
\label{amu_tot_babar}
\ee
and can be affected by the same additional systematic uncertainty
proposed above. Finally, the difference between $a_\mu^{HVP-LO}({\rm BaBar})$ 
and the experimental average \cite{FNAL:2021}  drops down to 
$\Delta a_\mu=a_\mu^{avrg}-a_\mu^{HVP-LO}({\rm BaBar})= 23.65 \pm 5.38$ 
and exhibits a statistical significance
of $4.78 \sigma$ not counting the systematic uncertainty effect.
Taking it into account, the significance may drop down to $4.35 \sigma$.
\subsection{The Different Muon HVP-LO Evaluations}
\label{hvp-compare}
\indentB
Figure \ref{Fig:amu_hvp_lo} collects recent evaluations of the 
leading order muon HVP. A numerical integration
of the annihilation and $R(s)$ data, appropriately completed by 
perturbative QCD calculations (see Table \ref{Table:T8}) yields the
entry displayed at top of the Figure, It is followed by  some reference
evaluations derived by LQCD Collaborations, namely 
 \cite{FmHPQCD_MILC_amu}, \cite{ETMC_amu_up}, \cite{RBC_UKQCD_amu},
\cite{BMW_amu_final}, \cite{Chakraborty2016}, \cite{Gerardin:2019rua};
the second of the RBC/UKQCD evaluations \cite{RBC_UKQCD_amu} relies on mixing
LQCD and dispersive information. This datum is just followed by the HVP-LO
dispersive evaluations from \cite{Keshavarzi_2019}, \cite{Davier_2019} and 
\cite{Jegerlehner_2017}. 
The bottom pair  of data points are the EBHLS$_2$  based evaluations of
the full HVP-LO derived in this study. It can be noted that the prefered EBHLS$_2$ 
evaluation
(90\% prob.), tagged by KLOE, is $5.30 \times 10^{-10}$ smaller than KNT19 
 \cite{Keshavarzi_2019} and $6.54 \times 10^{-10}$ smaller than DHMZ19
\cite{Davier_2019}. In contrast, the challenging evaluation (40\% prob.), 
tagged by BaBar, is only distant  by resp. $0.25 \times 10^{-10}$ and
$1.47 \times 10^{-10}$ from resp. KNT19 and DHMZ19. 

The difference
of our 2 evaluations ($5.05 \times 10^{-10}$) compared to their
respective accuracies ($2.9 \times 10^{-10}$), makes us reluctant
to propose a mixture of these or a common KLOE+BaBar fit evaluation.
Nevertheless, it shows that model dependence is not the main source
of disagreement between the various dispersive evaluations. 

%%  The gap between the most recent BMW evaluation \cite{BMW_amu_final}
%%  and  the estimates based on the Dispersion Relation Approach 
%%  ranges between $ 13 \times 10^{-10}$ and $20 \times 10^{-10}$;
%%  it  looks difficult to fill this gap only with  
%%  $e^+ e^-$ annihilation data below $\simeq 1$ GeV, an
%%  experimentally well explored energy region. 

The difference between the most recent BMW evaluation \cite{BMW_amu_final}
and those based on the various Dispersion Relation (DR)  Approaches 
ranges between $ 13 \times 10^{-10}$ and $20 \times 10^{-10}$;
it looks difficult to fill such a gap with only    
$e^+ e^-$ annihilation data below $\simeq 1$ GeV, an
experimentally well explored energy region.

\section{The EBHLS$_2$ Approach to the $[\pi^0,\eta,\etp]$ System}
\label{PS_mixing_1}
\indentB The mixing properties of the $[\pi^0,\eta,\etp]$
system underly the physics of  the light meson radiative decays as
well as the amplitudes for the $e^+e^- \ra P \gamma$ annihilations which are obviously 
tightly related.  The other important data involved in this issue, namely
the $P \gam \gam$ decays and 
the $V\gamma \etp$ couplings, are also part of the EBHLS$_2$ scope\footnote{ Actually, among 
the data traditionally used to address this topic, only the $J/\psi $ decay information remains 
outside the EBHLS$_2$ scope.}.

Phenomenological descriptions of the $[\pi^0,\eta,\etp]$
system were first based on using $U(3)$ symmetric $VP\gamma$ coupling 
expressions  enriched by  parametrizations 
of Nonet Symmetry breaking in both the pseudoscalar and vector sectors as
done in \cite{ODonnell}, for instance. A first HLS based model
including its anomalous sectors \cite{FKTUY} has provided a
unified framework which encompasses the $P \ra \gam \gam$ decays 
and the radiative decays of the form $VP\gamma$ \cite{rad}.
The  Effective Lagrangian approach, started long ago 
(see \cite{Schechter:1992iz}, for instance),
 has been pursued up to very recently (a comprehensive list of previous references
can be found in \cite{Escribano:2020}).

On the other hand, Chiral Perturbation Theory and its Extension (EChPT), 
originally formulated by Kaiser and Leutwyler \cite{leutw,leutwb},  allowed to 
fully address the $[\pi^0,\eta,\etp]$ mixing and gave rise 
to the singlet-octet basis description  in terms of 2 angles and 2 decay
constants. It was shown in \cite{WZWChPT} that this approach is naturally 
accomodated within a HLS framework with only one mixing angle provided that $SU(3)$
and Nonet Symmetry breakings are also accounted for within its Effective
Lagrangian. Beside the singlet-octet basis formulation, 
 another convenient formulation, known as quark flavor basis 
has also been proposed in \cite{feldmann_1,feldmann_2}; its properties 
and its relation with the singlet-octet formulation have been 
thoroughly reported in \cite{feldmann_3}. 

Finally, Isospin Symmetry breaking in the  $[\pi^0,\eta,\etp]$ system 
has also been considered and parametrized as $\eta$ and $\etp$
admixtures inside the physically observed $\pi^0$ meson \cite{leutw96,feldmann_3}.
Additional Isospin breaking effects have also been studied, generated by having different 
$u\overline{u}$ and $ d\overline{d}$ decay constants \cite{Kroll:2005}.

It happens that the parameters which substantiate  the singlet-octet \cite{leutw,leutwb}
and quark flavor basis parametrizations \cite{feldmann_1,feldmann_2,feldmann_3} 
can be accessed within the EBHLS$_2$ framework as reported above -- 
and in \cite{ExtMod7}. Its Lagrangian leads
to quite similar expressions and includes also Isospin Breaking contributions.
Moreover, the global fits performed and already referred to in the previous Sections
allow precise numerical determinations of the mixing parameters of
the $[\pi^0,\eta,\etp]$ system in both the singlet-octet and quark flavor bases.

In the following, one  reports on works performed in parallel with both the BHLS$_2$ framework 
previously defined   \cite{ExtMod7} and its EBHLS$_2$ Extension analyzed in the preceding 
Sections. However, the detailed analysis of the $[\pi^0,\eta,\etp]$ system exhibits
properties which make relevant a further analysis related with the kinetic breaking 
introduced above to account for
the $\tau$ dipion spectra, especially the Belle one, by far the most precise spectrum. 

  If the fate of EBHLS$_2$ versus BHLS$_2$ is
 tightly related with a forthcoming high statistics measurement  of the
dipion spectrum in the $\tau$ decay,  the analysis of the $[\pi^0,\eta,\etp]$ system
 nevertheless reveals constraints among the three terms of the kinetic breaking matrix $X_H$ which
should be addressed and which, certainly,  influence this picture.
\section{The Axial Currents from the EBHLS$_2$ Effective Lagrangian}
\label{Axial_currents}
\indentB
The main tool to address the $[\pi^0,\eta,\etp]$ mixing topic are the axial currents
which are derived from the pseudoscalar kinetic energy term Equation (\ref{kin12}) -- reminded
here for convenience~:
\be
 \displaystyle K =  {\rm Tr} \left [ \pa P_{bare} 
   X_A  \pa P_{bare}  X_A\right ]   + 
  2 ~ \{ {\rm Tr} \left [X_H \pa  P_{bare} \right ] \}^2  
%   \label{current1}
\ee
by the derivatives~:
\be
\begin{array}{ll}
 \displaystyle J_\mu^a  =  f_\pi \frac{\pa K}{\pa (\pa_\mu P^a_{bare})}~~,&
 ~ \left [
  P_{bare} =\sum_{a=0, \cdots 8} T_a  P^a_{bare}
 \right ]~~,
\end{array}
   \label{current2}
\ee
\noindent
with respect to the entries associated with each of the $U(3)$ basis matrices  normalized 
such that Tr$[T_aT_b]=\delta_{ab}/2$; the breaking matrix $X_A$ is given in 
Equation (\ref{kin2}) and $X_H$ can be found in Equation (\ref{kin10}).
They are given by (summation over repeated indices is understood)~:
\be
 \displaystyle J_\mu^a = 2f_\pi {\rm Tr} \left [ T_a
   X_A T_b  X_A\right ]\pa P_{bare}^b
 + 4 ~  {\rm Tr} \left [X_H \pa  P_{bare} \right ]    {\rm Tr} 
 \left [X_H T_a \right ] ( \delta_{a,0} + \delta_{a,3}  + \delta_{a,8}) ~~.
   \label{current3}
\ee

The axial currents relevant for our purpose are~:
  \be
  \left \{
\begin{array}{lll} 
J^{3}/f_\pi=&\displaystyle ~~~
\pa \pi^3_b +\frac{\Delta_A }{\sqrt{3}}  \left [\sqrt{2} \pa \eta^0_b+\pa \eta^8_b\right ] + 
\lambda_3 \left [\lambda_3 \pa \pi^3_b + \lambda_0 \pa \eta^0_b+ \lambda_8 \pa \eta^8_b \right]
\,,\\[0.5cm]
J^{0}/f_\pi=&\displaystyle D \pa \eta^0_b + G \pa \eta^8_b + \sqrt{\frac{2}{3}} \Delta_A ~\pa \pi^3_b + 
\lambda_0 \left [\lambda_3 \pa \pi^3_b + \lambda_0 \pa \eta^0_b+ \lambda_8 \pa \eta^8_b \right]
\,,\\[0.5cm]
J^{8}/f_\pi=&\displaystyle G \pa \eta^0_b + F \pa \eta^8_b + \frac{\Delta_A}{\sqrt{3}} ~ \pa \pi^3_b + 
\lambda_8 \left [\lambda_3 \pa \pi^3_b + \lambda_0 \pa \eta^0_b+ \lambda_8 \pa \eta^8_b \right]
\,,
 \end{array}
 \right .
 \label{current6}
\ee
where the subscript $b$ stands for {\it bare } and~:
 \be
% \hspace{-1.5cm}
\begin{array}{lll} 
 \displaystyle   D=\frac{z_A^2+2}{3} ,&
 \displaystyle   F=\frac{2z_A^2+1}{3} ,&
 \displaystyle   G=\frac{\sqrt{2}}{3} (1-z_A^2) \,,
 \end{array}
 \label{current5}
\ee
in terms of $z_A$, the $(s,\overline{s})$ entry of the $X_A$ breaking matrix.
%using the expressions for $A$, $B$ and $C$ given in Equations (\ref{kin5}).
Equation (\ref{kin20}) and the definitions given in Section \ref{thooft-kin} allow to express
the axial currents at first order in breakings in terms of the renormalized 
$R$-fields -- those which render canonical the PS kinetic energy term -- by~:
\be
\left \{
\begin{array}{ll} 
\displaystyle  J^3_\mu/f_\pi=
 \left (1+\frac{\lambda_3^2 }{2}\right ) \pa_\mu \pi^3_R
+  \left ( \frac{\lambda_3 \tlambda_0}{2} + \frac{\Delta_A}{\sqrt{6}}
\right ) \pa_\mu\eta_R^0
+  \left ( \frac{\lambda_3 \tlambda_8}{2} + \frac{\Delta_A}{2\sqrt{3}}
\right ) \pa_\mu\eta_R^8 \,, \\[0.5cm]
\displaystyle  J^0_\mu/f_\pi=
\left (
\frac{\Delta_A}{\sqrt{6}}+ \frac{\lambda_0 \lambda_3}{2}
\right )\pa_\mu \pi^3_R
+
\left (
z_A C +\frac{\lambda_0 \tlambda_0}{2}
\right ) \pa_\mu\eta_R^0
+
\left (
- z_A A 
+ \frac{\lambda_0 \tlambda_8}{2}
\right ) \pa_\mu\eta_R^8 \,,\\[0.5cm]
\displaystyle  J^8_\mu/f_\pi=
\left (
\frac{\Delta_A}{2\sqrt{3}}+ \frac{\lambda_3 \lambda_8}{2}
\right ) \pa_\mu \pi^3_R
+
 \left (
- z_A A  + \frac{\lambda_8 \tlambda_0}{2} 
 \right )  \pa_\mu\eta_R^0
+
 \left (
 z_A B + \frac{\lambda_8\tlambda_8}{2}
  \right )  \pa_\mu\eta_R^8  \,,
\end{array}
\right .
 \label{current10}
\ee
$\tlambda_0$ and $\tlambda_8$ having been given in Equations (\ref{kin14})
in terms of $A$, $B$, $C$, $\lambda_0$ and $\lambda_8$. The following matrix elements
 are of purpose for the $\eta-\etp$  mixing topic ($\pa_\mu  \ra iq_\mu$, 
 outgoing momentum)~:
\be
\left \{
\begin{array}{ll} 
\displaystyle  <0|J^0_\mu|\eta_R^0>= 
f_\pi \left \{~~  ~ z_A C +\frac{\lambda_0 \tlambda_0}{2}
\right\} iq_\mu (\eta_R^0) \equiv i f_0 ~ q_\mu (\eta_R^0)
 \,,\\[0.5cm]
\displaystyle  <0|J^0_\mu|\eta_R^8> =
f_\pi \left \{- z_A A  + \frac{\lambda_0 \tlambda_8 }{2}
\right\} iq_\mu (\eta_R^8) \equiv i b_0 ~q_\mu (\eta_R^8)
 \,,\\[0.5cm]
\displaystyle  <0|J^8_\mu|\eta_R^8> =
f_\pi \left \{~~~  z_A B 
+ \frac{\lambda_8\tlambda_8}{2}
\right\} iq_\mu (\eta_R^8) \equiv i f_8 ~ q_\mu (\eta_R^8)
 \,,\\[0.5cm]
\displaystyle  <0|J^8_\mu|\eta_R^0> =
f_\pi \left \{- z_A A 
+ \frac{\lambda_8 \tlambda_0}{2}
\right\} iq_\mu (\eta_R^0)  \equiv i b_8  ~ q_\mu (\eta_R^0)  \,,
\end{array}
\right .
 \label{current11}
\ee
which define the decay constants $f_0$, $f_8$, $b_0$ and $b_8$. One observes
that the kinetic breaking affects all these matrix elements; in order to
connect with \cite{WZWChPT}, one should identify the usual kinetic ('t~Hooft)
breaking term with $\lambda_0^2$.

\vspace{0.5cm}

Regarding the other fields, EBHLS$_2$ does not go  beyond the BKY 
breaking \cite{BKY, Heath} as their  $R$ renormalized
and physical states coincide. They relate to their bare partners 
by \cite{ExtMod3}~:
\be
\begin{array}{lll} 
\displaystyle 
\pi^\pm_b = \displaystyle \pi^\pm_R~~,&
\displaystyle 
K^{\pm}_b=
\frac{1}{\sqrt{z_A}} \left[ 1-\frac{\Delta_A}{4}\right ]
 K^{\pm}_R~~,&
\displaystyle 
K^{0}_b=\displaystyle \frac{1}{\sqrt{z_A}} \left[ 1+\frac{\Delta_A}{4}\right ] K^{0}_R  ~~.
\end{array}
 \label{other1}
\ee
This transformation to $R$ fields brings  the corresponding part of the kinetic 
energy term into canonical form \cite{ExtMod3}. The corresponding axial currents write~:
\be
\begin{array}{lll} 
\displaystyle 
J^{\pi^\pm}_\mu
=f_\pi \partial_\mu \pi^\mp_R~~,&  
\displaystyle 
J^{K^\pm}_\mu=\sqrt{z_A} 
\left[ 1+\frac{\Delta_A}{4}\right ]
f_\pi \partial_\mu K^\mp_R 
~~,& 
\displaystyle 
J^{K^0}_\mu=
\sqrt{z_A} \left[ 1-\frac{\Delta_A}{4}\right ]
f_\pi \partial_\mu\overline{K}^0_R ~~.
\end{array}
 \label{other2}
\ee
Using also the expression for $J^{3}_\mu$ given just above, one can use
the expectation values~:
\be
\displaystyle 
<0| J^P_\mu|P(q)>=i f_{P}~q_\mu~~~,~~ (P=~\pi^\pm,\pi^0,K^\pm,K_L/K_S)
\label{other3}
\ee
to derive~:
\be
\begin{array}{llll}
\displaystyle 
f_{\pi^0}  =  \left [ 1+\frac{\lambda_3^2}{2} \right ]  f_\pi ~~,&                     
\displaystyle 
f_{K^\pm} \equiv f_K
=\sqrt{z_A} \left [ 1+\frac{\Delta_A}{4} \right ]  f_\pi ~~,&
 \displaystyle 
f_{K^0}= \sqrt{z_A} \left [1-\frac{\Delta_A}{4}\right ] f_\pi
\end{array}
\label{other4}
\ee
and then, one gets $z_A=[f_K/f_\pi]^2$ up to Isospin breaking corrections.
Moreover, it should also be noted that, once $\lambda_3$ is floating, $f_{\pi^0}$  
may differ from $f_\pi (\equiv f_{\pi^+})$ by as much than  $\simeq$  2.5 \%, in 
line with the remarks in \cite{Kampf:2012pi0}. This
comes out of our fits which successfully involve, beside the $\tau$ dipion spectra,
 the $e^+ e^- \ra (\pi^0/\eta) \gam$ annihilation data and the widths of the
anomalous decays $\{P \ra \gam \gam, P\in (\pi^0,\eta,\etp)\}$. 

\section{The $\eta-\etp$ Mixing : The Octet-Singlet Basis Parametrization}
\label{octet-singlet}
\indent \indent
The axial current matrix elements in the two angle scheme write~:
\be
 <0|J^{0,8}_\mu|\eta/\eta^\prime>= i F^{0,8}_{\eta/\eta^\prime} q_\mu
 \label{current13}
\ee
in terms of the physical $\eta/\eta^\prime$ fields carrying a momentum $q_\mu$.
As usual, one defines the $ F^{0,8}$ couplings and the $\theta_{0,8}$ mixing angles by
\cite{Kaiser_2000, leutw,leutwb}~:
\be
\left \{
\begin{array}{lll} 
\displaystyle F^{8}_{\eta}& =~~~F^8 \cos{\theta_8} =& \displaystyle f_8  \cos{\theta_P} -b_8  \sin{\theta_P} \,,\\[0.5cm]
\displaystyle F^{8}_{\eta^\prime}& =~~~F^8 \sin{\theta_8} =& \displaystyle f_8  \sin{\theta_P} + b_8  \cos{\theta_P} \,,\\[0.5cm]
\displaystyle F^{0}_{\eta}& =- F^0 \sin{\theta_0} =& \displaystyle b_0  \cos{\theta_P} -f_0  \sin{\theta_P} \,,\\[0.5cm]
\displaystyle F^{0}_{\eta^\prime}& =~~~F^0 \cos{\theta_0} =& \displaystyle b_0  \sin{\theta_P} + f_0  \cos{\theta_P} \,,
\end{array}
\right .
 \label{current14}
\ee
the $f$'s and $b$'s being those given by  Equations (\ref{current11}).
The $\pi^3_R $ components of the physical $\eta$ and $\etp$ fields,
providing contributions of order ${\cal O}(\delta^2)$ in the matrix elements 
$ <0|J^{0,8}_\mu|\eta/\eta^\prime>$, are  discarded.
Using Equations (\ref{current11}) and (\ref{current14}) and the definitions for 
$A$, $B$ and $C$ given in Subsection \ref{diag_la}, one derives~:                              
\be
\left \{
\begin{array}{ll} 
\displaystyle [F^0]^2= [F^{0}_{\eta}]^2 + [F^{0}_{\eta^\prime}]^2
= f_0^2+b_0^2=\left [\frac{z_A^2+2}{3}+ \lambda_0^2 \right ] f_\pi^2  \,,
\\[0.5cm] 
\displaystyle [F^8]^2= [F^{8}_{\eta}]^2 + [F^{8}_{\eta^\prime}]^2
=f_8^2+b_8^2=
\left [\frac{2 z_A^2+1}{3}+ \lambda_8^2  \right ] f_\pi^2 ~~.
\end{array}
\right .
 \label{current15}
\ee
The no--BKY breaking limit is obtained by letting $z_A \simeq [f_K/f_\pi]^2 \ra 1$. The correspondence 
with others
\cite{leutw,feldmann_3,feldmann_2} becomes manifest, when using the following identities~:
\be
\left \{
\begin{array}{ll} 
\displaystyle \frac{z_A^2+ 2}{3}= \frac{2 z_A+1}{3}+ \frac{(z_A-1)^2}{3}
\\[0.5cm] 
\displaystyle \frac{2 z_A^2+ 1}{3}= \frac{4 z_A-1}{3}+ \frac{2}{3}(z_A-1)^2 
\end{array}
\right .
\label{current15b}
\ee
which provide~:
\be
\left \{
\begin{array}{ll} 
\displaystyle [F^0]^2=  \left \{\frac{2 z_A+1}{3}+ 
\left [\lambda_0^2 +\frac{(z_A-1)^2}{3} \right ] \right \} f_\pi^2~,
\\[0.5cm]
\displaystyle [F^8]^2=\left \{
\frac{4 z_A -1}{3}+ 
\left [\lambda_8^2 +\frac{2}{3}(z_A-1)^2 
\right ]
\right \}  f_\pi^2~.
\end{array}
\right .
 \label{current15c}
\ee

If one assumes $\lambda_8=0$,  these expressions coincide at 
leading order in breakings with their usual  EChPT analogs \cite{leutw,feldmann_3};
one should note the  $(z_A-1)^2$ dependent terms, which, even if
non--leading, are not negligible compared to the contributions provided
by the kinetic ('t~Hooft) breaking. Regarding the non--leading terms in the
$ [F^{0/8}]^2$, one can anticipate on our fit result analysis by mentioning that 
$\lambda_0^2 \simeq 8 \times 10^{-2}$ while the flavor breaking correction is governed by
$(z_A-1)^2 \simeq 9 \%$.

At  leading order in breakings, one also finds~:
\be
 \displaystyle F_0 F_8 \sin{\left(\theta_0-\theta_8\right )}=
 \left [ \frac{\sqrt{2}}{3} (z_A^2-1)
 -\lambda_0\lambda_8 \right ] f_\pi^2  \,,
\label{current16}
\ee
which vanishes when no breaking, leading to  $\theta_8=\theta_0=\theta_P$.
This expression can be rewritten~:
\be
 \displaystyle F_0F_8 \sin{\left(\theta_0-\theta_8\right )}=
 \frac{2\sqrt{2}}{3} (z_A-1) f_\pi^2 +  \left [
\frac{\sqrt{2}}{3}(z_A-1)^2   -\lambda_0\lambda_8 \right ] f_\pi^2  \,,
\label{current17}
\ee
which also coincides at  leading in breaking with its EChPT analog \cite{leutw,feldmann_3}, 
as soon as $\lambda_8$ -- which is not involved within EChPT -- is dropped out. 

\vspace{0.5cm}

The usual axial current  matrix elements in the 2--angle mixing scheme yield
the following expressions in terms of the singlet-octet mixing angle $\theta_P$
and of the BKY and kinetic breaking parameters~:
\be
\left \{
\begin{array}{ll} 
\displaystyle \tan{\theta_8}=   \tan{\left [ \theta_P+\Psi_8\right ]},
& \displaystyle \tan{\Psi_8}=\frac{b_8}{f_8}= -
\frac{2 z_A A - \lambda_8 \tlambda_0}{2 z_A B + \lambda_8 \tlambda_8}  \,,\\[0.5cm] 
\displaystyle \tan{\theta_0}= \tan{\left [\theta_P-\Psi_0 \right ]},
& \displaystyle \tan{\Psi_0}=\frac{b_0}{f_0}= -
\frac{2 z_A A - \lambda_0 \tlambda_8}{2 z_A C + \lambda_0 \tlambda_0} ~~.
\end{array}
\right .
 \label{current18}
\ee
Compared with  \cite{WZWChPT}, the expression for  $\tan{\theta_8}$
is recovered in the limit $\lambda_8 \ra 0$. Regarding  $\tan{\theta_0}$,
in the same limiting case, the leading order terms yield~:
$$ \tan{\Psi_0} = \frac{A}{C} \left[1 - \lambda_0^2 \frac{3(z_A+1)}{2 z_A (z_A+2)} \right]
=\frac{A}{C} [1 - 0.80\lambda_0^2]  \,, $$
which exhibits a behavior similar to the  nonet symmetry breaking coefficient $x$ 
defined in \cite{WZWChPT}.
\vspace{0.2cm}

Equations (\ref{current18}) allow to derive interesting expressions for 
the Kaiser-Leutwyler angles $\theta_0$ and $\theta_8$ in terms of the BHLS
model parameters. Discarding terms
of orders ${\cal O}[(z_A-1)^3]$ and  ${\cal O}[(z_A-1)^2 \delta]$ or 
higher\footnote{Let us remind that the $\lambda$ parameters introduced via the
generalized 't~Hooft term are treated as ${\cal O}(\delta^{1/2})$}, one gets~:
\be
\left \{
\begin{array}{ll} 
\displaystyle \theta_8 +\theta_0 = 2 \theta_P+
\frac{\sqrt{2}}{3} (z_A-1) 
\left [\frac{(z_A-1)}{3} - \lambda_0^2+ \lambda_8^2 \right ]    \,, \\[0.5cm]
\displaystyle \theta_8 -\theta_0 = -\frac{2\sqrt{2}}{3} (z_A-1)+ 
\frac{\lambda_0\lambda_8}{z_A} +
\frac{\sqrt{2}}{3} (z_A-1) \left [(z_A-1) + \lambda_0^2+ \lambda_8^2 \right ] ~~.
\end{array}
\right .
 \label{current18-2}
\ee
The expressions one can derive for $\theta_0$ and $\theta_8$ coincide
with those in  Equations (84) in \cite{feldmann_3} at leading order.
 Let us anticipate on the numerical information provided by our
 fits  to indicate that
$z_A-1 \simeq 0.3$ while the different squared $\lambda$ combinations
stand at the few percent level at most; therefore the breaking corrections affect
both the sum and the difference in a significant way.   

Equations (\ref{current18-2}) can be written~:
\be
\left \{
\begin{array}{ll} 
\displaystyle \theta_8  =  \theta_P -\frac{\sqrt{2}}{3} (z_A-1) [1 -\lambda_8^2] + 
\frac{\lambda_0\lambda_8}{2 z_A} + \frac{2\sqrt{2}}{9} (z_A-1)^2  \,, \\[0.5cm]
\displaystyle  \theta_0 =\theta_P +\frac{\sqrt{2}}{3} (z_A-1) [1 -\lambda_0^2] 
-\frac{\lambda_0\lambda_8}{2 z_A} -\frac{\sqrt{2}}{9} (z_A-1)^2  ~~.
\end{array}
\right .
 \label{current18-3}
\ee
One may note the symmetry between the expressions, symmetry  only spoiled by the term of order
 ${\cal O}[(z_A-1)^2]$. This shows that the departure from the 
 one mixing angle scheme only reflects
 the  breaking the $SU(3)$ and  Nonet (or kinetic) symmetries.
 
\section{The $\eta-\etp$ Mixing : The Quark Flavor Basis Parametrization}
\label{quark-flavor}
\indent \indent Beside the Octet-Singlet parametrization of the $\eta-\etp$ system
developped by \cite{Kaiser_2000, leutw,leutwb} and referred to  just above,  another
parametrization has been advocaded by \cite{feldmann_1,feldmann_2,feldmann_3}; this
challenging parametrization will be referred to either as Quark Flavor Basis
or FKS scheme. It looks worth analyzing how it shows up within the Broken HLS framework.
The axial currents relevant to figure out how the FKS parametrization arises within
EBHLS$_2$ are~:
\be
\left \{
\begin{array}{ll} 
\displaystyle J_\mu^q= \sqrt{\frac{2}{3}} J_\mu^0 + \sqrt{\frac{1}{3}} J_\mu^8  \,, \\[0.5cm]
\displaystyle J_\mu^s= \sqrt{\frac{1}{3}} J_\mu^0 -\sqrt{\frac{2}{3}} J_\mu^8   \,,
\end{array}
\right .
\label{current19}
\ee
in terms of the usual singlet and octet axial currents previously encountered.
Using the results collected in Equations (\ref{current10}), one can derive~:
\be
\left \{
\begin{array}{ll} 
\displaystyle  J^q_\mu/f_\pi=
\left [
\frac{\Delta_A}{2}+ \frac{\lambda_q \lambda_3}{2}
\right ] \pa_\mu \pi^3_R
+
\left [
\sqrt{\frac{2}{3}} + \frac{\lambda_q \tlambda_0}{2}
\right ] \pa_\mu\eta_R^0
+
\left [
\sqrt{\frac{1}{3}} + \frac{\lambda_q \tlambda_8}{2}
\right ] \pa_\mu\eta_R^8

 \,,\\[0.5cm]

\displaystyle  J^s_\mu/f_\pi=
\frac{\lambda_s \lambda_3}{2}
 \pa_\mu \pi^3_R
+
\left [
z_A \sqrt{\frac{1}{3}} + \frac{\lambda_s \tlambda_0}{2}
\right ]  \pa_\mu\eta_R^0
-
\left [
z_A \sqrt{\frac{2}{3}} - \frac{\lambda_s \tlambda_8}{2}
\right ] \pa_\mu\eta_R^8
 \,,
\end{array}
\right .
 \label{current20}
 \ee
where $\tlambda_0$ and $\tlambda_8$ have been given in Equations (\ref{kin14})
and where one has defined~:
\be
\begin{array}{ll} 
\displaystyle  \lambda_q = \sqrt{\frac{2}{3}} \lambda_0 + \sqrt{\frac{1}{3}} \lambda_8~~,&
\displaystyle  \lambda_s = \sqrt{\frac{1}{3}} \lambda_0 - \sqrt{\frac{2}{3}} \lambda_8~~,
 \end{array}
\label{current21}
\ee
in tight connection with the definitions (\ref{current19}).
The decay constants relevant in the FKS formulation are~:
\be
\left \{
\begin{array}{ll} 
\displaystyle  <0|J_\mu^q|\eta/\etp> =iq_\mu F^q_{\eta/\etp}  \,,\\[0.5cm]
\displaystyle  <0|J_\mu^s|\eta/\etp> =iq_\mu F^s_{\eta/\etp}  \,,
  \end{array}
\label{current22}
\right .
\ee
and the mixing angles are defined by \cite{feldmann_3}~:
\be
\left \{
\begin{array}{ll} 
\displaystyle  
F^q_{\eta} = ~~F_q \cos{\phi_q}~~,& F^q_{\etp} = F_q \sin{\phi_q} \,,\\[0.5cm]
F^s_{\eta} = -F_s \sin{\phi_s}~~,& F^s_{\etp} = F_s \cos{\phi_s}~~.
  \end{array}
\label{current23}
\right .
\ee
Using Equations (\ref{current21})  and the definition of the renormalized PS fields
in terms of their physical partners (see Equation (\ref{current12-1})), 
one can derive~:
\be
\left \{
\begin{array}{ll} 
\displaystyle  
F^q_{\eta}/f_\pi = - \left [
\sqrt{\frac{2}{3}} + \frac{\lambda_q \tlambda_0}{2}
\right ] \sin{\theta_P} + \left [
\sqrt{\frac{1}{3}} + \frac{\lambda_q \tlambda_8}{2}
\right ] \cos{\theta_P}  \,,\\[0.5cm]
\displaystyle  
F^q_{\etp}/f_\pi = ~~ \left [
\sqrt{\frac{2}{3}} + \frac{\lambda_q \tlambda_0}{2}
\right ] \cos{\theta_P} + \left [
\sqrt{\frac{1}{3}} + \frac{\lambda_q \tlambda_8}{2}
\right ] \sin{\theta_P} \,,\\[0.5cm]
\displaystyle  
F^s_{\eta}/f_\pi = - \left [
z_A\sqrt{\frac{1}{3}} + \frac{\lambda_s \tlambda_0}{2}
\right ] 
\sin{\theta_P} - \left [
z_A \sqrt{\frac{2}{3}} - \frac{\lambda_s \tlambda_8}{2}
\right ] \cos{\theta_P} \,,\\[0.5cm]
\displaystyle  
F^s_{\etp}/f_\pi =~~ \left [
z_A\sqrt{\frac{1}{3}} + \frac{\lambda_s \tlambda_0}{2}
\right ] 
\cos{\theta_P} - \left [
z_A \sqrt{\frac{2}{3}} - \frac{\lambda_s \tlambda_8}{2}
\right ] \sin{\theta_P} \,.
\end{array}
\label{current24}
\right .
\ee
From Equations (\ref{current23}) and (\ref{current24}), one derives~:
\be
\left \{
\begin{array}{ll} 
\displaystyle \tan{\phi_q}=   \tan{\left [ \theta_P+U_q\right ]},
& \displaystyle \tan{U_q}=  
\frac{\sqrt{\frac{2}{3}}+ \frac{\lambda_q \tlambda_0}{2}}
{\sqrt{\frac{1}{3}}+ \frac{\lambda_q \tlambda_8}{2}}~~
\simeq\sqrt{2} + 
\frac{3}{2z_A} \lambda_q\lambda_s
 \,, \\[0.9cm] 
\displaystyle \tan{\phi_s}=   \tan{\left [ \theta_P+U_s\right ]},
& \displaystyle \tan{U_s}=  
\frac{z_A \sqrt{\frac{2}{3}}- \frac{\lambda_s \tlambda_8}{2}}
{z_A \sqrt{\frac{1}{3}}+ \frac{\lambda_s \tlambda_0}{2}}\simeq \sqrt{2} -
\frac{3}{2 z_A} \lambda_q\lambda_s  \,,
\end{array}
\label{current25}
\right .
\ee
up to terms of order ${\cal{O}}(\delta^2)$. Using the definition 
of the FKS ideal mixing angle\footnote{ In our previous papers as well as below --
see Equations (\ref{AA25}) -- we prefered defining the ideal
mixing angle by $\theta_I=\arctan{1/\sqrt{2}}\simeq 35^\circ$ in more natural correspondence
with the $\rho^0-\omg $ ideal mixing angle.} 
\cite{feldmann_3}, $\theta_{FKS}= -\arctan{\sqrt{2}}= -54.7^\circ$ . Equations
(\ref{current25}) imply 
the following relationships~:
\be
\begin{array}{ll} 
\displaystyle [\phi_q -\phi_s]=\frac{ \lambda_q \lambda_s}{z_A}+{\cal{O}}(\delta^2)~~,&
\displaystyle [\phi_q +\phi_s]= 2[\theta_P-\theta_{FKS}]  \,,
+{\cal{O}}(\delta^2)
\end{array}
\label{current26}
\ee
which emphasizes the numerical closeness of the $\phi_q$ and $\phi_s$ FKS mixing angles.
It is worthwhile to go on by deriving additional expressions which can be compared to 
their partners in \cite{feldmann_1,feldmann_2,feldmann_3}. We have~:
\be
\left \{
\begin{array}{ll} 
\displaystyle [F_q]^2 = [F^q_\eta]^2 +[F^q_\etp]^2 = f_\pi^2 [ 1+\lambda_q^2] + {\cal{O}}(\delta^2)  \,,
\\[0.5cm] 
\displaystyle [F_s]^2 = [F^s_\eta]^2 +[F^s_\etp]^2 = f_\pi^2 [  z_A^2+\lambda_s^2] + {\cal{O}}(\delta^2)  \,,
\\[0.5cm] 
F^q_\eta F^s_\eta + F^q_\etp F^s_\etp=F_q F_s \sin{[\phi_q-\phi_s]} = 
f_\pi^2 \lambda_q \lambda_s + {\cal{O}}(\delta^2)  \,,
\end{array}
\right .
\label{current27}
\ee
where $F_s$ and $F_q$ and the FKS mixing angles are given by Equations (\ref{current23}) above.
It is worthwhile  remarking that the second Equation (\ref{current27}) can be rewritten~:
$$
\displaystyle [F_s]^2 =f_\pi^2[z_A^2+\lambda_s^2] =f_\pi^2[ 2 z_A -1 +\lambda_s^2 + (z_A-1)^2]~~,
$$ 
where the last term is second order in SU(3) breaking but not numerically
small compared to the $\lambda_0$ or $\lambda_s$ parameter squared values.

Then, using the definitions for our parameters (and cancelling out $\lambda_8$), 
it is obvious that 
the quantities given by Equations (\ref{current27}) coincide up to the 
${\cal{O}}(\delta^2)$ and  $(z_A-1)^2$ terms expected with the corresponding
 FKS expressions\footnote{See Equations (28-30) in \cite{feldmann_3}.}.
Moreover, it should be remarked that Equations (\ref{current25}) and 
(\ref{current26})  exhibit   the properties of the FKS 
mixing angles $\phi_q$ and $\phi_s$ emphasized in
\cite{feldmann_3} for instance. In particular, the single angle
$\phi$ occuring in the the FKS parametrization 
is $\phi=[\phi_q +\phi_s]/2$ which only depends on $\theta_P$, whereas
the difference $\phi_q - \phi_s$ is a pure
effect of the kinetic breaking mechanism defined in Subsection \ref{XA-BRK}.  

 It should thus be remarked that the non-vanishing character of 
 $[\phi_q -\phi_s]$ is {\it not} an Isospin breaking effect and  that
 $\phi_q=\phi_s +{\cal O}(\delta^2)$ rather implies either of 
 $\lambda_8=-\sqrt{2}\lambda_0$ or $\lambda_0=\sqrt{2}\lambda_8$.
 
\section{Further Constraining EBHLS$_2$}
\label{kroll_brk1}
\indentB
It can be of interest to identify additional constraints which could apply to
EBHLS$_2$ and highlight symmetry breaking effects not explicitly emphasized.
In the FKS approach, an important ingredient are some properties of axial
currents still not imposed to  EBHLS$_2$, namely the diagonal character (at
leading order) of the following matrix elements \cite{Kroll:2005}~:
\be
\begin{array}{llll}
\displaystyle  <0|J^a_\mu| \eta_a(p)> =ip_\mu f_a \delta_{ab}~~, 
&\displaystyle  |\eta_a(p)> =| a \overline{a}(p)>~~,
&\displaystyle J^a_\mu=\overline{a} \gamma_\mu \gamma_5 a~~, 
&\displaystyle \{a=u,d,s\}~~.
 \end{array} 
 \label{kroll-1}
\ee
which may look natural constraints to be plugged into our model where
one also works at order ${\cal O}(\delta)$.
The axial currents relevant for this purpose can readily be derived from those 
displayed in Equations (\ref{current6}) and (\ref{current19})~:
\be
\left \{
\begin{array}{lll} 
\displaystyle J_\mu^u= \frac{1}{\sqrt{2}} \left [J_\mu^q +  J_\mu^3 \right]~~,&
\displaystyle J_\mu^d= \frac{1}{\sqrt{2}} \left [J_\mu^q -  J_\mu^3 \right]~~,&
\displaystyle J_\mu^s= \sqrt{\frac{1}{3}} J_\mu^0 -\sqrt{\frac{2}{3}} J_\mu^8 ~~,
\end{array}
\right \}  ~~.
\label{kroll-2}
\ee
As one can identify the leading order term in the Fock expansion of the various  $ |\eta_a>$ 
states with  the following {\it bare} PS field combinations~:
\be
\left \{
\begin{array}{lll} 
\displaystyle |\eta_u> =|u \overline{u}> = ~~\frac{1}{\sqrt{2}} |\pi^0_{bare}> +\frac{1}{\sqrt{3}} 
			|\eta^0_{bare}> +\frac{1}{\sqrt{6}} |\eta^8_{bare}> ~~, \\[0.5cm]
\displaystyle  |\eta_d> =|d \overline{d}> = - \frac{1}{\sqrt{2}} |\pi^0_{bare}> +\frac{1}{\sqrt{3}} 
			|\eta^0_{bare}> +\frac{1}{\sqrt{6}} |\eta^8_{bare}> ~~, \\[0.5cm]
\displaystyle  |\eta_s> =|s \overline{s}> = ~~\frac{1}{\sqrt{3}} |\eta^0_{bare}> -\sqrt{\frac{2}{3}} |\eta^8_{bare}> ~~, 
\end{array}
\right .
\label{kroll-3}
\ee
the conditions imposed by Equation (\ref{kroll-1}) can be accessed within EBHLS$_2$. 
One thus gets\footnote{It happens that this $f_s$ differs from the $F_s$ defined in the
preceding Section; they are related by $f_s f_\pi =F_s^2$.}~:
\be
\left \{
\begin{array}{lll} 
\displaystyle \frac{1}{2} \left [ f_u+f_d \right]=\left [1 +\frac{\lambda_3^2+\lambda_q^2}{2} \right]f_\pi~~,&
\displaystyle \frac{1}{2} \left [ f_u-f_d \right]= \left [\lambda_3 \lambda_q +\Delta_A\right]f_\pi~~,&
\displaystyle f_s= \left [z_A^2 + \lambda_s^2\right]f_\pi
\end{array}
\right \}.
\label{kroll-4}
\ee
One should note that $(f_u+f_d)/2 \ne f_{\pi^0}$ if $\lambda_q$ does not identically 
vanish\footnote{The condition $\lambda_q=0$ implies that $\lambda_0$ and $\lambda_8$
are either simultanously non-vanishing  or simultaneously vanishing; we show just below 
that $\lambda_q=0 \Longrightarrow \lambda_3=0$.}, whatever is $\lambda_3$;
it should be emphasized that $\lambda_0$ is related with 
the so-called $\Lambda_1(\equiv \lambda_0^2)$ of 
 EChPT (see \cite{Kaiser_2000,leutw,leutwb,feldmann_3}). Moreover, the $z$ parameter
defined by Kroll  \cite{Kroll:2005} is~: 
\be
\displaystyle  z_{Kroll}= \frac{f_u-f_d}{f_u+f_d}=
\Delta_A+\lambda_3 \lambda_q  +{\cal O}(\delta^2)~~,
\label{kroll-5}
\ee
which exhibits the expected dependence upon 
 the Isospin breaking parameters of EBHLS$_2$ coming via the $X_A$ and $X_H$ matrices.

On the other hand, the $a\ne b$ matrix elements are (a factor $i  f_\pi p_\mu$ being understood)~:
\be
\left \{
\begin{array}{lll} 
\displaystyle <0|J^u_\mu | \eta_d (p)> =  \frac{1}{2} \left [ \lambda_q^2-\lambda_3^2 \right ]~~,&
\displaystyle <0|J^d_\mu | \eta_u (p)> =  \frac{1}{2} \left [ \lambda_q^2-\lambda_3^2 \right ] ~~,\\[0.5cm]
\displaystyle <0|J^u_\mu | \eta_s (p)> =  \frac{\lambda_s}{\sqrt{2}} \left [ \lambda_q+\lambda_3 \right ]~~,&
\displaystyle <0|J^d_\mu | \eta_s (p)> = \frac{\lambda_s}{\sqrt{2}} \left [ \lambda_q -\lambda_3  \right ] ~~,\\[0.5cm]
\displaystyle <0|J^s_\mu | \eta_u (p)> =  \frac{\lambda_s}{\sqrt{2}} \left [ \lambda_q+\lambda_3 \right ]~~,&
\displaystyle <0|J^s_\mu | \eta_d (p)> = \frac{\lambda_s}{\sqrt{2}} \left [ \lambda_q -\lambda_3  \right ] ~~,\\[0.5cm]
\end{array}
\right .
\label{kroll-6}
\ee
\noindent Three solutions\footnote{Actually, each of the solutions below is 2-fold degenerated;
indeed, as each physical quantity exhibits  a dependence only upon squares of the
$\lambda_i$, any solution   $\{\lambda_i=\lambda_i^0, i=0,3,8 \}$ carries the same physics
than its twin  $\{\lambda_i=-\lambda_i^0, i=0,3,8 \}$.} 
allow to exhaust the simultaneous vanishing of 
Expressions (\ref{kroll-6}); they are~:
\be
\left \{
\begin{array}{lll} 
 \displaystyle { \rm Solutions} ~~A_\pm~:  \lambda_s =0~~, \lambda_q= \pm \lambda_3 &\displaystyle 
 \Longrightarrow  \lambda_0= \pm  \sqrt{\frac{3}{2}} \lambda_3 = \sqrt{2} \lambda_8 ~~, \\[0.5cm]
{ \rm Solution~} ~~B~~~:  \displaystyle \lambda_s \ne 0~~, \lambda_q=  \lambda_3 =0 &\displaystyle 
 \Longrightarrow  \lambda_8= -\sqrt{2} \lambda_0~~,  \lambda_3 =0  ~~,  \\[0.5cm]
 \end{array}
\right .
\label{kroll-7}
\ee
not counting the trivial solution $T \equiv \{ [\lambda_0= \lambda_8= \lambda_3 =0]\}$, 
already known to be unable to accomodate satisfactorily our set of Reference data -- 
this statement is also valid for Solution B as shown in the first data column of
Table \ref{Table:T0} in connection with the account of the Belle dipion spectrum
\cite{Belle}.
In contrast, both Solutions $A_\pm$ are found to work well within our minimization
procedure. For these solutions, the 3 parameters of the kinetic 
breaking mechanism are no longer free -- as assumed in the preceding Sections in
line with the common belief -- 
but become algebraically related with each other.

So, it follows from the developments just stated that
imposing the Kroll conditions  (\ref{kroll-1}) is far from 
anecdotal; indeed,  any of the solutions (\ref{kroll-7}) which cancel out the matrix 
elements in Equations (\ref{kroll-6}), shows that  a non-vanishing 
$\lambda_0$ (the usual kinetic 't~Hooft determinant term) is possible {\it if and only if}
  $\lambda_8$ is  non-zero.
This statement -- valid if defining $X_H$ by Equation (\ref{kin10}) -- also applies
 if one prefers\footnote{This comes down  in dropping out 
the  products $\lambda_i \lambda_j$  for $i\ne j \in (0,3,8)$
in all the expressions given in the Sections above and in the Appendices.} 
defining  $X_H$  by Equation (\ref{kin11b}). It should also be stressed that
only one of the previously defined solutions can be valid; it could, hopefully,
identified by confronting  each  solution with the data.
 
\section{The $\pi^0-\eta-\etp$ Mixing~: Breaking of Isospin Symmetry }
\label{IB_brk}
\indentB
Another topic  relevant for the $\pi^0-\eta-\etp$ mixing
is the content of isospin zero mesons inside the physically 
observed $\pi^0$ wave-function; accounts of this can be found in \cite{feldmann_3,Kroll:2005} 
for instance. In standard ChPT approaches, the {\it physical} $\pi^0$ is expressed
in terms of the {\it bare} $\pi^0_{bare}$ field with admixtures of the {\it physical}
$\eta$ and $\etp$ mesons~:
\be
\displaystyle |\pi^0>=|\pi^0_{bare}> +\kappa |\eta>  +\kappa^\prime |\etp> + 
{\cal O}(\delta^2)~~,
\label{kroll-8}
\ee
the ${\cal O}(\delta)$ parameters $\kappa$ and $\kappa^\prime$  depending on
the light quark mass difference take respectively the
values $\kappa_0$ and $\kappa_0^\prime$ defind by  \cite{feldmann_3}~:
\be
\begin{array}{lll}
\displaystyle 
\kappa_0 =\frac{1}{2} \cos{\phi} \frac{m^2_{dd} -m^2_{uu} }{M_\eta^2 -M_\pi^2}~~,&
\displaystyle 
\kappa_0^\prime =\frac{1}{2} \sin{\phi} \frac{m^2_{dd} -m^2_{uu} }{M_\etp^2 -M_\pi^2}
~~,
\end{array}
\label{kroll-9}
\ee
up to higher order contributions\footnote{Another formulation \cite{leutw96} 
 in terms of the quark mass difference and of the mixing
 angle named here $\theta_P$ is reminded below; it
has been used in our previous studies \cite{ExtMod3,ExtMod4,ExtMod5,ExtMod7}
and its fit properties will be commented below. }. 
The quark mass term can be estimated
by\footnote{ In this expression, the subtracted electromagnetic contribution to 
kaon mass difference is estimated to 
$\Delta M_K = \Delta_\pi = M_{\pi^0}^2-M_{\pi^\pm}^2 =-1.24\times 10^{-3}$ GeV$^2$.
However, as discussed in \cite{Kroll:2005}, its exact magnitude is  rather
controversial; for instance, Moussallam \cite{Moussallam:1997}
rather yields $ \Delta M_K=k \Delta_\pi $ with $k=2 \div 3)$.} 
$m^2_{dd} -m^2_{uu}= 2 [M_{K^0}^2- M_{K^\pm}^2-M_{\pi^0}^2+M_{\pi^\pm}^2]
\simeq 1.03 \times 10^{-2}$ GeV$^2$ and $\phi$ is some approximate value derived
from the  $\phi_s$ and $\phi_q$ angles defined in Section \ref{quark-flavor}.
However, because $\phi_q -\phi_s \simeq\lambda_q \lambda_s +{\cal O}(\delta^2)$,
any  solution providing the vanishing of Equations (\ref{kroll-6}) automatically provides 
$\phi_s=\phi_q +{\cal O}(\delta^2)$.

On the other hand, Kroll has extended this formulation \cite{Kroll:2005} in order
to account for Isospin breaking effects not generated by the light quark mass 
difference~:
\be
\begin{array}{lll}
\displaystyle 
\kappa = \cos{\phi} \left [ \frac{m^2_{dd} -m^2_{uu} }{2(M_\eta^2 -M_\pi^2)} +z_{Kroll} \right ]~~,&
\displaystyle 
\kappa^\prime = \sin{\phi} \left [\frac{m^2_{dd} -m^2_{uu} }{2(M_\etp^2 -M_\pi^2)}+z_{Kroll} \right ]
~~,
\end{array}
\label{kroll-10}
\ee
where $z_{Kroll}$ is expressed in terms of the $f_u$ and $f_d$ decay constants
defined by Equation (\ref{kroll-1}) and expressed in the EBHLS$_2$ framework
 by Equation (\ref{kroll-5}).

\vspace{0.5cm}
In order to connect EBHLS$_2$ with the $\eta/\etp$ fractions inside the physically observed 
$\pi^0$, \cite{leutw96,Kroll:2005}, one needs the relation involving these and
$\pi^0_{bare}$. After some algebra, Equations (\ref{kin20}) and  (\ref{current12-1}) 
allow to derive an expression similar to Equation (\ref{kroll-8})~:
\be
\displaystyle
| \pi^0> =\left [ 1+\frac{\lambda_3^2}{2} \right ] |\pi^0_{bare}> +\varepsilon |\eta>
+\varepsilon^\prime |\etp > ~~,
 \label{kroll-11}
\ee
where the rescaling of the $\pi^0_{bare}$ term is specific of the kinetic breaking $X_H$
introduced in the EBHLS$_2$ Lagrangian and  $\varepsilon$ and $\varepsilon^\prime$ are
given by~:
\be
\left \{
\begin{array}{lll} 
\displaystyle
\varepsilon=~\hvar +\frac{\lambda_3 \lambda_q +\Delta_A}{2} \cos{(\frac{\phi_q+\phi_s}{2})}
-\frac{\lambda_3 \lambda_s}{2 z_A} \sin{(\frac{\phi_q+\phi_s}{2})}\,, \\[0.5cm]
\displaystyle
\varepsilon^\prime=
\hvar^\prime + \frac{\lambda_3 \lambda_q + \Delta_A}{2} \sin{(\frac{\phi_q+\phi_s}{2})}
+\frac{\lambda_3 \lambda_s}{2 z_A} \cos{(\frac{\phi_q+\phi_s}{2})}  \,,
\end{array}
\right .
 \label{kroll-18}
\ee
up to terms of order ${\cal O}(\delta^2)$, having used Equation (\ref{current26}) 
and defined $\theta_{FKS}=-\arctan{\sqrt{2}}$. 

As  $(\phi_q+\phi_s)/2$ is certainly a quite motivated expression for the FKS
parameter  $\phi$ \cite{feldmann_3,Kroll:2005}, the similarity of Equations (\ref{kroll-10}) 
and (\ref{kroll-18})  is striking, and even more if imposing $\lambda_s=0$ -- as requested
by any of the $A_\pm$ solutions (see  Equations (\ref{kroll-7})) --  which drops out
the last term in each of Equations (\ref{kroll-18}). The condition $\lambda_s=0$, indeed,
implies~:
\be
\left \{
\begin{array}{lll} 
\displaystyle
\varepsilon=~\hvar +\frac{\lambda_3 \lambda_q + \Delta_A}{2} \cos{\phi}
~~,&
\displaystyle
\varepsilon^\prime=
\hvar^\prime + \frac{\lambda_3 \lambda_q + \Delta_A}{2} \sin{\phi}
\end{array}
\right \}~~.
 \label{kroll-19}
\ee
with  $\phi=\phi_q=\phi_s$ up  ${\cal O}(\delta^2)$ terms.
Switching off the BKY ($\Delta_A$) and  kinetic breaking mechanisms
turns out to set $f_u=f_d$ and, then,
one expects recovering the  results usual in this limit \cite{feldmann_3,Kroll:2005}. Thus,
the following identifications~:
\be
\left \{
\begin{array}{lll} 
\displaystyle \hvar=
\frac{1}{2} \cos{\phi} \frac{m^2_{dd} -m^2_{uu} }{M_\eta^2 -M_\pi^2}~~~ (=\kappa_0) ~~~, %\\[0.5cm]
\displaystyle ~~\hvar^\prime=
\frac{1}{2} \sin{\phi} \frac{m^2_{dd} -m^2_{uu} }{M_\etp^2 -M_\pi^2}
~~~(=\kappa_0^\prime)
\end{array}
\right \}
 \label{kroll-20}
\ee
look motivated. However, because  additional singlets -- like a gluonium --
may contribute to the $\eta-\etp$ mixing, likely more inside the $\etp$ meson than 
inside  $\eta$, it is of concern to allow  for a departure from
the mere identification (\ref{kroll-20}), especially for the $\etp$ amplitude 
 term $\hvar^\prime$.
So, letting $\hvar$ and $\hvar^\prime$ floating independently provides a relevant
piece of information.

$\kappa_0$ and $\kappa_0^\prime$ are a common way to express Isospin breaking 
effects due to quark masses in the FKS picture; another way to proceed 
is proposed in \cite{leutw96} which has been used in our previous works.
This turns out to rely on quark masses, define~:
 \be
\left \{
\begin{array}{lll} 
\displaystyle
\varpi_0 = -\sqrt{2} \varepsilon_0 \cos{\theta_P} \tan{\delta_P}~~,&
\displaystyle
\varpi_0^\prime = \sqrt{2} \varepsilon_0 \sin{\theta_P} \cot{\delta_P}~~,&
\displaystyle
\varepsilon_0 =\frac{\sqrt{3}}{4} \frac{m_d -m_u}{m_s-\hat{m}}
\end{array}
\right \}
 \label{kroll-21}
\ee
and replace Equation (\ref{kroll-20}) by~: 
 \be
\begin{array}{lll} 
\displaystyle
\{ \hvar=\varpi_0~~~,&~~\hvar^\prime=\varpi_0^\prime\}~~.
\end{array}
 \label{kroll-22}
\ee
 The angle $\delta_P$ occuring in these
expressions,  defined in Equations (\ref{AA25}), is given by
$\delta_P=\theta_P -\theta_I$ where $\theta_I=\pi/2 -\theta_{FKS}$;
in this approach, the floating parameter is no longer
$m^2_{dd} -m^2_{uu}$, but $\varepsilon_0$.
Using the light quark masses masses from FLAG 2016 \cite{FLAG_2016}, 
$\varepsilon_0$ is expected around $\simeq 1.22\times 10^{-2}$. One can
anticipate on fit results and state that fitting with Equations (\ref{kroll-20})
or (\ref{kroll-21}) yields similar fit properties.

Finally, one should mention that the $z_{Kroll}$ dependence
in Equations (\ref{kroll-18}) and (\ref{kroll-19}) exhibits an unexpected 
difference compared to Equations (\ref{kroll-10})~: EBHLS$_2$ finds
a weight for $z_{Kroll}$ smaller by a factor of 2. Whether it is a
specific feature of EBHLS$_2$ is an open question. 

\section{The $\pi^0-\eta-\etp$ Mixing~: The EBHLS$_2$ Analysis}
\label{BHLS-mixing}
\indentB
In order to deal with the $\tau$ dipion spectra and the update of the muon HVP,
it was found appropriate to release at most the constraints on the model parameters
within the fit procedure; this also applies to the model parameters named
$\epsilon$ and $\epsilon^\prime$ which were left free to vary independently.

In order to compare with expectations, it is also 
 worthwhile to consider the case when $\epsilon=\kappa_0$ and 
$\epsilon^\prime=\kappa_0^\prime$ are imposed; 
this turns out to let the  parameter  $m^2_{dd} -m^2_{uu}$ floating  and
derive $\epsilon$ and $\epsilon^\prime$ by means of Equations (\ref{kroll-20})  
 which constrains  $\epsilon$ and $\epsilon^\prime$ to be likesign.
Instead, if $\epsilon$ and $\epsilon^\prime$
are floating independently, Equations (\ref{kroll-20}) allow for 
separate determinations of $m^2_{dd} -m^2_{uu}$  from the fitted $\eta$ and $\etp$ 
admixtures.
\begin{table}[!phtb!]
\begin{center}
\begin{minipage}{1.0\textwidth}
\hspace{1.cm}
{\scriptsize
\begin{tabular}{|| c  || c  | c | c | c || c | c ||}
\hline
\hline
\hhhv  \hhhw \hhhv  EBHLS$_2$ (BS)	& \hhhv Sol. F    &  \hhhv  Sol. $T/B$  & \hhhv Sol. $A_+$  & \hhhv Sol. $A_-$  & \hhhv Sol. $A_+$  & \hhhv Sol. $A_-$  \\
\hline
\hhhv  \hhhw 	&\multicolumn{4}{|c||}{ \hhhv $\epsilon=\kappa_0$ ~~~~ \& ~~~~ $\epsilon^\prime=\kappa_0^\prime$} 
& \multicolumn{2}{|c||}{ \hhhv  $\epsilon ~~\& ~~\epsilon^\prime $ free}  \\
\hline
 \hhhv NSK $\pi^+ \pi^-$  (127)         &   139   	&  134     &   132  	&  136    & 136	& 138 \\
\hline
 \hhhv KLOE $\pi^+ \pi^-$ (135)         &   137 	&  153 	     &    143  	&   139   & 141 	& 138 \\
\hline
 \hhhv  BESIII $\pi^+ \pi^-$ (60)       &  49   	& 48  	 &   47 	&   49	  &  48 	& 49	\\
\hline
 \hhhv	Spacelike $\pi^+ \pi^-$ (59)    &  61  		& 64	 &  61 	       &   59	  &  60	&  59\\
\hline
\hline
 \hhhv	$\tau$ (A+C) (66)               &   61   	& 75	 &  60	 	&    59	   & 65	&  61\\
\hline
 \hhhv	$\tau$ (B) (19)                 &   28   	& 53 	 &  32  	&  27	 &  35	&  28\\
\hline
\hline
 \hhhv	$\pi^0 \gam$ (112)              &   86  	&  98 	  &  86    	 &   94	  & 86	&  92\\
\hline
 \hhhv	$\eta \gam$  (182)              &  123  	& 132 	  &    131     	 &  120		& 125	& 120\\
\hline
 \hhhv	NSK $\pi^+ \pi^-\pi^0$ (158)      &  149   	& 158      & 154      & 150  & 	 150       &  149 \\
\hline
 \hhhv	BESIII $\pi^+ \pi^-\pi^0$ (128)    & 138 	&  138 &  138    & 	138	&  138	& 138\\
\hline
 \hhhv	$P\gam \gam$ \& $\etp V \gam$  ~~(5)    &  5  	   & 8 	&   8   &  9	&  4 & 7 \\
\hline
\hline
 \hhhv $\chi^2/N_{\rm pts}$ &    $1280/1366$  &  $1375/1366$      & 1309/1366 & 1289/1366  & 1292/1366 & 1286/1366\\
 \hhhw Probability      &         85.9\%      &  23.3  \%        & 70.6 \%    &82.5\% &  80.1  \% & 83.5\% \\
\hline
\hline
\end{tabular}
}  % \scriptsize
\end{minipage}
\end{center}
\caption {
\label{Table:T10} 
Selected individual $\chi^2/N_{\rm pts}$ values  in  EBHLS$_2$ fits versus the Kroll conditions 
(cf Equations (\ref{kroll-1})).   The first data column reports on the fit where the three $\lambda$'s vary  
independently, the others refer to the solutions defined in Equations (\ref{kroll-7}).
The leftmost 4 data columns assume Condition $C$ (see text),  whereas   Condition $C$ 
has been relaxed in the last two column fits.
The last lines display  the  global $\chi^2/N_{\rm pts}$ and probability for each fit.}
\end{table}

Furthermore, in the fits reported from now on, the polynomial 
$\delta P^\tau(s)$ is always second degree and, for completeness, one lets 
 the $\Delta_A$ Isospin breaking 
parameter floating, even if it  is not really significant -- never more than $2\sigma$.

The PDG value for the
ratio $f_K/f_\pi$ is included in the set of experimental 
data submitted to fit.  The Belle dipion spectrum is  included and 
one refers the reader to Section \ref{tau_spectra} for the specific consequences
this implies for the dipion spectra collected in the $\tau$ lepton decay.

To be the most comprehensive possible, several cases for the 
$[\lambda_0, \lambda_8,\lambda_3]$ triplet have been considered, 
namely, the $A_\pm$ and $B$ solutions defined in Equations (\ref{kroll-7}), as well
as the so-called trivial solution $\{T \equiv [\lambda_0= \lambda_8= \lambda_3 =0]\}$;
it has been found worthwhile to also consider the
 case when the three $\lambda$ parameters are left floating independently --
referred to hereafter as Solution F. Solution F is, actually, very similar to the fit
conditions of the previous Sections.
\vspace{0.5cm}

The main fits properties are gathered in Table \ref{Table:T10} and lead to the following
remarks~:
\begin{itemize}
\item  Regarding Solution $B$, the best fit returns 
$\lambda_0= (-0.01 \pm 36.26) \times 10^{-2}$,
which  clearly exhibits  convergence towards the trivial solution  
$\{T \equiv [\lambda_0= \lambda_8= \lambda_3 =0]\}$; therefore, there is no point
in distinguishing Solution $B$ from the trivial solution $T$ which 
is the one actually  reported.
\item With a minimum total $\chi^2$  larger by 60 to 95 units than the other
Solutions, Solution $T/B$ can be safely  discarded. 
\item When assuming condition 
$$C \equiv\{\hvar= \kappa_0, \hvar^\prime= \kappa_0^\prime\}$$
both Solutions $A_\pm$ return good probabilities. Solution $A_-$
is, however, clearly favored even if   $A_+$ exhibits a  reasonable goodness of fit.  Nevertheless, relaxing
Condition $C$, solutions $A_\pm$ exhibit practically the same fit probability.
This indicates that Condition $C$ is not a real constraint for Solution $A_-$
the total $\chi^2$ of which is almost unchanched ($\Delta \chi^2=3$). In contrast,
Condition $C$ exhibits a strong tension with   Solution $A_+$
and provides a strongly degraded fit as $\Delta \chi^2=17$ for only one parameter
less; however,  $\hvar$ and $\epspr$ become unlike signs when
relaxing Condition $C$ which is certainly inconsistent with Equations (\ref{kroll-20})
-- or Equations  (\ref{kroll-21}) -- and   with common expectations.

One observes, nevertheless,  that the decay information (at bottom in Table
\ref{Table:T10}) are better described\footnote{The $\pi^0 VP$ and $ \eta VP$
are hidden inside the $\pi^0 \gam$ and $\eta \gam$ annihilation cross sections
and are of comparable quality in both solution $A_\pm$ fits.}  
by solution $A_+$ than $A_-$. It should be noted that
most studies of the $[\pi^0,\eta,\etp]$ mixing properties just rely on the two-body 
decays with $P\gam\gam$ and $PV\gam$ couplings.
\end{itemize}
\begin{itemize}
\item Replacing Condition $C$ by 
$$C^\prime \equiv\{\hvar= \varpi_0, \hvar^\prime= \varpi_0^\prime\}$$
does not lead to substantial differences. Indeed, one gets 
$\chi^2/N_{\rm pts}=1303/1366$ and 74.0\% probability ($A_+$) or
$\chi^2/N_{\rm pts}=1286/1366$ and 83.8\% probability ($A_-$), e.g. Solution
$A_-$  remains prefered to $A_+$ by the data.

It should be noted that relaxing Condition $C$ (or $C^\prime$) leads to
likesign $\hvar$ and  $\hvar^\prime$ for Solution $A_-$, but to unlike
signs  for Solution $A_+$. As just noted, this could motivate discarding Solution 
$A_+$.

\item As could be expected, Solution $F$ is also good, benefiting from a larger
parameter freedom than  $A_\pm$ submitted to condition $C$ or not.

\end{itemize}
\begin{table}[!ptbh!]
\hspace{-0.9cm}
\begin{minipage}{0.8\textwidth}
\begin{center}
{\scriptsize
\begin{tabular}{|| c  || c  | c | c | c || c | c ||}
\hline
\hline
\hhhv  \hhhw \hhhv  EBHLS$_2$ (BS)	& \hhhv Sol. F    &  \hhhv  Sol. $T/B$  & \hhhv Sol. $A_+$  & \hhhv Sol. $A_-$  & \hhhv Sol. $A_+$  & \hhhv Sol. $A_-$  \\
\hline
\hhhv  \hhhw 	&\multicolumn{4}{|c||}{ \hhhv $\epsilon=\kappa_0$ ~~~~ \& ~~~~ $\epsilon^\prime=\kappa_0^\prime$} 
& \multicolumn{2}{|c||}{ \hhhv  $\epsilon ~~\& ~~\epsilon^\prime $ free}  \\
\hline
\hline
 \hhhv $g$                 &   $6.996\pm 0.002$       &  $7.052\pm 0.002$     & $6.536 \pm 0.002$    & $6.671\pm 0.001$ &  $7.069\pm 0.002$ & $6.954\pm 0.002$ \\
\hline
 \hhhv $a_{\rm HLS}$       &   $1.764\pm 0.001$       & $1.646\pm 0.001$      & $1.728 \pm 0.001$    & $1.765\pm 0.001$  & $1.752\pm 0.001$ & $1.766\pm 0.001$\\
\hline
 \hhhv $(c_3+c_4)/2$       &   $0.756 \pm 0.004$      & $0.739 \pm 0.003$     & $0.769\pm  0.004$    &  $0.742 \pm 0.003$ &  $0.769\pm 0.004$ &  $0.742\pm 0.003$ \\
\hline
 \hhhv $c_1-c_2$           &   $0.762 \pm 0.014$       &  $0.775\pm 0.012$     & $0.766 \pm 0.012$   &  $0.823 \pm 0.013$ &  $0.676 \pm 0.013$ & $0.809 \pm 0.013$\\
\hline
 \hhhv $10^2\times  z_3    $  & $-0.332 \pm 0.004$     &  $-0.364 \pm 0.030$   & $-0.372 \pm 0.004$  & $-0.354 \pm 0.004$ & $-0.345 \pm 0.004$&  $-0.339 \pm 0.004$\\
\hline
 \hhhv $10^2\times[m_{dd}^2- m_{uu}^2]$ 
 			  & $2.65 \pm 0.25$  &     $3.78 \pm 0.13 $          & $2.49 \pm 0.15$      &  $3.01 \pm 0.14$  & $\times$  &  $\times$\\
\hline
 \hhhv $10^2\times\epsilon $  & ${\bf 3.67 \pm 0.32}$   &  ${\bf 5.19}$       &  ${\bf 3.48 \pm 0.20}$ &  ${\bf 4.16 \pm 0.20}$ &  $2.28 \pm 0.30$  & $3.62 \pm 0.30$\\
\hline
 \hhhv $10^2\times\epspr $   &  ${\bf 0.93 \pm 0.09}$   &   ${\bf 1.34}$      &  ${\bf 0.86 \pm 0.05}$  & ${\bf 1.05 \pm 0.05}$  & $-1.20 \pm 0.30$ & $0.17 \pm 0.27$\\
\hline
 \hhhv $\theta_P $      &  $-15.89\pm 0.34$   &  $-15.36 \pm 0.28$      & $-16.63 \pm 0.30$  & $-15.78 \pm 0.28$  & $-16.63 \pm 0.30$ &  $-15.59 \pm 0.28$ \\
\hline
 \hhhv $z_A     $              & $1.417\pm 0.004$   &  $1.411 \pm 0.004$      & $1.429 \pm .004$     & $1.405\pm 0.004$  & $1.423\pm 0.004$ &  $1.406\pm 0.004$\\
\hline
 \hhhv $z_V$                  & $1.433 \pm 0.001$   &  $1.507 \pm 0.001$      & $1.463 \pm 0.001$    & $1.419 \pm 0.001$ & $1.436 \pm 0.001$ & $1.420 \pm 0.001$\\
\hline
 \hhhv $10^2\times\Delta_A $  &  $0.12\pm 5.09$     &   $10.93\pm 5.05$       & $-8.71 \pm 5.16$     & $12.65\pm 5.14$  & $-6.82\pm 5.23$ & $12.94\pm 4.91$\\
\hline
    \hhhv $\lambda_3$            &  $0.236 \pm 0.007$  & ${\bf \equiv 0}$     & ${\bf 0.212 \pm 0.007}$ & ${\bf -0.242 \pm 0.007}$ & ${\bf 0.197 \pm 008}$ & ${\bf -0.233 \pm 0.007}$\\
 \hline
 \hhhv $\lambda_0$            &  $0.152\pm 0.042$   & ${\bf \equiv 0}$        & $0.259 \pm 0.009$      & $0.295\pm 0.009$  & $0.241\pm 0.009$ &   $0.285\pm 0.009$\\
\hline
 \hhhv $\lambda_8$            &  $0.022\pm 0.023$      & ${\bf  \equiv 0}$  & ${\bf 0.183 \pm 0.006}$ & ${\bf 0.209\pm 0.006 }$  & ${\bf 0.170\pm 0.007 }$   & ${\bf 0.202\pm 0.006}$\\
\hline
 \hhhv $10^2 \times  \xi_0 $  &  $-7.237\pm 0.019$     &  $-5.130 \pm 0.030$  & $-0.022 \pm 0.027$    & $-3.114\pm 0.020$  & $-7.809\pm 0.019$ &  $-6.838\pm 0.018$\\
\hline
 \hhhv $10^2 \times  \xi_3$ &   $2.231 \pm 0.155$     &  $-3.598 \pm 0.072$   & $3.155 \pm 0.129$    & $3.034 \pm 0.148$ & $0.599 \pm 0.136$ &  $1.496 \pm 0.150 $\\
\hline
\hline
 \hhhv $\chi^2/N_{\rm pts}$      &    $1280/1366$      &  $1375/1366$      & $1309/1366$   & $1289/1366$ & $1292/1366$ &  $1286/1366$\\
 \hhhw Probability               &        85.9\%       &  23.3 \%          & 70.6\%        & 82.5\%   & 80.1\% &   83.5\%\\
\hline
\hline
\end{tabular}
}	%{\scriptsize
\end{center}
\end{minipage}
\caption {
\label{Table:T11}
Model parameter values from the  fits performed within  EBHLS$_2$ under the various  configurations
defined in the text. Angles are in degrees, $[m_{dd}^2- m_{uu}^2]$ in GeV$^{2}$. 
Values written boldface are not floating but derived from the floating 
parameters  through Equations given in the main text; 
their uncertainties are derived likewise and take into account the parameter 
covariance matrix.}
\end{table}

The model parameter values returned by the various fits are displayed in 
Table \ref{Table:T11}. One can remark that the specific HLS model parameters 
do not much vary depending on the solution examined; this is indeed so 
for $g$, $a (\equiv a_{HLS})$, $z_3$, $c_1-c_2$  and for\footnote{In our fits, $c_3=c_4$
is imposed \cite{ExtMod3,ExtMod7}.} $(c_3+c_4)/2$. This is also observed for
the BKY breaking parameters $z_A$ and $z_V$ whereas $\Delta_A$ is clearly not
significant. In contrast $\xi_0$ undergoes surprisingly a large change
when going from $A_+$ to $A_-$.  The value for $\xi_3$ strongly depends
on requiring or not Condition $C$ but are alike for Solutions $A_+$ and $A_-$
in the former case.

The parameter equivalent to the so-called $\Lambda_1$ 
\cite{Kaiser_2000,leutw,feldmann_3} 
($\Lambda_1 =\lambda_0^2$) is found in the range from 6.5\% ($A_+$) to 8.5\% ($A_-$).
However, it should be stressed that, once assuming the Kroll conditions (\ref{kroll-1}), 
 it cannot come alone as reflected by Equations (\ref{kroll-7}) and determined by our fits.
 Their numerical values are marginally affected by Condition $C$ or by choosing $A_+$ or 
 $A_-$ -- up to the sign for $\lambda_3$.
 
 Therefore, an important piece of information should be stressed~: 
Because of the strict relation between $\lambda_3$ and $\lambda_0$
 -- and hence $\Lambda_1$ -- the Kroll conditions (\ref{kroll-1}) implies
 that the pion form factor in the $\tau$ decay fulfills $F^\tau_\pi(0)=
 1-\lambda_3^2/2$ and, then, is no longer unity as inferred at the beginning of the
 present study.
 
\section{ Side Results from Fits}
\label{side-results}                                      

%\begin{table}[!pt!]
\begin{table}[!ptbh!]
\hspace{+0.5cm}
\begin{minipage}{0.8\textwidth}
\begin{center}
{\scriptsize
\begin{tabular}{|| c  || c  | c  | c || c  | c ||}
\hline
\hline
\hhhv  \hhhw \hhhv  EBHLS$_2$ (BS)	& \hhhv Sol. F         & \hhhv Sol. $A_+$  & \hhhv Sol. $A_-$   & \hhhv Sol. $A_+$  & \hhhv Sol. $A_-$\\
\hline
\hhhv  \hhhw 	&\multicolumn{3}{|c||}{ \hhhv $\epsilon=\kappa_0$ ~~~~ \& ~~~~ $\epsilon^\prime=\kappa_0^\prime$} 
& \multicolumn{2}{|c||}{ \hhhv  $\epsilon ~~\& ~~\epsilon^\prime $ free}  \\
\hline
 \hhhv $\theta_P $ (deg.)               &  $-15.89\pm 0.34$       & $-16.63 \pm 0.30$  & $-15.78 \pm 0.28$    &  $-16.63 \pm 0.30$ & $-15.59 \pm 0.28$\\
\hline
\hline
 \hhhv $\theta_0$ (deg.)                &   $-6.35\pm 0.47$     & $-8.04\pm 0.39$     & $-8.05\pm 0.33$       & $-7.95 \pm 0.39$    & $-7.71 \pm 0.34$ \\
\hline
 \hhhv $\theta_8$ (deg.)                &   $-24.55\pm 0.30$    & $-24.44\pm 0.25$    & $-22.83\pm 0.27$ 	& $-24.50 \pm 0.25$ & $-22.77 \pm 0.28$\\
\hline
 \hhhv $\theta_0-\theta_8$ (deg.)       &   $18.21\pm 0.24$     & $16.45\pm 0.23$     & $14.85\pm 0.24$ 	&  $16.59\pm 0.23$  & $15.13\pm 0.24$ \\
\hline
 \hhhv $F_0/f_\pi$                      &  $1.166 \pm 0.006$    & $ 1.190 \pm 0.003$  & $1.190 \pm 0.003$	& $ 1.184 \pm 0.003$  & $ 1.187 \pm 0.003$\\
 \hline
 \hhhv $F_8/f_\pi$                      &  $1.293 \pm 0.003$    & $ 1.315 \pm 0.003$  & $1.302 \pm 0.003$	& $1.309 \pm 0.003$ & $1.302 \pm 0.003$\\
 \hline
 \hhhv FKS $\phi$  (deg.)               &  $38.85 \pm 0.35$     & $ 38.96 \pm 0.27$   & $38.08 \pm 0.29$	& $38.09 \pm 0.30$ & $39.15 \pm 0.27$\\
 \hline
 \hhhv  $\phi_q-\phi_s$  (deg.)      &  $0.39 \pm 0.18$      & $ \equiv 0$         & $\equiv 0$			&$ \equiv 0$         & $\equiv 0$	\\
\hline
 \hhhv $F_q/f_\pi$                      &  $1.008 \pm 0.007$    & $ 1.050 \pm 0.003$  & $1.066 \pm 0.004$	& $ 1.044 \pm 0.003$ & $ 1.061 \pm 0.004$\\
 \hline
 \hhhv $F_s/f_\pi$                      &  $1.418 \pm 0.005$    & $ 1.428 \pm 0.004$  & $1.405 \pm 0.004$	&  $1.423 \pm 0.004$ &  $1.406 \pm 0.004$\\
 \hline
 \hhhv $F_s/f_K$                        &  $1.192 \pm 0.003$    & $ 1.198 \pm 0.003$  & $1.181 \pm 0.002$	& $1.195 \pm 0.002$ & $1.182 \pm 0.002$\\
\hline
\hline
 \hhhv $f_K/f_\pi$                      &  $1.190 \pm 0.002$    & $ 1.193 \pm 0.002$  & $1.189 \pm 0.002$	&   $1.191 \pm 0.002$ &  $1.190 \pm 0.002$ \\
\hline
\hline
 \hhhv $10^{10}\times a_\mu(HLS)$ &   $572.52 \pm 1.02$      & $571.84 \pm 0.98$    & $572.44 \pm 0.98$		&  $575.00  \pm 0.95$ &   $572.59  \pm 0.99$\\
\hline
 \hhhv $\chi^2/N_{\rm pts}$      &    $1280/1366$         & $1309/1366$   & $1289/1366$ 	& $1292/1366$  &  $1286/1366$	\\
 \hhhw Probability               &        85.9\%          & 70.6\%        & 82.5\%       	& 80.1\% &    83.5\%\\
\hline
\hline
\end{tabular}
}		 %\scriptsize
\end{center}
\end{minipage}
\caption {
\label{Table:T12}
Singlet-Octet and Quark Flavor bases mixing parameter values derived from fits performed within  
EBHLS$_2$ under the $F$ and $A_\pm$  configurations defined in the text. For the configuration
$F$, $\phi =(\phi_q+\phi_s)/2$ whereas $\phi=\phi_q ~(=\phi_s)$ for the $A_\pm$ solutions.
Correspondingly, the contributions of the HLS channels for $\sqrt{s}\le 1.05$ GeV to the HVP are 
also given
in each case; they can be compared to Table \ref{Table:T7}. The main fit properties are reminded 
at the bottom end of the Table.
}
\end{table}
\indentB
Table \ref{Table:T12} collects our main results, mostly related to the 
$\pi^0-\eta-\etp$ mixing parameter evaluations. However, it is worthwhile
to include some topical pieces of information
which deserve special emphasis.
\begin{itemize}
\item The various estimates for $f_K/f_\pi$ displayed in Table \ref{Table:T12} 
 nicely compare to LQCD determinations, namely \cite{FLAG_2016}
$1.195 \pm 0.005$ and  \cite{ETMC_fK} $1.1995 \pm 0.0044$.
\end{itemize}

%\begin{table}[!pt!]
\begin{table}[!ptbh!]
\hspace{+0.5cm}
\begin{minipage}{0.8\textwidth}
\begin{center}
{\scriptsize
\begin{tabular}{|| c  || c  || c  | c | c  || c ||}
\hline
\hline
\hhhv  \hhhw \hhhv   & \hhhv EBHLS$_2$ avrg.	   & \hhhv FKS 98       & \hhhv EF 05       & \hhhv EGMS 15b       & \hhhv OU 17\\
\hline
\hline
 \hhhv $\theta_0$    &  $-8.04\pm 0.39 \pm 0.00 $   &  $-9.2 \pm 1.7$	 & $-2.4\pm 1.9$        & $-6.9 \pm 2.4$   & $\times$ \\
\hline
 \hhhv $\theta_8$    &   $-23.64\pm 0.30 \pm 0.27$  & $-21.2\pm 1.6 $    & $-23.8\pm 1.4$       & $-21.2 \pm 1.9$ & $\times$\\
\hline
\hhhv $F_0/f_\pi$    &  $ 1.190 \pm 0.003 \pm 0.000$   & $ 1.17 \pm 0.03$   & $1.29 \pm 0.04$   & $ 1.14 \pm 0.05$  & $\times$\\
 \hline
 \hhhv $F_8/f_\pi$   &  $1.309 \pm 0.003 \pm 0.007$ & $ 1.26 \pm 0.04$   & $1.51 \pm 0.05$      & $1.27 \pm 0.02$ & $\times$\\
 \hline
 \hhhv  $\phi$    &  $38.52 \pm 0.29 \pm 0.44$   & $ 39.3 \pm 1.0$    & $41.4 \pm 1.4$          & ($38.3 \pm 1.6$) & $39.8 \pm 2.2 \pm 2.4$\\
\hline
 \hhhv $F_q/f_\pi$   &  $1.058 \pm 0.004 \pm 0.008$ & $ 1.07 \pm 0.02$   & $1.09\pm 0.03$       & $ 1.03 \pm 0.04$ & $ 0.960 \pm 0.037 \pm 0.046$\\
 \hline
 \hhhv $F_s/f_\pi$   &  $1.417 \pm 0.004 \pm 0.012$  & $ 1.34 \pm 0.06$  & $1.66 \pm 0.06$      &  $1.36 \pm 0.04$ &  $1.363 \pm 0.27 \pm 0.006$\\
 \hline
 \hhhv $F_s/f_K$     &  $1.190 \pm 0.003 \pm 0.009$  & $\times$          & $\times$             & $\times$         & $1.143 \pm 0.023 \pm 0.005$\\
\hline
\hline
\end{tabular}
}		 %\scriptsize
\end{center}
\end{minipage}
\caption {
\label{Table:T13}
Mixing parameters in the Singlet-Octet and Quark Flavor bases from various sources. 
The EBHLS$_2$ evaluations displayed are the average values 
derived for Solutions $A_+$ and $A_-$ assuming Condition $C$, whereas the second uncertainty is half their difference; the 
original  $A_+$ and $A_-$ are given in Table \ref{Table:T12}. The data derived by other groups are 
 FKS~98 \cite{feldmann_2},  EF~05 \cite{Escribano:2005}, EGMS~15b \cite{Escribano:2015yup} and the LQCD 
 results OU~17 \cite{Ottnad:2017bjt}; the number within parentheses is from EMS~15 \cite{Escribano:2015}.
 Angles are expressed in degrees.
}
\end{table}

\begin{itemize}
\item The pion and kaon charge radii given in Table 5 of \cite{ExtMod7}
remain unchanged within the EBHLS$_2$ framework; they were observed in fair accord with expectations.  
\item  The values derived for the muon HVP contribution $a_\mu(HLS)$ of the (6) annihilation channels
embodied inside the EBHLS$_2$ framework and integrated up to $\sqrt{s}= 1.05$ GeV are also shown
and can be compared with the corresponding information in Table  \ref{Table:T7}. The reference
evaluation reported  there from  a fit using a least constrained EBHLS$_2$ variant was~:
$$  a_\mu(HLS, \sqrt{s}= 1.05~{\rm GeV}) = [571.97\pm 0.95] \times 10^{-10}~~,$$
which -- accidentally -- coincides with the average value derived using  $A_+$ and $A_-$
under Condition $C$. In this case,  the EBHLS$_2$ variants fulfilling the Kroll
conditions (\ref{kroll-1}) and  Condition $C$ do not depart from the average  estimate by more than 
$\simeq \pm 0.3 \times 10^{-10}$; this can be conservatively taken as the model uncertainty
affecting our evaluation of  $a_\mu(HLS)$ as, moreover, taking into account the mixing 
properties of the $\pi^0-\eta-\etp$ discussed
in Section \ref{IB_brk}, it looks natural to impose Condition C to EBHLS$_2$.

Finally, as noted in the preceeding Section \ref{BHLS-mixing}, the closeness observed between 
 Solutions F and $A_-$ has led  to conclude that Condition C is a intrinsic feature of 
 Solution $A_-$, a nice property not shared by $A_+$; this  leads to favor Solution $A_-$ over 
 Solution $A_+$.
\end{itemize}

Releasing, for completeness, Condition C exhibits  interesting results concerning  $\epsilon$
 and $\epsilon^\prime$. In this case, Solution $A_-$ returns likesign $\epsilon$
 and $\epsilon^\prime$ -- as expected from  Equations (\ref{kroll-20})-- whereas
 Solution $A_+$ returns unlike sign values and a significant shift\footnote{A closer look
 at the various channel contributions indicates that this excess entirely comes from the
 anomalous channels, in particular from the $3 \pi$  annihilation which then contributes
  $[46.30\pm 0.36]\times 10^{-10}$ whereas the solution reported in Table \ref{Table:T7}
 only yields $[44.22 \pm 0.32] \times 10^{-10}$.}  upward of 
 $\Delta a_\mu(HLS) =574.83-571.97=2.86$  in units of $10^{-10}$. The unlike
 sign character of  $\epsilon$ and $\epsilon^\prime$,  contradicting 
  the  expected properties of the  $\pi^0-\eta-\etp$ mixing, also disfavors Solution
  $A_+$ over Solution $A_-$.
   
\section{ Evaluations of the $\pi^0-\eta-\etp$ Mixing Parameters}
\label{Mixing-param}
\indentB
Table \ref{Table:T12} displays the parameter values derived by fitting our set
of data  within EBHLS$_2$ under the various solutions to Equations (\ref{kroll-1}).
We have found interesting to also produce the results derived assuming the $\lambda_i$
unconstrained (the so-called Solution F). One can remark a fair stability of the
usual mixing parameters as the spread of values is very limited  for each of them.

It is, of course, worth comparing our results with other determinations. For this purpose,
we have selected a limited  set of data and refer the reader to the corresponding papers to track
back to former references; the comparison can be easily performed by looking at
Table  \ref{Table:T13}.

In order to ease the comparisons, the first data column in Table \ref{Table:T13} 
displays the averages of the 
values derived using Solutions $A_+$ and $A_-$ under Condition $C$
which can be found in Table \ref{Table:T12}; half their difference 
is given as an estimate of
the systematic uncertainty and shown as the second error. 

The agreement is clearly satisfactory with  FKS~98 \cite{feldmann_2} -- based on meson decays 
involving $P\gam \gam$ and $J/\psi$ decays to $\eta$ and $\etp$. EF~05 \cite{Escribano:2005}
produces several parameter values depending on the information implemented. For instance,
using also  the $P \ra \gam \gam$ and  $J/\psi \ra (\eta/\etp) \gam$ decays only, together 
with the ChPT prediction $F_8=1.28 f_\pi$, Escribano and Fr\`ere derive~:
\be
\left \{
\begin{array}{llll}
\displaystyle \theta_8 =(-22.2 \pm 1.8)^\circ~~,&
\displaystyle \theta_0 =(-8.7 \pm 2.1)^\circ~~,&
\displaystyle F_0 /f_\pi = 1.18 \pm 0.04
\end{array}
\right \}~~,
\label{escribano-1}
\ee
in very good agreement with FKS~98 and EBHLS$_2$. Introducing, in addition, a 
parametrization\footnote{The $\etp V \gam$ couplings are explicitely involved
in the data of the EBHLS$_2$ bunch, the  $\eta V \gam$ couplings are treated
as part of the $e^+ e^- \ra \eta \gam$ annihilation cross-sections.}
of the coupling constants $(\eta/\etp)V\gam$, where $V=\rho^0,\omg,\phi$, they can use the 
corresponding tabulated decays widths to produce the numbers displayed in the third data column
 of Table \ref{Table:T13}. As for the Singlet-Octet parameters, the comparison with others
 is not as satisfactory, nevertheless, the Quark Flavor scheme parameters compare reasonably
 well.

 Analyzing the asymptotic behavior of the 
 $\eta/\etp$ meson transition form factors $F_{(\eta/\etp) \gam^* \gam} (Q^2)$ and using
 the Pad\'e  Approximant method, EMS~15 \cite{Escribano:2015}  derive 2 solutions;
 those based on the asymptotics of  $F_{\eta \gam^* \gam}(Q^2)$ is in good accord
 with our results and the value for the $\phi$ angle is displayed in Table \ref{Table:T13}.
 The solution based on the $F_{\etp \gam^* \gam}(Q^2)$ asymptotics, improved soon
 after,  is given in  Table \ref{Table:T13} under the tag  EGMS~15b \cite{Escribano:2015yup};
 their evaluations are in good accord with ours as well as with those in FKS ~98. On the other hand,
 they also obtain ~:
 $$ \phi_q = [39.6 \pm 2.3]^\circ~~~{\rm and}~~~\phi_s = [40.8 \pm 1.8]^\circ ~~,$$
 \noindent which are consistent with $ \phi_q = \phi_s$ at a $1 \sigma$ level.
Finally, the ETM Lattice QCD  Collaboration has derived the numbers given in the last
data column tagged OU~17 \cite{Ottnad:2017bjt}. Our results are consistent with these LQCD
evaluation at the $\simeq 1 \sigma$ level.

One more piece of information can be of interest  which could mimic higher order effects.
Using Solution $F$ which slightly
violates the Kroll Conditions, $\phi_q$ and $\phi_s$ become slightly different; they allow to 
derive~:
\be
\displaystyle \frac{\phi_q-\phi_s}{\phi_q + \phi_s} =[0.50 \pm 0.24] \times 10^{-2} ~~~<< 1
\label{sol_F}
\ee
as expected. 
%\begin{table}[!pt!]
\begin{table}[!ptbh!]
\hspace{+1.2cm}
\begin{minipage}{0.8\textwidth}
\begin{center}
{\scriptsize
\begin{tabular}{|| c  || c  | c  | c || c  | c ||}
\hline
\hline
\hhhv  \hhhw \hhhv  EBHLS$_2$ (BS)	& \hhhv Sol. F         & \hhhv Sol. $A_+$  & \hhhv Sol. $A_-$   & \hhhv Sol. $A_+$  & \hhhv Sol. $A_-$\\
\hline
\hhhv  \hhhw 	&\multicolumn{3}{|c||}{ \hhhv $\epsilon=\kappa_0$ ~~~~ \& ~~~~ $\epsilon^\prime=\kappa_0^\prime$} 
& \multicolumn{2}{|c||}{ \hhhv  $\epsilon ~~\& ~~\epsilon^\prime $ free}  \\
\hline
\hline
 \hhhv $10^2\times[m_{dd}^2- m_{uu}^2]$ 
 			           	& $2.65 \pm 0.25$      & $2.49 \pm 0.15$    &  $3.01 \pm 0.14$ 	   &  $\times$ &  $\times$ \\
\hline
 \hhhv $10^2\times\epsilon $       	& $ 3.67 \pm 0.32$     &  $ 3.48 \pm 0.20$  &  $ 4.16 \pm 0.20$     & $2.28 \pm 0.30$  & $3.62 \pm 0.30$ \\
\hline
 \hhhv $10^2\times\epspr   $       	& $0.93 \pm 0.09$      &  $0.86 \pm 0.05$   & $1.05 \pm 0.05$     & $-1.20 \pm 0.30$ & $0.17 \pm 0.27$ \\
\hline
 \hhhv $10^2\times\varepsilon $       	& $ 4.92 \pm 0.37$     &  $5.80 \pm 0.31$   &   $ 1.24 \pm 0.32$    &  $ 4.33 \pm 0.34$ &  $ 0.95 \pm 0.36$\\
\hline
 \hhhv $10^2\times\varepsilon^\prime$   & $1.94 \pm 0.26$      &  $2.66 \pm 0.18$   & $-1.30 \pm 0.23$    & $ 0.40 \pm 0.28$ &  $ -2.00 \pm 0.32$  \\
\hline
 \hhhv  $f_u/f_\pi$			& $1.070 \pm 0.015$     &$1.131 \pm 0.009$  &$1.020 \pm 0.005$     &  $1.114 \pm 0.009$ & $1.020 \pm 0.005$ \\
\hline
 \hhhv  $f_d/f_\pi$			& $1.006 \pm 0.006$     &$1.014 \pm 0.005$   &$1.170 \pm 0.012$    & $1.012 \pm 0.005$ & $1.157 \pm 0.012 $ \\
\hline
 \hhhv  $10^2\times z_{Kroll}$	  	& $3.24 \pm 0.95$ 	& $5.86 \pm 0.58$    & $-7.49 \pm 0.72$    &  $5.13 \pm 0.58$   & $-6.86 \pm 0.69$ \\
\hline
\hline
 \hhhv  \hhhw $\chi^2/N_{\rm pts}$      &    $1280/1366$         & $1309/1366$   & $1289/1366$    & $1292/1366$ & $1286/1366$\\
 \hhhw Probability               &        85.9\%          & 70.6\%        & 82.5\%         & 80.1\% &  83.5\% \\
\hline
\hline
\end{tabular}
} 	%{\scriptsize
\end{center}
\end{minipage}
\caption {
\label{Table:T14}  Isospin breaking effects within EBHLS$_2$ Using Condition $C$
to relate $\hvar$ and $\epspr$. See text for definitions and notations. The entry
for $[m_{dd}^2- m_{uu}^2]$ is in GeV$^2$.
The main fit properties are reminded at the bottom end of the Table.
}
\end{table}
\section{Isospin Breaking Effects in the  $\pi^0-\eta-\etp$ System}
\label{IB-kroll}
\indentB
Table \ref{Table:T14} collects the main EBHLS$_2$ results related to Isospin breaking 
effects in the $[\pi^0,\eta,\etp]$ mixing. 
In contrast with Section \ref{Mixing-param},  the parameter values 
returned by the different solutions may be very different and, then, 
averaging can often be misleading. On the other hand, to our knowledge, there is 
very limited number of external evaluations of these parameters to compare with.

Regarding  the $[m_{dd}^2- m_{uu}^2]$ evaluations, they are all much larger 
than the estimates based on meson masses we sketched above; whether this is due 
to unaccounted for higher order corrections is unclear; in this case, one may expect
the fit to take them effectively into account to accomodate the data. 
Related with this, fits performed
using Conditions $C^\prime$ ($\hvar=\varpi_0$ and $\epspr=\varpi_0^\prime$)
return  the following piece of information~:
\be
\left \{
\begin{array}{lll}
\displaystyle {\rm Solution} ~~ A_+~~: &\epsilon_0=[2.02 \pm 0.11]\times 10^{-2}~~&,~~
\displaystyle {\rm Prob.~~~} 74.0\% ,~~\\[0.5cm]
\displaystyle {\rm Solution} ~~A_-~~: & \epsilon_0=[2.39 \pm 0.11]\times 10^{-2}~~&,~~
\displaystyle {\rm Prob.~~~} 83.4\%  ,~~
\end{array}
\right .
\label{leut96}
\ee
while the quark mass estimate expects $\epsilon_0 \simeq 1.2\times 10^{-2}$.
So, the picture looks somewhat confusing and may indicate than our evaluations
for $[m_{dd}^2- m_{uu}^2]$ and   $\epsilon_0$   absorb higher order (or
other kinds of)  effects to accomodate the data.

The issue just raised obviously propagates to the evaluations for the $\eta$ and $\etp$
fractions inside the physical $\pi^0$. Here also the values for $\hvar$ and $\epspr$
are also found much larger than expected.
Related with this, Kroll \cite{Kroll:2004} quoted  an estimate for
 $\hvar= [3.1 \pm 0.2]\times 10^{-2}$ coming from a  ratio 
 of\footnote{We don't know about an update of this old result.}  $\Psi(2S) \ra J/\psi P$
 decay widths, in line with our own findings.
 
 Regarding $z_{Kroll}$, our $A_+$ and $A_-$ evaluations are consistent which each other
 up to the sign -- which is the key feature of these solutions;
  its absolute magnitude is found in the $[6\div 8] \% $ range. 
 Finally $f_u$ and $f_d$ are found very close to $f_\pi$ when considering 
 their uncertainty ranges.
 
 Stated otherwise, the picture in the realm of Isospin breaking effects involved
 in the $\pi^0-\eta-\etp$ system provided by phenomenology is somewhat confusing.

\section{Summary and Conclusions}
\label{conclusions}
Three main topics have been addressed in the present paper~: The treatment of $\tau$
dipion spectra, the update of the HVP-LO using global fit methods and the 
mixing properties of the $[\pi^0,\eta,\etp]$ system showing up in the EBHLS$_2$
framework.

\begin{itemize} 
\item Regarding the $\tau$ dipion spectra~:

In the previous version of the broken HLS model -- named BHLS$_2$ \cite{ExtMod7} --
the difficulty of the basic solution (BS) to satisfactorily address  
 the dipion spectrum collected by the Belle Collaboration \cite{Belle}
was noted; it was partly compensated 
by  the Primordial Mixing (PM) of the vector fields  which led to
the so-called Reference Solution (RS). However,  the treatment of the
Belle spectrum -- which carries a statistics larger by a factor of $\simeq 50$ than
Aleph \cite{Aleph} or Cleo \cite{Cleo} -- deserves improvement.
On the other hand, the analysis of the {\it lineshape} of the three $\tau$
dipion spectra clearly shows that there is no tension among them   
 -- as already noted in a previous study \cite{ExtMod2} -- nor with the other 
channels embodied inside our HLS framework, except for the spacelike 
spectra \cite{NA7}.

However, the present analysis clearly shows that the assumption which fits the best 
simultaneously the whole EBHLS$_2$ reference data set -- including the Belle
spectrum -- is slightly more involved than a mere rescaling. 
It is found  that the kinetic breaking mechanism\footnote{A kinetic breaking effect
going beyond the usual 't~Hooft determinant term -- which only provides a correction 
to the singlet term -- has  been proposed by other authors using different Lagrangians,
see  \cite{Masjuan} for instance. } defined 
in Section \ref{thooft-kin} allows  for a fair description of each of 
 the Aleph, Belle and Cleo  dipion spectra and, likewise, for  the whole physics 
 channels included inside the EBHLS$_2$ framework, in particular the pion form factor
 $F_\pi^e(s)$ in the spacelike and timelike regions. 

On the other hand, the relevance of a kinetic breaking term -- involving 
simultaneously  components along the $T_0$, $T_3$ and $T_8$
basis matrices of the canonical  Gell-Mann $U(3)$ algebra -- is also strengthened
by considering  properties related with the $[\pi^0,\eta,\etp]$  system
 as it comes  inside the EBHLS$_2$ framework. This   led us to examine the consequences 
 following from imposing conditions to matrix elements of the axial currents
 as expressed by Kroll \cite{Kroll:2005}~:
 $$<0|J_\mu^q |[q^\prime \overline{q^\prime}](p)>=ip_\mu f_q \delta_{q q^\prime }~~,
       \{ [q \overline{q}], q=u,d,s\}$$ 
 where the various $J_\mu^q$ are the axial currents associated with the leading 
 $[q \overline{q}]$
 terms  occuring in the Fock expansion of the  $[\pi^0,\eta_0,\eta_8]$ bare fields.

Within the EBHLS$_2$ context, these conditions relate the mixing properties 
of the  $[\pi^0,\eta,\etp]$ system
and the $\tau$ dipion spectrum because of the $\pi^0$ meson. More precisely, it is
proven in Section \ref{kroll_brk1}  that  the solutions satisfying the Kroll 
Conditions written 
just above generate non-trivial  correlations between  $F_\pi^\tau(s=0)$ and the 
$\Lambda_1$ parameter traditionally included in EChPT to break the $U(3)$ symmetry of
the chiral Lagrangian \cite{Kaiser_2000,leutw,feldmann_3}
via a sole singlet term $\Lambda_1/2 \pa_\mu \eta_0\pa^\mu \eta_0$.  

As a matter of fact, the Kroll Conditions imposed to the EBHLS$_2$ Lagrangian
relate the breaking of $U(3)$ symmetry in the PS sector and   the violation of CVC 
in the $\tau$ decay which explains the observed  Belle spectrum;  
this CVC violation is invisible
 in the Aleph and Cleo spectra because of their lower statistics but our fits
 illustrates that Aleph and Cleo absorb it quite naturally as obvious
 from Table \ref{Table:T0}.

 It is clear that this unexpected property  deserves confirmation and a 
 forthcoming high  statistics  $\tau$ dipion spectrum is welcome to answer this question.
 On the other hand, the picture which emerges from EBHLS$_2$ indicates that
 using $\tau$ data to estimate the Isospin breaking effects involved in  
 $F_\pi^e(s)$ is not straigthforward outside a global fit context.

\end{itemize}

\begin{itemize}
\item The EBHLS$_2$ update of the muon HVP-LO raises several topics~:
	\begin{enumerate}
	\item Using the EBHLS$_2$ model, 
	one examines the two recently published data samples of interest
	in the HLS energy range ($\le 1.05$ GeV).
	
	The BESIII $e^+ e^- \ra \pi^0\pi^+ \pi^-$  cross-section  \cite{BESIII_3pi}
	is important as it doubles the statistics covering this annihilation
	channel.  Once the energies of this spectrum are appropriately\footnote{
	It is shown that this results in mere energy shifts, however different 
	in the $\omg$ and $\phi$  peak regions; both values are found consistent with 
	BESIII expectations. A possible origin for this difference is discussed in 
	Appendix \ref{fred_isr}.}  recalibrated to match the energy scale of the 
	($> 50$) data samples already included in our standard sample set, it is 
	shown that the EBHLS$_2$  framework leads to fairly good  global fit properties. 
	
	The SND Collaboration running on the new VEPP-2000 Facility has produced a new
	spectrum \cite{SND20}  for the $e^+ e^- \ra \pi^+ \pi^-$  cross-section covering 
	the HLS energy range which may allow to readdress the KLOE--Babar controversy.
	
	{\bfEPJC
	Indeed, comparing the SND spectrum properties with those of the samples
	already belonging to our Reference Benchmark gives the opportunity to 
	re-emphasize our sample analysis method.
	 
	 Substantially, our approach is based on a few salient properties~: 
	 {\bf (i)} one sticks to using in fits only the uncertainty information provided by each 
	 experiment and refrain from using any additional input like error inflation factors; 
	 {\bf (ii)} one treats canonically the normalization uncertainty \cite{ExtMod5};
	 {\bf (iii)} preliminary fits allow to identify  the Reference Benchmark data samples
	 by their satisfactory fit properties; the Reference Benchmark is found to
	  include more than 90\% of the available data samples covering all the channels
	 addressed by EBHLS$_2$.  
	 
	 Then, any newcomer   sample is appended to the statistically consistent
	 Reference Benchmark within a global fit: If its fit quality is satisfactory, 
	 it becomes part of the  Reference Benchmark; otherwise, having detected 
	 inconsistencies between the newcomer and the Reference Benchmark samples, 
	 one discards the newcomer, preserving this way the statistical 
	 consistency of the Reference Benchmark.
	}
	
	The outcome can be summarized as follows~: Naming   ${\cal H}$ the set of
	all reference data samples except for the dipion spectra from KLOE and BaBar, 
	it is shown that the most consistent combinations one can define are 
	${\cal H}_K =\{{\cal H} + {\rm KLOE}\}$ and
	 ${\cal H}_B =\{{\cal H} + {\rm BaBar}\}$, the goodness of fit  clearly favoring 
	  ${\cal H}_K$ compared to ${\cal H}_B$; moreover, the goodness of fit for
	 each of ${\cal H}_K$ and ${\cal H}_B$ is much better than those
	 for ${\cal H}_{KB} =\{{\cal H} + {\rm KLOE + BaBar} \}$. To deal with the
	 muon HVP-LO issue, this remark has led us 
	 to perform separate analyses for ${\cal H}_K$ and  ${\cal H}_B$ and
	  avoid using ${\cal H}_{KB}$ which returns a poor probability and is found to produce significant 
	 biases compared to either of  ${\cal H}_K$ and ${\cal H}_B$.
	 This is further commented below.
	 
	 On the other hand, when a new data sample covering the $e^+ e^- \ra \pi^+ \pi^-$ annihilation
	 channel is published, the issue is always to re-examine whether it may favor one among
	 the  ${\cal H}_K$ and  ${\cal H}_B$ sample combinations or not. It has been found previously
	 that the dipion spectra referred to above as NSK and Cleo-c do not substantially 
	 modify the fit picture of either of  the ${\cal H}_K$ and ${\cal H}_B$ combinations; the
	 BESIII sample -- recently  corrected \cite{BESIII-cor} -- is reported to rather favor
	 ${\cal H}_K$ over ${\cal H}_B$, but nothing really conclusive. The question
	 is thus whether the SND spectrum \cite{SND20} modifies the picture. The main results
	 of our study are  gathered\footnote{The $0.60\div 0.71$ GeV  region of the BaBar spectrum
	 has been eliminated from the fit to keep close to what SND suggests \cite{SND20} whereas
	 no cut has been applied  in the global fit involving the KLOE spectra; this obviouly
	 enhances the probability of the  (SND20+Babar) combination.}
	  in Figure \ref{SND20_vs_all}; the fit properties displayed
	 therein  indicate a better consistency with the ${\cal H}_K$ combination over 
	 the ${\cal H}_B$ one, however, there is still something unclear with the reported
	 SND uncertainty information -- or its dealing with -- which, for now, 
	 leads  to use it only to estimate systematics.

	 To end up with this topic, one should note that our reference set of data samples
	 contains 1366 pieces of information. Beside the data samples covering the 6 annihilation
	 channels already listed and the $\tau$ dipion spectra, one finds the partial widths for the
	 $P \ra \gam \gam$ decays and the $VP\etp$ couplings not involved in the listed annihilation
	 channels; the PDG value for the ratio of decay constants  $f_K/f_\pi$ is also included.
	 Stated otherwise, EBHLS$_2$ treats consistently the largest set of data and physics channels
	 ever submitted to a unified description in the non-perturbative QCD region. Global fits
	 have been performed under various conditions and return probabilites in the range of 80\% to 
	 90\% for the ${\cal H}_K$ set combination and $\simeq 40\%$ for the  ${\cal H}_B$  combination;
	 the results based on  ${\cal H}_K$ are commented in Section \ref{global-fits} 
	 and their results gathered in Table \ref{Table:T3}, those based on the  ${\cal H}_B$  combination
	 are the matter of Subsection \ref{HVP-BaBar}.
	 
	 \item Regarding model dependence of the HLS estimates for the muon HVP-LO~:
	 
	 In order to  figure out possible model dependence effects, the most appropriate
	 approach  is to compare the information derived from our fits with the corresponding
	 information derived by others using  so-called "direct numerical integration" methods 
	 -- which  are also far from being free of assumptions.
	 
	 Table \ref{Table:T1}  collects the numerical  estimates  for $a_\mu(\pi \pi)$
	 over the range $s \in [0.35,0.85]$ GeV$^2$
	 derived by the KLOE Collaboration itself \cite{Anastasi:2017eio} for the different
	 data samples they published (KLOE08, KLOE10, KLOE12) and their combination (KLOE85).
	 Including each of these samples as single representative of the $\pi \pi$ channel
	 within the EBHLS$_2$ fitting procedure, one gets the numbers displayed in the
	 second data column with fit properties shown in the third data column. Except for KLOE08
	 which yields a poor goodness of fit, each "experimental" central value is distant
	 from its  EBHLS$_2$ analog by only a fraction of the relevant $\sigma_{exp}$; moreover, the
	 gain in precision by performing
	 global fitting is here especially striking as the uncertainty of the fitted
	 $a_\mu(\pi \pi)$ values is significantly smaller than their corresponding $\sigma_{exp}$'s. 
	 %{\it A contrario},
	 %the results for KLOE08 should also be emphasized as they clearly illustrate how
	 %keeping poorly fitted data samples  generates uncontrolled biases. 
	 
	 Comparing different methods of combining data  is the subject of Table \ref{Table:T6}.
	 This illustrates that, besides the selection of data samples, the way to deal with the
	 reported normalization uncertainty is a much more significant source of bias than the
	 choice of a model  even if ours, by correlating  different channels with $\pi \pi$, 
	 allows for a much improved uncertainty for the $\pi \pi$  contribution -- which is
	  just the purpose for promoting global fit methods.
	 
	 \item 
	 Evaluations of the muon HVP~:  KLOE versus BaBar.
	 
	 The matter of Subsection \ref{HVP-HLS} is to deal with various
	 estimates for the HVP-LO derived from EBHLS$_2$ under various fit conditions. 
	  Table \ref{Table:T7} displays  specifically our results concerning the
	  energy region up to 1.05 GeV and has to be completed with information
	  given in Table \ref{Table:T8} to derive the full HVP-LO. The content 
	  of Table \ref{Table:T7} is associated with using for the fits what was named above 
	  in this Section the $\{ {\cal H} + {\rm KLOE}\}$  sample set. Similarly, 
	  Subsection \ref{HVP-BaBar} provides the analog evaluation based on 
	  using the $\{ {\cal H} + {\rm BaBar} \}$  sample set. One gets for the muon HVP-LO~:
	  \be
	  \left \{
	  \begin{array}{lll}
	  \displaystyle
	  \{ {\cal H} + {\rm KLOE}\} \Longrightarrow 
a_\mu^{HVP-LO} &= 687.48 \pm 2.93_{fit} + \left [^{+2.31}_{-0.69} \right ]_{syst}~~, ~~ 90 \% {\rm ~~Prob.}\\[0.5cm]
	  \displaystyle
	  \{ {\cal H} + {\rm BaBar} \} \Longrightarrow 
a_\mu^{HVP-LO} &= 692.53 \pm 2.95_{fit} + \left [^{+2.31}_{-0.69} \right ]_{syst}~~, ~~ 40 \% {\rm ~~Prob.}
	  \end{array}
	  \right .
	  \nonumber
	  \ee
	  in units of $10^{-10}$. These are displayed together with other estimates in Figure \ref{Fig:amu_hvp_lo}.
	  One observes the strong effect of using $\{ {\cal H} + {\rm BaBar} \}$ preferably to $\{ {\cal H} + {\rm KLOE}\}$
	  despite the better goodness of fit of the latter set. One should note that the $\{ {\cal H} + {\rm BaBar} \}$
	  evaluation of the HVP-LO differs from the KNT19 evaluation \cite{Keshavarzi_2019} by only $0.42\times 10^{-10}$.
	  However, taking into account the $5.47\times 10^{-10}$ difference between the BaBar and KLOE based evaluations,
	  it may look hasardeous to perform any kind of combination of these. 
	  
	  Nevertheless, it looks interesting to quote the results derived from a fit based on
	  solely the ${\cal H}$  sample set; indeed,  ${\cal H}$  only
	  includes the NSK, Cleo-c and BESIII samples as representatives of the $\pi^+ \pi^-$ 
	  annihilation channel for which there is a commonly  shared consensus. One thus gets~:
	  \be
	  \begin{array}{lll}
	  \displaystyle
	   \{ {\cal H} \}~~~ \Longrightarrow 
a_\mu^{HVP-LO} &= 689.43 \pm 3.08_{fit} + \left [^{+2.31}_{-0.69} \right ]_{syst}~~, ~~ 91 \% {\rm ~~Prob.}
	  \end{array}
	  \nonumber
	  \ee
	  \noindent from a fit which also returns $\chi^2/N_{pts}=1137/1231$. This evaluation, 
	  just midway between the 
	  $ \{{\cal H} + {\rm KLOE}\}$ and $\{{\cal H} + {\rm BaBar}\}$ estimates, still 
	  benefits from a very good uncertainty
	  and from a probability as good as those of the  $ \{{\cal H} + {\rm KLOE}\}$  fit.
	  
	  Compared to the average experimental value \cite{FNAL:2021} for $a_\mu$
	    and taking into account the systematic uncertainties, we find for the difference 
	    $\Delta a_\mu=a_\mu^{exp}-a_\mu^{pheno.}$ a significance greater than $5.3 \sigma$ (KLOE) or
	  $4.4 \sigma$ (BaBar). It is worthwhile noting 
	  that the difference of these evaluations is {\it not} a model effect
	  but a pure reflection  of the tension between the BaBar and KLOE evaluations distant
	  by $ 1.9\sigma_{fit}$ from each other.

	 Regarding the hiatus between the  LQCD evaluation \cite{BMW_amu_final} for the muon HVP-LO and 
	 any of the evaluations based on dispersive methods shown in Figure \ref{Fig:amu_hvp_lo}, it looks uneasy yielding
	 a missing  $\delta a_\mu \simeq (10\div 20) \times 10^{-10}$ from annihilation data below 
	 $\simeq 1$ GeV. 	 
	\end{enumerate}
	
    \item Regarding the $[\pi^0,\eta,\etp]$ mixing properties~:
    
    	The EBHLS$_2$ Lagrangian provides a convenient framework to also examine the mixing properties of the 
	$[\pi^0,\eta,\etp]$ system.
    	As this Lagrangian allows to derive the various axial currents, it is possible to construct  explicitely 
	the parametrizations in the so--called octet-singlet and quark-flavor bases. It is found that, at leading order
	in breakings, one recovers the known expressions -- compare to \cite{Kaiser_2000,feldmann_3} for instance
	-- somewhat generalized to also include  the $\lambda_8$ and $\lambda_3$ terms. 
	
	Related with this, it has been found worthwhile to examine in detail how the Kroll Conditions reminded 
	at the beginning of this Section can be fulfilled by the  EBHLS$_2$ Lagrangian. It is found that
	2 solutions -- named $A_\pm$ -- among the 4 possible ones lead to fair  descriptions
	of our whole reference set of data. The $A_+$ and $A_-$ solutions  return similar fit parameter 
	values and the $A_-$ solution is slightly favored compared\footnote{In case 
	 the kinetic breaking term given by Equation (\ref{kin11}) is replaced
	by Equation (\ref{kin11b}),  the Kroll Conditions yield a  unique non-trivial solution.} to $A_+$.
	
	However, an unexpected aspect shows up~:  the kinetic breaking term of the PS fields which is 
	usually 
	a determinant term leading  to solely a PS singlet contribution $\pa_\mu \eta_0 \pa_\mu \eta_0$ 
	cannot come alone and should be complemented by quadratic terms involving also  the 
	$\pi^0$ and $\eta_8$ field derivatives. It follows herefrom that the Kroll Conditions  generate
	 a violation of CVC in the dipion spectrum  of the $\tau$ lepton decay, as already noted. One may expect
	that these conclusions are not specific to the broken HLS modellings.
	
	Using the fit results derived by running the $A_+$ and $A_-$ solutions to the Kroll Conditions,
	the octet-singlet and quark flavor basis parametrization of the $[\pi^0,\eta,\etp]$ mixing  
	are computed  (see Table \ref{Table:T12}) and compared with  other available estimates
	(see Table \ref{Table:T13}). A good agreement is observed with the other estimates with
	however, here also, better precision for the EBHLS$_2$ evaluations.   The Isospin breaking
	effects which can affect the $[\pi^0,\eta,\etp]$ system \cite{Kroll:2005} are also derived
	(see Table \ref{Table:T14}), but here, there is little external information to 
	compare with and conclude.
\end{itemize}
%KKKKK

%\section*{Acknowledgments}
%\indent \indent
%XXXXXXXXXXXXXXXXXX

\clearpage
\section*{\Large{Appendices}}
\appendix
\section{The $AAP$ and $VVP$ Anomalous sectors} 
\label{AAP-VVP}
\subsection{The $AAP$ Lagrangian}
\indent \indent The $AAP$ Lagrangian is given by~:
\be
\displaystyle 
{\cal L}_{AAP}= -\frac{3 \alpha_{em}}{\pi f_\pi}(1-c_4) 
~\epsilon^{\mu \nu \alpha \beta} \pa_\mu A_\nu \pa_\alpha A_\beta \mathrm{Tr} \left [ Q^2 P \right ]~~,
\label{AA20}
\ee
where $Q$ is the quark charge matrix and $P$ denotes the $U(3)$ symmetric matrix
of the bare pseudoscalar fields. Let us define~:
\be
\begin{array}{lll} 
\displaystyle 
\lambda_0^\prime= \left[ \lambda_3 +   \lambda_0 \sqrt{\frac{2}{3}}  ~\frac{5 z_A^2+1}{3 z_A^2} \right]~~,
&\displaystyle 
\lambda_8^\prime=  \left[\lambda_3 +   
\lambda_8 \frac{1}{\sqrt{3}} \frac{5 z_A^2-2}{3 z_A^2}\right] 
\end{array}
\label{AA22}
\ee
and the angle $\delta_P$~:
\be
\left \{
\begin{array}{ll}
\sin{\theta_P}= \displaystyle \frac{1}{\sqrt{3}}
\left ( \cos{\delta_P} + \sqrt{2} \sin{\delta_P}
\right ) \\[0.5cm]
\cos{\theta_P}= \displaystyle \frac{1}{\sqrt{3}}
\left ( \sqrt{2} \cos{\delta_P} -  \sin{\delta_P}
\right ) 
\end{array}
\right . 
\label{AA25}
\ee
which measures the departure
from ideal mixing ($\theta_I=\arctan{1/\sqrt{2}}\simeq 35^\circ$)~:
$ \delta_P =\theta_P - \theta_I$. 
Defining~:
\be
\left \{
\begin{array}{lll} 
\displaystyle g_{\pi^0\gam \gam}=&\displaystyle  ~~\frac{1}{6} \left \{
1 - \frac{5}{6}\Delta_A + \frac{\lambda_3}{2} \left[\lambda_3-\lambda_0^\prime-\lambda_8^\prime\right ]
\right \} \\[0.5cm]
~& +\displaystyle 
 \frac{\hvar}{6\sqrt{3}}
\left \{\frac{5z_A-2}{3z_A}\cos{\theta_P}-\sqrt{2}\frac{5z_A+1}{3z_A}\sin{\theta_P}
\right\} \\[0.5cm]
~ & \displaystyle + \frac{\hvar^\prime}{6\sqrt{3}}\left \{\frac{5z_A-2}{3z_A}\sin{\theta_P}+
\sqrt{2}\frac{5z_A+1}{3z_A}\cos{\theta_P}\right\} \,,\\[0.6cm]   %%%%%%%%%%%%%%%%%%
\displaystyle g_{\eta \gam \gam}=&\displaystyle -\frac{\hvar}{6}
+\frac{\cos{\delta_P}}{36 z_A} \left \{ -2\sqrt{2} + \sqrt{3} \lambda_0 \lambda_0^\prime -\sqrt{6} \lambda_8 \lambda_8^\prime
- \frac{5 z_A^2+4}{3 z_A^2} \lambda_0 \lambda_8
\right  \} \\[0.5cm]
~&+\displaystyle \frac{\sin{\delta_P}}{36} \left \{
 3 \Delta_A-10 + \sqrt{6}\lambda_0 \lambda_0^\prime
+ \sqrt{3}\lambda_8 \lambda_8^\prime +\sqrt{2} \frac{10 z_A^2 -1}{3 z_A^2} \lambda_0 \lambda_8
\right \} \,, \\[0.5cm] %%%%%%%%%%%%%%%%%%
\displaystyle g_{\etp \gam \gam}=&\displaystyle -\frac{\hvar^\prime}{6}
\displaystyle 
-\frac{\cos{\delta_P}}{36} \left \{ 
 3 \Delta_A-10 + \sqrt{6}\lambda_0 \lambda_0^\prime
+ \sqrt{3}\lambda_8 \lambda_8^\prime +\sqrt{2} \frac{10 z_A^2 -1}{3 z_A^2} \lambda_0 \lambda_8
\right \} \\[0.5cm]
~&\displaystyle     + \frac{\sin{\delta_P}}{36 z_A} \left \{ 
-2\sqrt{2} + \sqrt{3} \lambda_0 \lambda_0^\prime -\sqrt{6} \lambda_8 \lambda_8^\prime
- \frac{5 z_A^2+4}{3 z_A^2} \lambda_0 \lambda_8\right \}\,,
\end{array}
\right . 
\label{AA26}
\ee
the coupling constants of the physical pseudoscalar fields to a photon pair,
 $\pi^0 \gam \gam$, $\eta \gam \gam $ 
and $\eta^\prime \gam \gam$,
are given by~: 
\be
\displaystyle  G_{P^0 \gam \gam} = -\frac{3 \alpha_{em}}{\pi f_\pi} (1-c_4) g_{P^0 \gam \gam} ~~.
\label{AA24}
\ee

\subsection{The $VVP$ Lagrangian}
\indent \indent The $VVP$ Lagrangian is given by~:

\be
\displaystyle 
{\cal L}_{VVP}= -\frac{3 g^2}{4 \pi^2 f_\pi}~c_3 
~\epsilon^{\mu \nu \alpha \beta} \mathrm{Tr} \left [ \pa_\mu V_\nu \pa_\alpha V_\beta   P \right ]
~~~, ~~~~ \displaystyle C=-\frac{N_c g^2 c_3}{4 \pi^2 f_\pi}\,.
\label{BB1}
\ee

\subsubsection{The $VV\pi$ Lagrangians}
\indent \indent The $VV\pi$ Lagrangians relevant for our phenomenology 
 are given by~:

\be
\hspace{-1.cm}
\begin{array}{ll}
\displaystyle
{\cal L}_{VVP}(\pi^\pm)=&
\displaystyle
\frac{C}{2} \epsilon^{\mu \nu \alpha \beta} \Biggl\{
 \left [
 \left ( 1+\frac{2 \xi_0+\xi_8}{3} \right )
\partial_\mu \omg_\nu^I
+\frac{\sqrt{2}}{3} (\xi_0-\xi_8)
\partial_\mu \phi^I_\nu 
 \right] \times \left[ \partial_\alpha \rho^+_\beta \pi^- +\partial_\alpha \rho^-_\beta \pi^+
 \right]\Biggr\}
 \end{array}
\label{AC1}
\ee
and~:
\be
\hspace{-1.cm}
\begin{array}{ll}
\displaystyle
{\cal L}_{VVP}(\pi^0)=&
\displaystyle \frac{C}{2}
\epsilon^{\mu \nu \alpha \beta} \Biggl\{ 
G_0 \partial_\mu \rho^I_\nu \partial_\alpha \omg^I_\beta 
+G_1 \left [2 \partial_\mu \rho^-_\nu \partial_\alpha \rho^+_\beta +
 \partial_\mu \rho^I_\nu \partial_\alpha \rho^I_\beta +
 \partial_\mu \omg^I_\nu \partial_\alpha \omg^I_\beta
\right]  \Biggr. \\[0.5cm]
~~& \displaystyle \hspace{2. cm} \Biggl . + G_2  \partial_\mu \Phi^I_\nu \partial_\alpha \Phi^I_\beta
+ G_3   \partial_\mu \rho^I_\nu  \partial_\alpha \Phi^I_\beta  \Biggr\}~\pi^0
 \end{array}
\label{AC2}
\ee
where~:
\be
\left\{
%\hspace{-1.cm}
\begin{array}{ll}
\displaystyle G_0=\left [
1-\frac{\lambda_3^2}{2} + \frac{2 \xi_0+\xi_8}{3} +\xi_3 \right] \,,\\[0.5cm]
\displaystyle G_1 = -\frac {1}{4\sqrt{3}} \left [\sqrt{3} \Delta_A 
+\lambda_3 (\sqrt{2} \lambda_0 + \lambda_8) \right] 
+\frac{1}{2} \left [ \hvar^\prime \cos{\delta_P} -  \hvar  \sin{\delta_P} \right ]\,,
\\[0.5cm]
\displaystyle G_2 = 
-\frac{\lambda_3}{2 z_A^2\sqrt{6}} \left [\lambda_0 -\sqrt{2} \lambda_8\right] 
-\frac{1}{z_A \sqrt{2}} \left [ \hvar^\prime \sin{\delta_P} +  \hvar  \cos{\delta_P} \right ]\,,
\\[0.5cm]
\displaystyle G_3 = \frac{\sqrt{2}}{3} (\xi_0-\xi_8)~~.
\end{array}
\right .
\label{AC3}
\ee
As actually, one imposes $\xi_0=\xi_8$, one has $G_3 =0$.
\subsubsection{The $VV\eta$ Lagrangian}
\indent \indent The $VV\eta$ Lagrangian is given by~:
\be
\hspace{-1.cm}
\begin{array}{ll}
\displaystyle
{\cal L}_{VVP}(\eta)=&
\displaystyle \frac{C}{2}
\epsilon^{\mu \nu \alpha \beta} \Biggl\{
K_1 \partial_\mu \rho^-_\nu \partial_\alpha \rho^+_\beta +
 K_2 \partial_\mu \rho^I_\nu \partial_\alpha \rho^I_\beta +
 K_3  \partial_\mu \omg^I_\nu \partial_\alpha \omg^I_\beta +
 K_4  \partial_\mu \Phi^I_\nu \partial_\alpha \Phi^I_\beta  \Biggr. \\[0.5cm]
 ~~& \displaystyle \hspace{2. cm} \Biggl . + 
 K_5  \partial_\mu \omg^I_\nu  \partial_\alpha \Phi^I_\beta +
 K_6 \partial_\mu \rho^I_\nu \partial_\alpha \omg^I_\beta  \Biggr\}~\eta~~.
 \end{array}
\label{AD1}
\ee

Defining~:
\be
\left\{
%\hspace{-1.cm}
\begin{array}{ll}
\displaystyle H_1= \frac{1}{12 z_A} ~\kappa_0 \kappa_8 \,,\\[0.5cm]
\displaystyle H_2 = \frac {1}{12} \left [ \kappa_0^2 -6
 \right]\,, \\[0.5cm]
\displaystyle H_3 = \frac{\sqrt{2}}{12 z_A^3} 
\left [ \kappa_8^2 -2   z_A^2 (3+2 \xi_0+4 \xi_8) \right] \,,
\end{array}
\right .
\label{AD2}
\ee
where~:
\be
\displaystyle \kappa_0= \sqrt{2} \lambda_0 +  \lambda_8
~~~~ {\rm and}~~~~ 
  \kappa_8=  \lambda_0 - \sqrt{2} \lambda_8\,,
\label{AD3}
\ee
the $VV\eta$ couplings are~:
\be
\hspace{-1.cm}
\left\{
\begin{array}{ll}
\displaystyle K_1= 2 \left [ H_1 \cos{\delta_P} + H_2 \sin{\delta_P}
 \right] \,,\\[0.5cm]
\displaystyle K_2 = H_1 \cos{\delta_P} + (H_2 - \xi_3) \sin{\delta_P}
 \,, \\[0.5cm]
\displaystyle K_3 = H_1 \cos{\delta_P} + 
\left[ H_2 - \frac{2 \xi_0+\xi_8}{3} \right] \sin{\delta_P}
 \,, \\[0.5cm]
\displaystyle K_4 = H_3 \cos{\delta_P} + \frac{\sqrt{2}}{z_A} H_1 \sin{\delta_P}
 \,, \\[0.5cm]
\displaystyle K_5 = 
-\frac{(\xi_0 -\xi_8)}{3 z_A} \left [ 2 \cos{\delta_P} + z_A \sqrt{2} \sin{\delta_P}
 \right]\,,
 \\[0.5cm]
\displaystyle K_6 = 
\frac{\lambda_3 \kappa_8}{2 z_A\sqrt{3}}  \cos{\delta_P}
+\frac{1}{2 \sqrt{3}} \left[\sqrt{3} \Delta_A + \lambda_3 \kappa_0
\right] \sin{\delta_P} - \hvar ~~.
\end{array}
\right .
\label{AD4}
\ee
\subsubsection{The $VV\etp$ Lagrangian}
\indent \indent The $VV\etp$ Lagrangian is given by~:
\be
\hspace{-1.cm}
\begin{array}{ll}
\displaystyle
{\cal L}_{VVP}(\etp)=&
\displaystyle \frac{C}{2}
\epsilon^{\mu \nu \alpha \beta} \Biggl\{
K^\prime_1 \partial_\mu \rho^-_\nu \partial_\alpha \rho^+_\beta +
 K^\prime_2 \partial_\mu \rho^I_\nu \partial_\alpha \rho^I_\beta +
 K^\prime_3  \partial_\mu \omg^I_\nu \partial_\alpha \omg^I_\beta +
 K^\prime_4  \partial_\mu \Phi^I_\nu \partial_\alpha \Phi^I_\beta  \Biggr. \\[0.5cm]
 ~~& \displaystyle \hspace{2. cm} \Biggl . + 
 K^\prime_5  \partial_\mu \omg^I_\nu  \partial_\alpha \Phi^I_\beta +
 K^\prime_6 \partial_\mu \rho^I_\nu \partial_\alpha \omg^I_\beta  \Biggr\}~\etp \,,
 \end{array}
\label{AE1}
\ee

the $VV\etp$ couplings are~:
\be
\hspace{-1.cm}
\left\{
\begin{array}{ll}
\displaystyle K^\prime_1= 2 \left [- H_2 \cos{\delta_P} + H_1 \sin{\delta_P}
 \right] \,,\\[0.5cm]
\displaystyle K^\prime_2 = -(H_2 - \xi_3)\cos{\delta_P} + H_1  \sin{\delta_P}
  \,,\\[0.5cm]
\displaystyle K^\prime_3 = -\left[ H_2 - \frac{2 \xi_0+\xi_8}{3} \right]  \cos{\delta_P} + H_1 \sin{\delta_P}
 \,, \\[0.5cm]
\displaystyle K^\prime_4 = H_3 \sin{\delta_P} - \frac{\sqrt{2}}{z_A} H_1 \cos{\delta_P}
 \,, \\[0.5cm]
\displaystyle K^\prime_5 = 
-\frac{(\xi_0 -\xi_8)}{3 z_A} \left [- z_A \sqrt{2} \cos{\delta_P} +2 \sin{\delta_P}
 \right]
 \,,\\[0.5cm]
\displaystyle K^\prime_6 = 
-\frac{1}{2 \sqrt{3}} \left[ \sqrt{3} \Delta_A + \lambda_3 \kappa_0 \right]  \cos{\delta_P}
+ \frac{\lambda_3 \kappa_8}{2 z_A\sqrt{3}}  
\sin{\delta_P} -\hvar^\prime~~.
\end{array}
\right .
\label{AE2}
\ee
\section{The $APPP$ and $VPPP$ Anomalous sectors} 
\label{APPP-VPPP}
\subsection{The APPP Lagrangian}
\indent \indent The $APPP$ Lagrangian is given by~:
\be
\displaystyle 
{\cal L}_{APPP}= D
~\epsilon^{\mu \nu \alpha \beta} A_\mu
\mathrm{Tr} \left [Q \pa_\nu P \pa_\alpha   P \pa_\beta   P\right ]
~~, ~~ \displaystyle 
D = -i\frac{N_c e}{3 \pi^2 f_\pi^3} \left[1-\frac{3}{4}(c_1-c_2+c_4)\right]~~.
\label{CC1}
\ee
Limiting oneself to the Lagrangian pieces relevant for our purpose, 
it can be written~:
\be
\displaystyle 
{\cal L}_{APPP}= D
\epsilon^{\mu \nu \alpha \beta} A_\mu
\left \{
g_{\gamma \pi^0} \partial_\nu \pi^0 +
g_{\gamma \eta} \partial_\nu \eta +
g_{\gamma \etp} \partial_\nu \etp 
\right \} \partial_\alpha \pi^-  \partial_\beta \pi^+ 
\,,
\label{CC2}
\ee
 in terms of fully renormalized PS fields. Defining~:
\be
\displaystyle \kappa_0= \sqrt{2} \lambda_0 +  \lambda_8
~~~~ {\rm and}~~~~ 
\label{CC3}  \kappa_8=  \lambda_0 - \sqrt{2} \lambda_8\,,
\ee
the couplings can be written~:
\be
\hspace{-1.cm}
\left \{
\begin{array}{ll}
\displaystyle g_{\gamma \pi^0} = -\frac{1}{4} 
\left [ 1 - \frac{\Delta_A}{2} -\frac{\lambda_3}{2\sqrt{3}} \left(\kappa_0 
+\sqrt{3}\lambda_3 \right )
-\hvar\sin{\delta_P} +\hvar^\prime \cos{\delta_P}
\right] \,,\\[0.5cm]
 \displaystyle g_{\gamma \eta} = -\frac{\kappa_0 + \sqrt{3} \lambda_3}{24 z_A} 
 \left[ \kappa_8 \cos{\delta_P}
+ z_A  \kappa_0 \sin{\delta_P}
\right] +\frac{\sin{\delta_P}}{4} \left( 1 -\frac{\Delta_A}{2} \right) + 
\frac{\hvar}{4}\,,\\[0.5cm]

\displaystyle g_{\gamma \etp} = ~~ \frac{\kappa_0 + \sqrt{3} \lambda_3}{24 z_A} \left[z_A
\kappa_0 \cos{\delta_P} - \kappa_8   
 \sin{\delta_P}
\right] - \frac{\cos{\delta_P}}{4} \left( 1 -\frac{\Delta_A}{2} \right) +
\frac{\hvar^\prime}{4}~~.
\end{array}
\right.
\label{CC4}
\ee

\subsection{The VPPP Lagrangian}
\label{VVPLag}
\indentB The $VPPP$ Lagrangian is given by~:
\be
\displaystyle 
{\cal L}_{VPPP}= E
~\epsilon^{\mu \nu \alpha \beta} 
\mathrm{Tr} \left [V_\mu \pa_\nu P \pa_\alpha   P \pa_\beta   P\right ]
~~, ~~ \displaystyle 
E = -i\frac{N_c g}{4 \pi^2 f_\pi^3} \left[c_1-c_2-c_3\right]~~.
\label{DD1}
\ee
Its relevant part can be rewritten~:
\be
\displaystyle 
{\cal L}_{VPPP}= E
~\epsilon^{\mu \nu \alpha \beta} \sum_{V = (\rho^I,\omg^I,\phi^I)}
V_\mu   \left \{
g_{V \pi^0} \partial_\nu \pi^0 +
g_{V \eta} \partial_\nu \eta +
g_{V \etp} \partial_\nu \etp 
\right \} \partial_\alpha \pi^-  \partial_\beta \pi^+ 
\,,
\label{DD2}
\ee
in terms of ideal vector fields and fully renormalized PS fields.
The corresponding couplings are~:
\be
\hspace{-1.cm}
\left \{
\begin{array}{ll}
\displaystyle g_{\rho^I \pi^0} =~~ \frac{1}{4} 
\left[ \frac{\Delta_A}{2}  + \frac{1}{2\sqrt{3}} \lambda_3  \kappa_0 
+\hvar \sin{\delta_P}-\hvar^\prime \cos{\delta_P}
\right] \,,\\[0.5cm]
 \displaystyle g_{\rho^I  \eta} = -\frac{\kappa_0}{24 z_A} \left[\kappa_8  \cos{\delta_P}
+ z_A  \kappa_0\sin{\delta_P}
\right] +\frac{\sin{\delta_P}}{4} \left( 1+\xi_3 \right) \,,\\[0.5cm]
 \displaystyle g_{\rho^I  \etp} = ~~ \frac{\kappa_0}{24 z_A} \left[z_A
\kappa_0 \cos{\delta_P} 
- \kappa_8    \sin{\delta_P}
\right] - \frac{\cos{\delta_P}}{4} \left( 1+\xi_3 \right)\,,
\end{array}
\right.
\label{DD3}
\ee
and~:
\be
\hspace{-1.cm}
\left \{
\begin{array}{ll}
\displaystyle g_{\omg^I \pi^0} = -\frac{3}{4} \left [1 
- \frac{\lambda_3^2}{2} + \frac{2 \xi_0 +\xi_8}{3} \right] \,,\\[0.5cm]

 \displaystyle g_{\omg^I  \eta} = -\frac{\lambda_3 \sqrt{3}}{8 z_A} 
 \left[ \kappa_8  \cos{\delta_P} + z_A  \kappa_0\sin{\delta_P}
\right] - \frac{3 \Delta_A}{8} \sin{\delta_P} + \frac{3 }{4} \hvar \,,\\[0.5cm]

 \displaystyle g_{\omg^I  \etp} = ~~\frac{\lambda_3 \sqrt{3}}{8 z_A} \left[z_A
\kappa_0 \cos{\delta_P} - \kappa_8    \sin{\delta_P}
\right] + \frac{3 \Delta_A}{8} \cos{\delta_P}+ \frac{3 }{4} \hvar^\prime~~.
\end{array}
\right.
\label{DD4}
\ee

Finally, we also have~:
\be
\hspace{-1.cm}
\begin{array}{lll}
\displaystyle g_{\phi^I \pi^0} = -\frac{\sqrt{2}}{4} \left ( \xi_0-\xi_8 \right )~~,&
\displaystyle g_{\phi^I \eta} = 0~~,&
\displaystyle g_{\phi^I \etp} = 0~~.
\end{array}
\label{DD5}
\ee
It should be reminded that the condition $F_\pi^e(0)=1+{\cal O}(\delta^2)$ leads to $\xi_0=\xi_8$.
\section{The $e^+e^- \ra \pi^0\pi^+\pi^-$ Cross-section}
\label{3pionMod}
\indent \indent  The amplitude for the  $\gamma^* \ra \pi^0\pi^+\pi^-$ transition
involves most of the FKTUY Lagrangian pieces; it can be written~:
\be
 \displaystyle
 T(\gamma^* \ra \pi^0\pi^+\pi^-)=T_{APPP}+T_{VPPP}+T_{VVP}~~,
\label{eq2-22}
\ee
 labeling each term  by the particular piece of the FKTUY Lagrangian
from which it originates. As already noted, because  $c_3=c_4$ is assumed,
there is no $T_{AVP}$ piece.

The $T_{APPP}$ contribution to the full  $T(\gamma^* \ra \pi^0\pi^+\pi^-)$   is~:
\be
\hspace{-1cm}
\begin{array}{lll}
\displaystyle T_{APPP}= C_{APPP}
\left [1  -G(\delta_P) \right ]
\epsilon_{\mu \nu \alpha \beta} ~~\epsilon_\mu(\gamma) p_0^\nu, 
p_-^\alpha p_+^\beta ,&  \displaystyle
C_{APPP}= -\frac{ie}{4 \pi^2 f_\pi^3}\left [ 1-\frac{3}{4} (c_1-c_2+c_4) \right ]~~,
\end{array}
\label{eq2-23}
\ee
where $\epsilon_\mu(\gamma)$ is the off-shell photon polarisation vector, the
other notations being obvious. One has also defined~:
\be
\begin{array}{llll}
\displaystyle G(\delta_P)=  \left [\frac{\Delta_A}{2}+
\epsilon \sin{\delta_P} -\epsilon^\prime \cos{\delta_P} \right ] + \frac{\lambda_3}{2\sqrt{3}}
(\sqrt{3}\lambda_3 +\sqrt{2}\lambda_0+\lambda_8)
\epo
\end{array}
\label{eq2-24}
\ee

Three pieces are coming from the $VPPP$~:
\be
\hspace{-0.5cm}
\begin{array}{ll}
\displaystyle T_{VPPP}=& \displaystyle C_{VPPP}
\left [ \sum_{V=\rho,\omg,\phi} \frac{F^e_{V\gamma}(s)}{D_V(s)}  g_{V\pi}^R (s) \right ]
\epsilon_{\mu \nu \alpha \beta} ~~\epsilon_\mu(\gamma) p_0^\nu p_-^\alpha p_+^\beta\,,
\end{array}
\label{eq2-25}
\ee
where the renormalized vector couplings   $g_{V\pi}^R (s)$ to 3 pions have been
derived using the vector relation~:
\be
g_{V \pi}^R (s) ={\cal R }(s) g_{V \pi}^I~,
\label{eq2-25b}
\ee
${\cal R }(s)$ being the matrix given in Eq. (38) or in Eq. (43) of the BHLS$_2$
companion paper \cite{ExtMod7}
 depending on whether the Primordial Breaking is discarded or not. The components
of the $g_{V \pi}^I$ vector which refers to the coupling of the ideal vector field combinations
are~:
\be
\hspace{-0.5cm}
\left \{
\begin{array}{ll}
\displaystyle g_{\rho \pi}^I=\frac{1}{4} 
\left [\frac{\Delta_A}{2}+ \epsilon \sin{\delta_P} -\epsilon^\prime \cos{\delta_P} 
+\frac{\lambda_3}{2\sqrt{3}} (\sqrt{2}\lambda_0+\lambda_8) \right ]\,,\\[0.5cm]
\displaystyle g_{\omg \pi}^I=-\frac{3}{4} 
\left [1 +\frac{2\xi_0+\xi_8}{3} -\frac{\lambda_3^2}{2}
\right ]\,,\\[0.5cm]
\displaystyle g_{\phi \pi}^I=-\frac{\sqrt{2}}{4} (\xi_0-\xi_8) ~~.
\end{array}
\right .
\label{eq2-25c}
\ee

The $V-\gamma$ amplitudes $F^e_{V\gamma}(s)$ and the inverse 
$\rho$  propagators have been constructed
in Section 11 of \cite{ExtMod7}. The inverse propagators for the  $\omg$ 
and $\phi$ mesons have been discussed and defined in  Section 9 of the same Reference. 
One has also defined~:
\be
\displaystyle
C_{VPPP}= -\frac{3ige}{4 \pi^2 f_\pi^3} 
\left [c_1-c_2-c_3 \right ]\epo
\label{eq2-26}
\ee

 The $VVP$ Lagrangian piece in Equation (\ref{AC2}) given in terms of 
 ideal vector fields has to be reexpressed in terms of their renormalized partners 
 as developped in Section 12 of \cite{ExtMod7} -- see Equations (70-75) therein.
The simplest way to write down $T(\gamma^* \ra \pi^0 \pi^+ \pi^-)$  in a way
easy to code within our global fit procedure is displayed just below.

One first defines the $H_i(s)$ functions~:
 \be
\hspace{-1cm}
\left \{
\begin{array}{llll}
\displaystyle  H_0(s) =1  \,, & \displaystyle H_1(s) =  \frac{1}{D_\rho(s_{+-})} +
\frac{1}{D_\rho(s_{0+})} + \frac{1}{D_\rho(s_{0-})}\,,~\\[0.5cm]
\displaystyle H_2(s)=\displaystyle \frac{1}{D_\rho(s_{+-})}~~,&
\displaystyle H_3(s)=\displaystyle
\widetilde{\alpha} (s_{+-}) 
\left [\frac{1}{D_\rho(s_{+-})} - \frac{1}{D_\omg(s_{+-})} \right ]\,, 
\end{array}
\right.
\label{eq2-27}
\ee
where $s$ is the incoming squared energy and the $s_{ij}$'s indicate the invariant mass squared of the
corresponding outgoing $(i,j)$ pairs; the tilde mixing angles are those defined by Equation (43) in
\cite{ExtMod7}. $T_{VVP}$ depends on the 3 functions $(H_i(s), ~i=1~\cdots 3)  $
with the $s$-dependent coefficients $F_i(s)$ given below.

Collecting all terms, the full amplitude writes~:
 \be
%\hspace{-1cm}
\displaystyle T(\gamma^* \ra \pi^0 \pi^+ \pi^-)=
\left [F_0(s) H_0(s)+ C_{VVP}\sum_{i=1\cdots 3}  F_i(s) H_i(s) \right ]
\epsilon_{\mu \nu \alpha \beta} ~~\epsilon_\mu(\gamma) p_0^\nu p_-^\alpha p_+^\beta\,,
\label{eq2-28}
\ee
with~:
 \be
 \displaystyle C_{VVP}= -i \frac{3 e g m^2 }{8 \pi^2 f_\pi^3} (1+\Sigma_V) c_3\epo
\label{eq2-29}
\ee
In this way,  to write down the full amplitude, the various $F_i(s)$ functions only depend on
the incoming off-shell photon energy squared $s$; the dependence upon the various sub-energies $s_{ij}$
is, instead,  only carried by the $H_i(s)$ functions as clear from Equations (\ref{eq2-27}). One has~:
\be
%\hspace{-1.cm}
\left \{
\begin{array}{llll}
\displaystyle F_0(s)= C_{APPP} \left [1-G(\delta_P)\right ]
+C_{VPPP}\left [\frac{F^R_{\rho\gamma }(s)}{D_\rho(s)}  g_{\rho\pi}^R (s)
+\frac{F^R_{\omg\gamma}(s)}{D_\omg(s)}  g_{\omg\pi}^R (s)+
\frac{F^R_{\phi\gamma}(s)}{D_\phi(s)}  g_{\phi\pi}^R (s)
\right ]
\,,\\[0.5cm]
\displaystyle F_1(s) = \displaystyle
\widetilde{\alpha}(s)\frac{F^R_{\rho\gamma}(s)}{D_\rho(s)}
+ \left [1-\frac{\lambda_3 ^2}{2}
\right ]\left [1+\frac{2\xi_0+\xi_8}{3} \right ]\frac{F^R_{\omg\gamma}(s)}{D_\omg(s)}
+\left [\frac{\sqrt{2}}{3} (\xi_0-\xi_8)  +\widetilde{\gamma}(s)\right ]
\frac{F^R_{\phi\gamma}(s)}{D_\phi(s)}
\,,\\[0.5cm]
\displaystyle F_2(s) = \displaystyle
\left [\epsilon^\prime \cos{\delta_P}-
\epsilon \sin{\delta_P} - \frac{\Delta_A}{2} 
-\frac{\lambda_3}{2\sqrt{3}} (\sqrt{2} \lambda_0+\lambda_8)
\right ] \frac{F^R_{\rho\gamma}(s)}{D_\rho(s)}
+ 2 \xi_3  \frac{F^R_{\omg\gamma}(s)}{D_\omg(s)}
\,, \\[0.5cm]
 \displaystyle F_3(s) = \displaystyle
\frac{F^R_{\rho\gamma}(s)}{D_\rho(s)}
\,,
\end{array}
\right.
\label{eq2-30}
\ee
where $\widetilde{\alpha}(s)=\alpha(s)+\psi_\omg$ and $\widetilde{\gamma}(s)=\gamma(s)+\psi_0$, to 
possibly keep track of the Primordial Breaking \cite{ExtMod7}, $\alpha(s)$ and $\beta(s)$
being the angles generated by the Dynamical Mixing mechanism \cite{ExtMod3,ExtMod7}.

A global fit to all cross-sections
but $e^+e^- \ra \pi^0 \pi^+\pi^-$ allows to yield the relevant  parameters
with a good approximation;  then,  having at hand all ingredients  defining the $F_i(s)$'s,
a first minimization run \cite{ExtMod5} including the $e^+e^- \ra \pi^0 \pi^+\pi^-$ cross-section
can be performed to also derive a first estimate for $c_1-c_2$. The output of this {\sc minuit}
minimization run is then used as input for a next minimization step. This initiates an iteration
procedure involving all cross-sections which is carried on up to convergence --
when some criterion is met, e.g. generally, $\Delta \chi^2 \le 0.3$ -- for the global $\chi^2$
of all the processes involved in the EBHLS$_2$ procedure. 

This method  converges in a couple of minimization steps \cite{ExtMod5}. What makes such a
 minimization procedure unavoidable is that the $e^+e^- \ra \pi^0 \pi^+\pi^-$ cross-section expression implies
to integrate over the $s_{ij}$'s at each step. This is obviously  prohibitively computer time consuming for a negligible gain.
Hence, at each step, one starts by tabulating  coefficient functions,
exhibited in the next expression between curly brackets~:
\be
\hspace{-1cm}
\left .
\begin{array}{ll}
\displaystyle
\sigma(e^+e^- \ra \pi^0 \pi^+\pi^-,s)=&  \displaystyle
\frac{\alpha_{\rm em} ~s^2}{192 \pi^2}
\displaystyle
\times
\left [\left \{ \int G dx dy \right \} |F_0(s)|^2 \right. \crn[0.3cm]&
\left.\hspace*{-3.5cm}
+ C_{VVP}^2 \sum_{i,j=1\cdots 3}  F_i(s) F_j^*(s)
\left  \{ \int G H_i H_j^* dx dy \right  \}
\right .
 \\[0.3cm]
~~~&  \displaystyle  \left . \hspace*{-3.5cm}
+C_{VVP} \sum_{i=1\cdots 3} \left ( F_0(s) F_i^*(s) \left  \{ \int G H_i^* dx dy\right  \}
\displaystyle
+ F_0^*(s) F_i(s) \left  \{ \int G H_i dx dy\right  \}  \right )
\right ]
\end{array}
\right .
\label{eq2-31q}
\ee
\be
\phantom{xxx}
\label{eq2-31}
\ee
and these tables are used all along the next step. Equation (\ref{eq2-31}) uses
the Kuraev-Silagadze parametrization \cite{Kuraev}
and its kernel $G(x,y)$ function; these are reminded in Appendix H of \cite{ExtMod3}.
\section{Energy Shift of Resonances and Secondary ISR Photons}
\label{fred_isr}
\indentB
For the BESIII data sample \cite{BESIII_3pi} of concern here, 
photon radiation effects have been
unfolded  by utilizing the PHOKHARA
code~\cite{Czyz:2013xga}, which also includes second-order photon
emission\footnote{For a discussion of the radiative correction tools
see~\cite{Rodrigo:2001kf,Actis:2010gg}.}. In the leading ISR process,
$\sqrt{s}=3.773$ GeV is the energy at which the primary ISR photon is
emitted. Here we address the emission of a second photon near the
 $\phi$ and $\omega$ resonances, where the corresponding radiation
effect appears resonance-enhanced.

The effect of a photon radiated off a Breit-Wigner (BW) resonance is
well known from the $Z$ resonance physics at LEP
I~\cite{Barroso:1987ae,Bardin:1989qr,Beenakker:1989ec,Jegerlehner:1991ed}. A
specific analysis for the process $e^+e^- \to \pi^+\pi^-$ may also be
found in~\cite{Hoefer:2001mx,Gluza:2002ui}. The leading effect is
due to initial state radiation, where the observed cross-section is
given by a convolution\footnote{Here one only considers the ISR part at
order $\cal{O}(\alpha)$ which yields the main shift between the
observed and the physical peak cross-sections. For a more complete
discussion of photon radiation effects see e.g. Eq.~(39) ff
in~\cite{Hoefer:2001mx}. The final state radiation piece contributing
to the $\pi^+\pi^-\pi^0$ cross-section has been taken into account in
the BES III measurement.}~:
 
\be
\sigma^{\rm obs}(s)=\int_0^{1} dk
\rho_{ini}(k)\;\sigma_{phys}(s(1-k))\:,
\label{fred-1}
\ee
 where $k=E_\gamma/E_{e}$ is
the energy emitted by the photon in units of the electron or positron
energy $E_{e}$ at which the resonance is formed and $s=4E_e^2$. The
photon spectral function is 

\be
\rho_{ini}(k)=\beta k^{\beta-1}\:(1+
\delta_1^{v+s}+ \dots) +\delta_1^h + \cdots\:,
\label{fred-2}
\ee
 with $\beta=\frac{2\alpha}{\pi}\:(L-1)$, $L=\ln \frac{s}{m_e^2}$ and the
photon radiation corrections for the virtual + soft (v+s) and the hard
(h) parts read~:
\ba
\delta_1^{v+s}=\frac{\alpha}{\pi} \left(\frac32
\,L+\frac{\pi^2}{3}-2 \right) \semis \delta_1^h =
\frac{\alpha}{\pi}\:(1-L)(2-k)\:.
\label{fred-3}
\ea

Let us consider the narrow resonances $\omega$ and $\phi$ which are well
parametrized by the Breit-Wigner (BW) formula~:
\be
\sigma_{\rm
  BW}(s)=\frac{12\pi\Gamma_e\Gamma_f}{M_R^2}\frac{s}{(s-M_R^2)^2+M_R^2\Gamma_R^2}\:,
\label{fred-4}
\ee
which has its maximum at $\sqrt{s}=M_R\:(1+\gamma^2)^{\frac14}$ with
$\gamma=\Gamma_R/M_R$.  The peak value is
$\sigma_{max}=\frac{12\pi\Gamma_e\Gamma_f}{M_R^2\Gamma_R^2}\:(1+\frac14
\gamma^2)$ and the half maximum locations read
$(\sqrt{s})_\pm=M_R\:(1+\frac38 \gamma^2)\pm
\frac{\Gamma_R}{2}\:(1-\frac18 \gamma^2)$. The leading photon
radiation modifies these resonance parameters to
\ba
\sqrt{s}_{\rm max}&=& M_R+\frac{\beta \pi \Gamma_R}{8}-\frac14
\gamma^2\:M_R\:,\\
\rho &=&\frac{\sigma^{\rm obs}}{\sigma_{\rm BW}}=\left(\frac{\Gamma_R}{M_R}\right)^\beta\:(1+\delta_1^{v+s})\:,\\
\sqrt{s}_+-\sqrt{s}_-&=&\Gamma_R\:
\left [1-\frac{\pi}{2}\beta\gamma-\frac{5}{8} \gamma^2
+(\frac{\pi}{4}\beta +\frac{\beta}{2}\ln
2)(1+\beta +\frac{\pi}{4}\beta) \right ] \:.
\label{fred-5}
\ea
The physical resonance appears wider with a reduced peak cross-section and
the position of the observed peak is shifted towards higher energies.
Considering the emission of a second photon, $s'=s(1-k)$ in the
convolution formula (\ref{fred-1}) turns into $s''=s'(1-k_2)$ where $s'=s(1-k_1)\sim
M_R^2$, i.e. $s=M_R^2$ in the relations listed before. This second photon
shifts the peak energy of the $\omega$ and $\phi$ by $\delta E$ as given in
Table~\ref{Table:T_shifts}. 

The last data column here displays, correspondingly, the
values for the shifts $\delta E^{\omg}_{BESIII}$ and $\delta E^{\phi}_{BESIII}$
returned by a global EBHLS$_2$ fit involving all data and, especially for the
3$\pi$ channel, both the NSK and the BESIII samples; these fit values differ from
the corresponding calculated ones by only $1\sigma$.

\begin{table}[!h!]
\centering
\caption{
The energy shifts in keV of the resonance locations by local ISR photon
  emission at $s=M_R^2$. We adopted PDG resonance parameter values (in MeV).
  The values for $(\delta E)^{\rm fit}$ are obtained in a global fit
 involving all 3$\pi$ data samples (BESIII+NSK). 
  \vspace{0.5cm}
  }
\label{Table:T_shifts}
\begin{tabular}{|c|rrrr|}
\hline\noalign{\smallskip}
 & $M_R$ & $\Gamma_R$ & $\delta E$& $(\delta E)^{\rm fit}$   \\   
\noalign{\smallskip}\hline\noalign{\smallskip}
  $\omega$        & 782.65&  8.49  & -403   & -486 $\pm$ 72\\
  $\phi$          &1019.46&  4.26  & -213   & -135 $\pm$ 59\\
\noalign{\smallskip}\hline
\end{tabular}
\end{table}

Interestingly the calculated leading order shifts quite
well reproduce the shifts found by optimizing the fits of the
BESIII data  (about a $1\sigma$ distance only for both signals). 
Therefore, the question is whether BESIII has included (or not)
the corresponding corrections when unfolding the raw data from radiation
effects.

%\section{XXXX}
%\label{XX}
%\indent

%\newpage
                   \bibliographystyle{h-physrev}

                     \bibliography{cov19_save}
\end{document}